\documentclass{aa}  

 \pdfoutput=1

\usepackage{graphicx}
\usepackage{txfonts}
\usepackage{lscape}
\usepackage{orcidlink}
\usepackage{hyperref}
\hypersetup{
colorlinks   = true,
urlcolor     = blue,
linkcolor    = blue,
citecolor   = blue
}
\usepackage{rotating}
\usepackage{natbib}
\bibpunct{(}{)}{;}{a}{}{,}

\begin{document}

\title{Lyman continuum leaker candidates among highly ionised, low-redshift dwarf galaxies selected from \ion{He}{II}}
\titlerunning{LCE candidates from \ion{He}{II}}

\author{
\href{https://orcid.org/0000-0002-8719-8069}{A. U. Enders}
\inst{1,2}\,\orcidlink{0000-0002-8719-8069}
\and
\href{https://orcid.org/0000-0001-5126-5365}{D. J. Bomans}
\inst{1,2,3}\,\orcidlink{0000-0001-5126-5365}
\and
\href{https://orcid.org/0000-0002-8173-3438}{A. Wittje}
\inst{1}\,\orcidlink{0000-0002-8173-3438}
}

\institute{Ruhr University Bochum, Faculty of Physics and Astronomy, Astronomical Institute (AIRUB), Universit\"atsstrasse 150, 44801 Bochum, Germany\\
\email{\href{mailto:enders@astro.rub.de}{enders@astro.rub.de}}
\and
Ruhr Astroparticle and Plasma Physics (RAPP) Center, 44780 Bochum, Germany
\and
Ruhr University Bochum, Research Department, Plasmas with Complex Interactions, 44780 Bochum, Germany
}

\date{Received October 7 2022 / Accepted January 31 2023}

\abstract
{Contemporary research suggests that the reionisation of the intergalactic medium (IGM) in the early Universe was predominantly realised by star-forming (proto-)galaxies (SFGs). Due to observational constraints, our knowledge on the origins of sufficient amounts of ionising Lyman continuum (LyC) photons and the mechanisms facilitating their transport into the IGM remains sparse. Recent efforts have thus focussed on the study of local analogues to these high-redshift objects.}
{We aim to acquire a set of very low-redshift SFGs that exhibit signs of a hard radiation field being present. A subsequent analysis of their emission line properties is intended to shed light on how the conditions prevalent in these objects compare to those predicted to be present in early SFGs that are thought to be LyC emitters (LCEs).}
{We used archival spectroscopic SDSS DR12 data to select a sample of low-redshift \ion{He}{II}~4686 emitters and restricted it to a set of SFGs with an emission line diagnostic sensitive to the presence of an active galactic nucleus, which serves as our only selection criterion. We performed a population spectral synthesis with \textsc{fado} to reconstruct these galaxies' star-formation histories (SFHs). Utilising the spectroscopic information at hand, we constrained the predominant ionisation mechanisms in these galaxies and inferred information on ISM conditions relevant for the escape of LyC radiation.}
{Our final sample consists of eighteen ionised, metal-poor galaxies (IMPs). These low-mass ($6.2\leq\log\left(M_\star/\mathrm{M}_\sun\right)\leq8.8$), low-metallicity ($7.54\leq\log{\left(\mathrm{O}/\mathrm{H}\right)}+12\leq8.13$) dwarf galaxies appear to be predominantly ionised by stellar sources. We find large \ion{[O}{III]}~5007/\ion{[O}{II]}~3727 ratios and \ion{[S}{II]}~6717,6731/H$\alpha$ deficiencies, which provide strong indications for these galaxies to be LCEs. At least 40\% of these objects are candidates for featuring cosmologically significant LyC escape fractions $\gtrsim10\%$. The IMPs' SFHs exhibit strong similarities and almost all galaxies appear to contain an old (>1~Gyr) stellar component, while also harbouring a young, two-stage ($\sim$10~Myr and <1~Myr) starburst, which we speculate might be related to LyC escape.}
{The properties of the compact emission line galaxies presented here align well with those observed in many local LCEs. In fact, our sample may prove as an extension to the rather small catalogue of local LCEs, as the extreme interstellar medium (ISM) conditions we find are assumed to facilitate LyC leakage. Notably, all of our eighteen candidates are significantly closer ($z<0.1$) than most established LCEs. If the inferred LyC photon loss is genuine, this demonstrates that selecting SFGs from \ion{He}{II}~4686 is a powerful selection criterion in the search for LCEs.}


\maketitle

\section{Introduction}\label{sec_intro}

About 12.8~Gyr ago, at a redshift of $z\sim6$, the last major phase transition of our Universe's intergalactic medium (IGM) from neutral gas to ionised plasma was complete \citep[e.g.][]{Fan06,McGreer15}. In the period preceding this, known as the Epoch of Reionisation (EoR), the young galaxies at that time must have contained a population of objects that have leaked large amounts of Lyman continuum (LyC) photons ($\lambda<912~\AA$) into intergalactic space in order to account for this transition.

Precisely which astrophysical objects exhibit properties that qualify them as LyC emitters (LCEs, often also referred to as LyC leakers) is a topic of debate in contemporary astrophysical research. While in the low-redshift Universe active galactic nuclei (AGN) readily provide the energy required to explain the IGM's ionised state \citep{Haardt12}, their number density drops steeply with increasing redshift \citep[e.g.][]{McGreer13,Kulkarni19}. Thus, unless a signficant population of low-luminosity AGN existed in the early Universe \citep{Madau15}, star-forming galaxies (SFGs) must have provided the majority of ionising radiation during this era \citep{Robertson15,Mitra18}. Here, some researchers advocate that luminous, massive SFGs play a dominant role \citep{Sharma16,Naidu20,MarquesChaves21,MarquesChaves22a}, but many arguments seem to be in favour of the faint, low-mass class of dwarf galaxies (DGs), as they are the most abundant class of galaxy with properties that may facilitate the transport of LyC photons into the IGM \citep[e.g.][]{Wise14,Finkelstein19}. Ultimately though, the magnitude of their contribution as well as the mechanisms driving the loss of ionising radiation into the IGM remain a topic of debate.

In part, this uncertainty is owed to the poorly understood intrinsic properties of these galactic systems, which in conjunction need to result in a Lyman continuum escape fraction ($f_\mathrm{esc}\left(\mathrm{LyC}\right)$) of some 10 -- 20 \% to provide a plausible scenario for reionisation \citep[e.g.][]{Ouchi09,Robertson13,Stark16,Naidu20}. In young galaxies, most of the intrinsic LyC flux is provided by OB stars, and the total amount of high energy photons can be elevated by an overabundance of these sources due to intense starburst episodes, but likewise due to variations of the initial mass function \citep[IMF, e.g.][]{Hopkins06,Wise09}. As studies of integrated galaxy properties strongly hinge on current stellar models, one needs to take into account that for low-metallicity systems, these models are backed by few to no observations \citep[the latter certainly being true for Population III stars, whose impact on a galaxy's LyC output is significant,][]{Schaerer02,Schaerer03} and consequently introduce a high amount of uncertainty. Parameters governing stellar evolution such as rotation and binarity introduce further uncertainty, as they may significantly alter the spectral energy distribution (SED) of a given star, particularly in the ultra-violet (UV) regime \citep[e.g.][]{Eldridge09}. Additionally, other sources capable of boosting a galaxy's LyC output need to be taken into consideration, with the most prominent contenders being Wolf-Rayet (WR) stars \citep{Schaerer96,Smith02}, high-mass X-ray binaries \citep[HMXBs,][]{Garnett91,Sana12,Schaerer19}, and fast radiative shocks \citep{Garnett91,Dopita96,Thuan05,Plat19}. This list can be expanded by including other sources such as post-AGB stars and the nuclei of planetary nebulae \citep{Binette94}, but these bear less relevance here due to the insufficient evolutionary timescale of these systems.

\begin{figure*}
\sidecaption
\includegraphics[width=12cm]{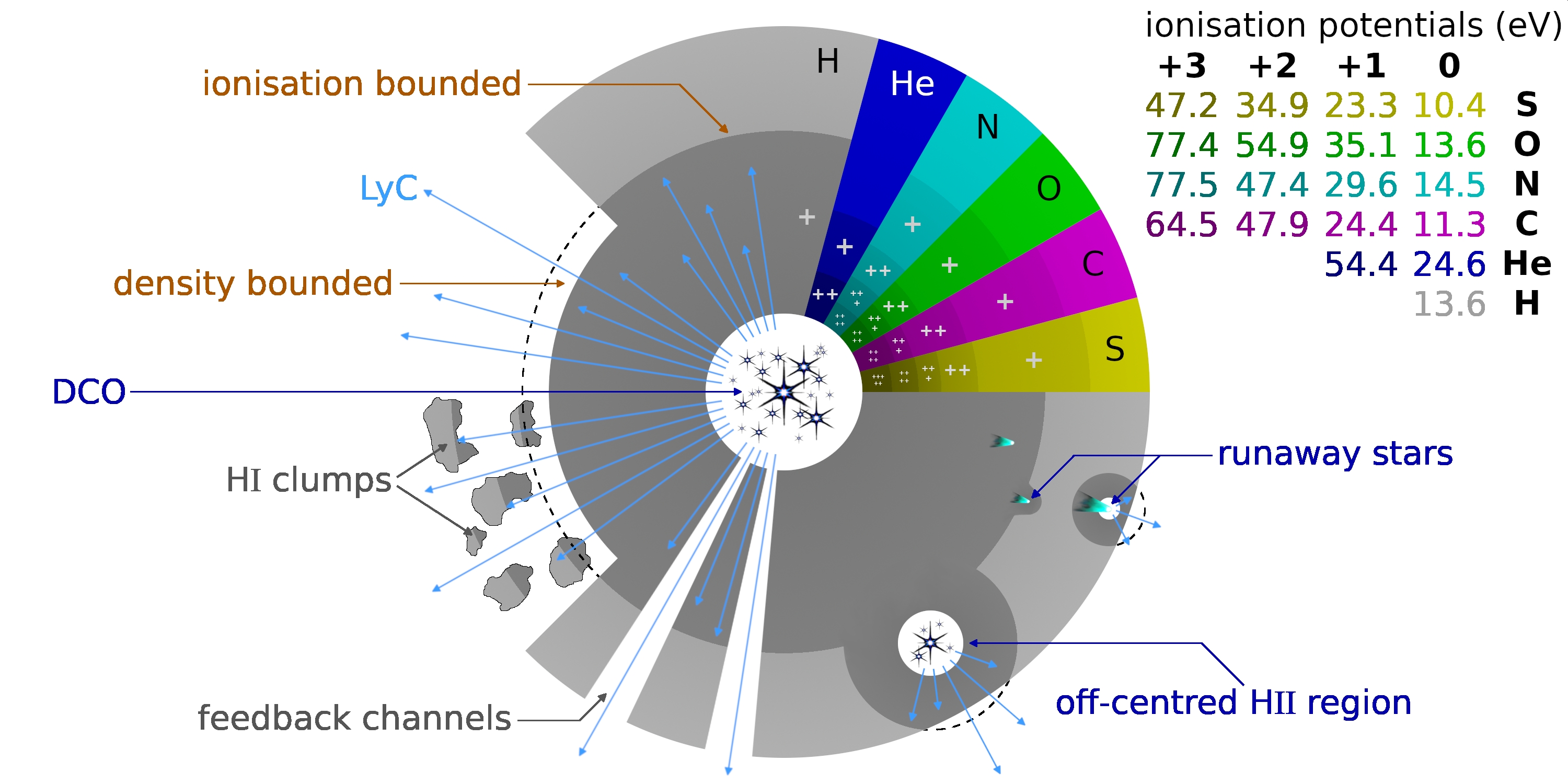}
\caption{Annotated sketch highlighting several scenarios present in an \ion{H}{II} region around a dominant central object (DCO). Light blue arrows represent LyC photons, where an arrow ending within the extent of the \ion{H}{II} region indicates absorption, and an arrow exceeding the region's extent represents escape. The top left quadrant depicts the extent of \ion{H}{I} and \ion{H}{II} for the density- and ionisation-bounded textbook scenarios, whereas the bottom left quadrant is intended to highlight the role of a more complex gas geometry on LyC escape. The bottom right quadrant serves as a reminder that the spatial distribution of ionising sources further influences the observed LyC flux. Finally, the top right quadrant illustrates that the regions occupied by different ions vary in size, mediated by their respective ionisation potentials (an overview of the ionisation energies of a few select atoms is given in the table at the top right).}
\label{fig_lychii}
\end{figure*}

Likewise, our insights regarding the mechanisms facilitating the transport of ionising LyC radiation into the IGM remain sparse. Here, the distribution of the interstellar medium (ISM) in different stages of ionisation certainly plays a significant role in LyC escape, and several plausible models backed by observations have been suggested. Such scenarios include almost entirely density-bounded SFGs \citep{Nakajima14,deBarros16}, a low-density ISM permeated by optically thick HI clumps \citep[the picket-fence model,][]{Heckman11,Gronke16}, or an optically thick ISM riddled by low-density, highly ionised cavities and tunnels carved by supernovae (SNe) and galactic winds \citep{Zackrisson13,Behrens14}. Especially the latter scenario introduces a strong viewing angle dependence, as the ionised tunnels would then have a strong directional preference, mediated by the host galaxy's gravitational potential gradient and thus, its morphology \citep{Zastrow13,Cen15}. We provide a sketch of different LyC escape scenarios in Fig. \ref{fig_lychii}, for illustrative purposes.

Generally speaking, studying the properties of early galaxies remains a challenging observational task, owed to the apparent faintness and the low spatial resolution that can be achieved with present-day telescopes for such high-redshift objects. In particular, the intrinsically faint UV regime of the SED is heavily affected by extinction, and beyond redshifts of $z\sim4$, the detection of LyC photons is rendered highly unlikely due to absorption by neutral hydrogen on the line of sight \citep{Madau14}.

In order to gain further insights into the characteristics of early SFGs, a promising approach to circumvent these problems is achieved by studying local counterparts which share similar properties. Notably, the starbursts among the luminous compact galaxies \citep[LCGs,][]{Izotov11} and compact SFGs \citep[CSFGs, ][]{Izotov21a}, characterised by H$\beta$ equivalent widths (EWs) of EW(H$\beta$)>$100~\AA$, have many commonalities to high-$z$ SFGs, such as low metallicities, low stellar masses, and large specific star-formation rates (sSFRs). The first results from observations with the \textit{James Webb Space Telescope} suggest similar findings for galaxies well in the EoR, further reinforcing the argument that the study of local analogues offers many insights into the high-$z$ Universe \citep[e.g.][]{Trussler22,Endsley22,Schaerer22b,Topping22,Rhoads22}.

If SFGs were the dominant driver of reionisation, we can reasonably expect to find LCEs among such populations of local analogues. Indeed, studies of nearby, low-metallicity DGs have revealed high-ionisation nebular emission lines, such as \ion{O}{III]}, \ion{C}{IV}, or \ion{He}{II} \citep[e.g.][]{Senchyna17,Berg19}, requiring energies well in excess of those necessary to ionise hydrogen ($13.6~\mathrm{eV}$). Not only is this another link to several high-$z$ SFGs \citep[e.g.][]{Sobral15,Stark15,Mainali17}, but also a strong motivator for the search for LCEs among these objects, as a hard radiation field is evidently present. We would like to point out that this connection has been acknowledged before, for instance by \citet{Berg19} and \citet{PerezMontero20}.

Over the last few years, a significant, albeit small sample of low-$z$ galaxies with direct detection of LyC leakage has been compiled  \citep{Bergvall06,Leitet13,Borthakur14,Leitherer16,Izotov16a,Izotov16b,Izotov18a,Izotov18b,Wang19,Malkan21,Izotov21b,Flury22a}. A particularly high detection rate of LyC escape was achieved by \citet{Izotov16a,Izotov16b} among the so-called green pea galaxies \citep[GPs,][]{Cardamone09}, which have been shown to be a subset of the LCGs \citep{Izotov11}.

Whereas many LCEs do show nebular \ion{He}{II}~4686 emission, recent works have demonstrated the absence of a correlation between spectral hardness (as traced by \ion{He}{II}~4686/H$\beta$) and $f_\mathrm{esc}\left(\mathrm{LyC}\right)$ \citep{MarquesChaves22b}. This does not imply, however, that the inverse is true, and drawing a sample of SFGs from high-ionisation emission lines may yield an appreciable fraction of LCEs, which we intend to explore in this paper. This idea is further reinforced by the fact that several spectra of high redshift LCE candidates feature intense high-ionisation UV emission lines \citep[e.g.][]{Schaerer22a,Naidu22,Saxena22}.

It is to be noted that high-ionisation lines are often difficult to reproduce in spectral modelling of SFGs. Emission of \ion{He}{II}~4686, anti-correlated with metallicity \citep{Schaerer19} and utilised in this work, appears to be particularly challenging. Often assumed to be a spectral signature of WR stars, \cite{Shirazi12} found no evidence of WR stars being present in $\sim40\%$ of their \ion{He}{II}~4686 emitting SFG sample. Similar situations arise for example in the works of \cite{Stasinska15} and \cite{Schaerer19}, who find that current stellar models cannot reproduce the observed \ion{He}{II}~4686 intensities, and the inclusion of non-stellar sources such as HMXBs or shocks are necessary to bring their model SEDs in line with observations.

In this work, we present our findings from legacy data of the Sloan Digital Sky Survey \citep[SDSS,][]{Eisenstein11}, adding a total of 18 SFGs to the list of high-$z$ analogues, selected from their \ion{He}{II}~4686 emission lines. Inferred from indirect tracing methods, virtually all of these galaxies are promising candidates to expand the list of local LCEs, and their recent star-formation histories (SFHs) may provide an explanation as to why their LyC radiation can escape into the IGM.

The paper is structured as follows: in Sect. \ref{sec_data}, we describe the dataset and sample selection, while the analysis methods are covered in Sect. \ref{sec_ana}. We present the properties we derived for the galaxies we selected in Sect. \ref{sec_prop}, and summarise our results and augment them by a few concluding remarks in Sect. \ref{sec_conc}.
Throughout this paper, we assume a flat $\Lambda$CDM cosmology with $H_0=67.4~\mathrm{km}~\mathrm{s}^{-1}~\mathrm{Mpc}^{-1}$ and $\Omega_m=0.315$ \citep{Planck20}.

\section{Data}\label{sec_data}

\subsection{Dataset}

In order to obtain a sample of SFGs with \ion{He}{II}~4686 emission, we used the Baryon Oscillation Spectroscopic Survey \citep[BOSS,][]{Dawson13} dataset from the SDSS's third run \citep[SDSS-III,][]{Eisenstein11} in their twelfth data release (DR12), the final data release in the SDSS-III.
The choice of this dataset is motivated by an advantage particular to DR12, as opposed to the more recent data releases, which is found in the augmentation by \citet{Thomas13}. They have published a value-added catalogue derived from fits of model spectra to the data, in which they provide refined line fluxes for the most prominent emission lines amongst other spectroscopic properties. This allows for a fairly efficient scouring of the database for the objects of our interest, whose properties are most readily derived from emission line ratios.

It is to be noted that DGs can be expected to be somewhat underrepresented in this dataset. This is evidently owed to the primary science goal of the BOSS, which has an intentional bias towards luminous, and hence massive galaxies \citep{Dawson13}. However, their observation runs allowed for a fraction of ancillary targets deviating from their main science goal \citep{Dawson13}, among which we deemed it likely to find suitable candidates. Previous works have demonstrated this -- \citet{Guseva17} have used this catalogue to spectroscopically identify 287 metal-deficient DG candidates, and \citet{Yang17} extracted a sample of analogues to high-$z$ Lyman~$\alpha$ (Ly$\alpha$) emitters by photometric selection (dubbed 'blueberry galaxies').

\subsection{Sample selection}

Typically, studies concerning SFGs utilise the presence of \ion{He}{II}~4686 emission as an exclusion criterion \citep[e.g.][]{Izotov11,Guseva17}, as this line is indicative of the presence of an AGN. However, not all \ion{He}{II}~4686 emitters have a significant AGN contribution to their radiation field \citep{Shirazi12}, and as we are primarily interested in finding LyC leaking galaxies, helium with its ionisation potential of $54.4\ \mathrm{eV}$ is chosen as an excellent indicator for the presence of a hard radiation field.

Based on SDSS DR7 data, \citet{Shirazi12} have shown that an AGN contribution to the \ion{He}{II}~4686 emission is negligible (<10\%) if a line ratio relation of
\begin{equation}\label{eqn_SB12}
\log\left(\frac{\ion{He}{II}~4686}{\mathrm{H}\beta}\right)\leq-1.22+\frac{1}{8.92\log\left(\ion{[N}{II]}~6583/\mathrm{H}\alpha\right)+1.32}
\end{equation}

is satisfied. We expect this restriction to be sufficient to ensure the purity of our SFG sample with respect to AGN contamination and thus draw our final sample from Eq. \ref{eqn_SB12} alone. To ensure credibility in the measured emission line fluxes, following \citet{Thomas13}, we required all objects to have an amplitude over noise (AoN) >2 in each of the utilised emission lines.

We note that as opposed to LCGs \citep{Izotov11}, the CSFG sample compiled in \citet{Izotov21a} by selection allows for the presence of \ion{He}{II}~4686 emission. Thus, at least a part of our sample is a priori likely to be a subset of the CSFGs.

\section{Data analysis}\label{sec_ana}

\subsection{Spectral analysis}\label{sec_specana}

After our initial sample selection, we analysed the SDSS spectra using \textsc{fado} \citep{Gomes17}. In brief, this code performs a population spectral synthesis by fitting a set of simple stellar populations to a galaxy's spectrum. \textsc{fado} then reconstructs the SFH from these fits, while computing a multitude of derived galaxy properties. This principle method is well established and has been successfully applied in the past to study the physical properties of galaxies \citep[e.g][]{CidFernandes07}. We selected \textsc{fado} as our synthesis code of choice, since as opposed to other commonly used tools such as \textsc{starlight} \citep{CidFernandes05}, \textsc{fado} also accounts for the contribution of nebular continuum emission, which can be assumed to be substantial in emission line galaxies such as those studied in this work. While the overall agreement between \textsc{starlight} and \textsc{fado} appears to be adequate, the study by \citet{Cardoso19} particularly emphasises the better performance of \textsc{fado} for (starbursting) galaxies with strong nebular emission.

Prior to the analysis, we corrected the input spectra for foreground extinction assuming a \citet{Cardelli89} extinction curve. We assumed a standard ratio of total to selective extinction of $R_V\equiv A_V/E(B-V)=3.1$. For each object, we adopted the V-band extinction values as listed in the NED\footnote{\href{https://ned.ipac.caltech.edu/}{https://ned.ipac.caltech.edu/}}, taken from \citet{Schlafly11}. Afterwards, the spectra were de-redshifted using the values provided on the website of the SDSS\footnote{\href{https://www.sdss.org/}{https://www.sdss.org/}}.

As for our choice of the stellar populations, we adopted the models from \citet{Bruzual03} in their latest version\footnote{\href{https://www.bruzual.org/bc03/Updated\_version\_2016/}{https://www.bruzual.org/bc03/Updated\_version\_2016/}} assuming a universal \citet{Kroupa01} IMF. We note that this choice is somewhat arbitrary, and the implied evolutionary parameters may be different for the stellar populations present here. In particular, as we find our objects to be compact, star-forming galaxies, binary stellar evolution may impact the galaxies' SEDs, which is covered in the BPASS models by \citet{Stanway18}. Likewise, as stated above, the results are sensitive to the choice of IMF, and it has been suggested that high-$z$ SFGs may be governed by a top-heavy IMF \citep[e.g.][]{Steinhardt22}, which would have significant implications for LyC leakage \citep{Wise09}. As this paper is of an exploratory nature and observations of deeper optical spectra for future analysis have already been scheduled, we postpone this to a follow-up study and limit this analysis to the models by \citet{Bruzual03}.

In principle, these models cover a range of metallicities extending to $2.5~Z_\sun$, spanning 220 unequally spaced time steps from $0.1~\mathrm{Myr} - 20~\mathrm{Gyr}$. In order to limit the range of free parameters and therefore improve the accuracy of the age estimates, we calculated preliminary oxygen abundances as outlined below (Sect. \ref{sec_abund}) from the emission line fluxes provided by \citet{Thomas13}. With the derived metallicities from the emission lines analysis (values in the range of $Z/Z_\sun=0.07 - 0.22$), we thus limit the set of stellar populations to a total of 600 model spectra with values of $Z/Z_\sun \in \left[0.005,0.02,0.2\right]$ and stellar ages of $\tau_\star=0.1~\mathrm{Myr} - 15~\mathrm{Gyr}$. 

Using this set of models, we then performed the spectral synthesis by fitting the spectral range of $3400 - 9000~\AA$. As we expect our objects to contain very little dust, we initially assumed negligible internal extinction, that is $A_\lambda\sim0$. We then consider the spectral fits by \textsc{fado} as a reasonable approximation for the best (extinction-corrected) solution; in particular, the emission line fluxes found this way should be adequately corrected for underlying stellar absorption. The (emission) flux ratios of H$\alpha$, H$\gamma$ and H$\delta$ relative to H$\beta$ then are compared to the expected values from theory, where we consider the Case B values tabulated in \citet{Storey95} for $t_\mathrm{e}=15\,000~\mathrm{K}$ and $n_\mathrm{e}=100~\mathrm{cm}^{-3}$ a reasonable match for the bulk of the galaxies studied here. Following \citet{Osterbrock06}, we then determine the extinction coefficient $c\left(\mathrm{H}\beta\right)$ as the mean of the extinction coefficients found for the individual line ratios, and subsequently correct the spectra assuming an SMC-like extinction law \citep[][$R_V=2.74$]{Gordon03}. In the cases where we find a slightly negative mean value for $c\left(\mathrm{H}\beta\right)$, we assume no internal extinction, that is $c\left(\mathrm{H}\beta\right)=0$.

Following this, we then performed an anewed population spectral synthesis with the dereddened spectra, leaving all other parameters unchanged.
During this process, \textsc{fado} performed Gaussian fits for up to 51 prominent emission lines ranging from \ion{[Ne}{V]}~3426 to \ion{[Fe}{II]}~8617. After visual inspection of the spectra, we found these fits to be in very good agreement to the line fluxes observed, and hence used these values for each subsequent analysis in this work. The line fluxes along with their corresponding EWs and the derived extinction coefficients are listed in the Appendix in Table \ref{tab_lines}.

For the sake of completeness, we note that \textsc{fado} occassionally would shift the peak of a measured emission line by $\sim1~\AA$ bluewards. Our visual inspection of the spectra suggested that this offset had no impact on the measured fluxes, and we verified this via manual line fitting using \textsc{iraf}'s \citep{Tody86} \texttt{splot}, finding excellent agreement between the values obtained.

While \textsc{fado} models a detailed SFH for each galaxy, we additionally calculated the star-formation rate (SFR) one would infer from H$\alpha$ for the sake of comparability with other studies. Here, we follow \cite{Kennicutt12}, thus $\log\mathrm{SFR}\left(\mathrm{H}\alpha\right)=\log L\left(\mathrm{H}\alpha\right)-41.27$.
As our analysis (Sect. \ref{sec_prop}) reveals that the galaxies in our final sample exhibit many similarities in the properties we considered, we further produced a stack of their spectra and analysed it in the same way as the individual objects.

\subsection{Abundance determination}\label{sec_abund}

To determine the metallicities of the objects considered here, we largely follow the prescriptions as outlined in \citet{PerezMontero17}.
For the determination of the oxygen abundance, we adopt the so-called 'direct' method, as the auroral \ion{[O}{III]}~4363 line is detected in all of our objects. In brief, we first calculate the electron temperature $t_{\mathrm{e}}\left(\ion{[O}{III]}\right)$ in the high-excitation zone from the \ion{[O}{III]} lines. The low-excitation zone electron temperature $t_{\mathrm{e}}\left(\ion{[O}{II]}\right)$ then is inferred from $t_{\mathrm{e}}\left(\ion{[O}{III]}\right)$ and the electron density $n_\mathrm{e}$, using the calibration of \citet{Haegele06}. The latter is calculated from the $\ion{[S}{II]}~6717,6731$ lines whenever possible and otherwise assumed to have a value of $100\ \mathrm{cm}^{-3}$. This value is comparable to those found in the LCGs of \citet{Izotov11}, who report median electron densities between $90~\mathrm{cm}^{-3}$ and $180~\mathrm{cm}^{-3}$. These values, along with \ion{[O}{III]}~4959,5007 and \ion{[O}{II]}~3727=\ion{[O}{II]}~3726,3729 fluxes then are used to determine the fractions of singly and doubly ionised oxygen relative to ionised hydrogen.

As the galaxies studied here are highly ionised, we also consider an ionisation correction factor (ICF) compensating for the neglection of $\mathrm{O}^{3+}$ in this calculation. Again following \citet{PerezMontero17}, we first calculate the abundance of doubly ionised helium $y^{2+}$ from \ion{He}{II}~4686 and a weighted mean of the singly ionised helium abundance $y^+$ from \ion{He}{I}~4471,5876,6678,7065, under the assumption that the intrinsic attenuation is small and the optical depth function $f_\lambda(n,t,\tau)$ \citep{Olive04} can be approximated as $f_\lambda\approx1$ \citep{PerezMontero17}. We note that for a handful of objects, the $y^+$ values derived from \ion{He}{I}~6678 would deviate from the values obtained from the other \ion{He}{I} lines by up to 12 orders of magnitude. In those cases, we excluded these unphysical values from the calculations and limit our evaluation to \ion{He}{I}~4471,5876,7065. Using the calibrations from \citet[][their Appendix A]{PerezMontero20}, $y^+$ and $y^{2+}$ then are converted into $\mathrm{ICF}\left(\mathrm{O}^{+} + \mathrm{O}^{2+}\right)$, and the total oxygen abundance is calculated as $\mathrm{O}/\mathrm{H}=\mathrm{ICF}\left(\mathrm{O}^{+} + \mathrm{O}^{2+}\right)\times\left(\mathrm{O}^{+}/\mathrm{H}^{+}+\mathrm{O}^{2+}/\mathrm{H}^{+}\right)$.

For our preliminary oxygen abundance evaluation (see above, Sect. \ref{sec_specana}), this was not possible for nine objects due to the absence of reliable measurements of various required emission line fluxes in the catalogue of \citet{Thomas13}. To determine the oxygen abundance, we then used the empirical relation between oxygen abundance and $t_{\mathrm{e}}\left(\ion{[O}{III]}\right)$ as calibrated in \citet{PerezMontero21}. In case both methods were not applicable, we adopted the strong-line method by \citet{Pagel79}, also known as the R23 method. As this diagnostic is double-valued and does not allow a conclusive oxygen abundance determination in the range of $8.0\la \log\left(\mathrm{O}/\mathrm{H}\right)+12\la 8.3$ \citep{PerezMontero17}, we first computed a rough estimate of the metallicity from $\ion{[N}{II]}~6583/\mathrm{H}\alpha$ \citep{PerezMontero09}, and consequently applied the upper or lower branch of the R23 method.

From the values obtained during the ICF calculations outlined above, we also determine the helium abundance, assuming that $\mathrm{He}/\mathrm{H}=\left(\mathrm{He}^{+}+\mathrm{He}^{2+}\right)/\mathrm{H}^+=y^++y^{2+}$.
The nitrogen abundance relative to oxygen, again following \citet{PerezMontero17}\footnote{We note that the prescription in \citet[][their Eq. 51]{PerezMontero17} includes a typo in the second-to-last summand, where $0.687 t_\mathrm{e}\left(\ion{[O}{II]}\right)^{-1}$ should be a subtrahend rather than a factor.}, was calculated from \ion{[N}{II]}~6583, \ion{[O}{II]}~3727, and $t_\mathrm{e}\left(\ion{[O}{II]}\right)$, assuming that $\mathrm{N}^+/\mathrm{N}=\mathrm{O}^+/\mathrm{O}$.

\subsection{LyC leakage diagnostics}\label{sec_lycdiag}

To this date, direct measurements of the LyC flux shortwards of $912\ \AA$ are the only fully reliable diagnostic for LyC leakage and derivations of $f_\mathrm{esc}\left(\mathrm{LyC}\right)$. However, significant efforts to relate other observable quantities to LyC escape have been made in the last few years. Particularly promising ist the study of Ly$\alpha$, whose line profile is related to the ISM structure \citep[e.g.][]{Gronke16}, and the peak separation of the double-peaked Ly$\alpha$ line emerging from a clumpy ISM may relate to $f_\mathrm{esc}\left(\mathrm{LyC}\right)$ \citep{Verhamme15}. \citet{Izotov20} provide a calibration relating the Ly$\alpha$ escape fraction $f_\mathrm{esc}\left(\mathrm{Ly\alpha}\right)$ to $f_\mathrm{esc}\left(\mathrm{LyC}\right)$, but caution this is a preliminary relation inferred from low sample statistics.

In case of LyC escape being realised through ionised low density tunnels in the ISM, several FUV metal absorption lines may display residual flux at their line centre \citep{Heckman01}. As this residual flux then directly relates to the covering fraction of neutral gas, they can be used to infer a prediction for $f_\mathrm{esc}\left(\mathrm{LyC}\right)$ \citep{Chisholm18}.

These diagnostics, however, require information from the ultraviolet portion of the spectrum, a wavelength range that for objects in the low-redshift regime is accessible only to the \textit{Cosmic Origins Spectrograph} (COS) aboard the Hubble Space Telescope (HST). No such data is available for our objects, and even the \ion{Mg}{II}~2796,2803 lines that provide strong evidence for the possibility of LyC escape as they trace an optically thin medium \citep{Chisholm20} were not or not reliably measured here.

In the optical regime, there are a few emission line diagnostics that relate to conditions favourable for LyC escape. A promising approach to look for LyC leakers is to select galaxies with large $\mathrm{O}_{32}=\ion{[O}{III]}~5007/\ion{[O}{II]}~3727$ flux ratios, which indicates optically thin gas with a high degree of ionisation \citep{Jaskot13}, and using it as a selection criterion has lead to several successful detections of genuine LyC leakers \citep[e.g.][]{Izotov16a,Izotov16b,Izotov18a,Izotov18b}. \citet{Chisholm18} even provide a tentative calibration relating this line ratio to $f_\mathrm{esc}\left(\mathrm{LyC}\right)$, but as \citet{Jaskot19} note, a high $\mathrm{O}_{32}$ flux ratio alone is an insufficient LyC leakage diagnostic, as the line ratio can be modulated by the structure of the ISM. \citet{Nakajima20} confirm that such an elevated line ratio is indeed a necessary, yet not sufficient condition for a galaxy to be a LyC leaker. Still, the results presented in \citet{Flury22b} show that along with the (related) parameter $\Sigma_\mathrm{SFR,H\beta}$, $\mathrm{O}_{32}$ to date is the best optical diagnostic relating to $f_\mathrm{esc}\left(\mathrm{LyC}\right)$. A similar pre-selection of LyC leaker candidates may be possible with the \ion{[Ne}{III]}~3869/\ion{[O}{II]}~3727 ionisation parameter diagnostic \citep{Levesque14}, which has the additional benefit of being less affected by internal reddening.

Another characteristic many LyC leakers seem to share is a deficiency in \ion{[S}{II]}~6717,6731/H$\alpha$. Sulphur with its rather low ionisation potential of $10.36\ \mathrm{eV}$ causes these lines to predominantly originate in the partially ionised zone beyond the edge of the fully ionised gas, which is less pronounced in density-bounded nebulae and thus correlates with LyC leakage \citep{Alexandroff15}.
\citet{Wang19} have demonstrated this as a powerful selection tool for LyC leaker candidates and expanded upon this diagnostic in \citet{Wang21} with the sample of the recent Low-Redshift LyC Survey \citep{Flury22a}, where they provide a calibration in the \ion{[O}{III]}~5007/H$\beta$ vs. \ion{[S}{II]}~6717,6731/H$\alpha$ parameter space for typical emission line galaxies. A deviation in \ion{[S}{II]}~6717,6731/H$\alpha$ from this ridge line (denoted as $\Delta\ion{[S}{II]}$) statistically correlates with a larger $f_\mathrm{esc}\left(\mathrm{LyC}\right)$, even though it does not allow for an accurate determination of this parameter.

In conclusion, there is no definite, independent method to accurately predict a galaxy's $f_\mathrm{esc}\left(\mathrm{LyC}\right)$ as of now, but the application of a multitude of diagnostics correlated with ionising photon loss as outlined above provide strong selection criteria for genuine LyC leakers. Here, we make use of the $\mathrm{O}_{32}$ based calibration by \cite{Chisholm18} to obtain an estimate of $f_\mathrm{esc}\left(\mathrm{LyC}\right)$, that is
\begin{equation}\label{eqn_fescO32}
f_\mathrm{esc}\left(\mathrm{LyC}\right)=\left(0.0017\pm0.0004\right)\mathrm{O}_{32}^2 + \left(0.005\pm0.007\right) \, .
\end{equation}
An alternative calibration can be found in \cite{Izotov18a}, yielding significantly larger values of $f_\mathrm{esc}\left(\mathrm{LyC}\right)$ in galaxies with large $\mathrm{O}_{32}$.

\section{Sample properties}\label{sec_prop}

\subsection{Global sample characteristics}\label{sec_globalchar}

\begin{figure}
\centering
\includegraphics[width=\hsize]{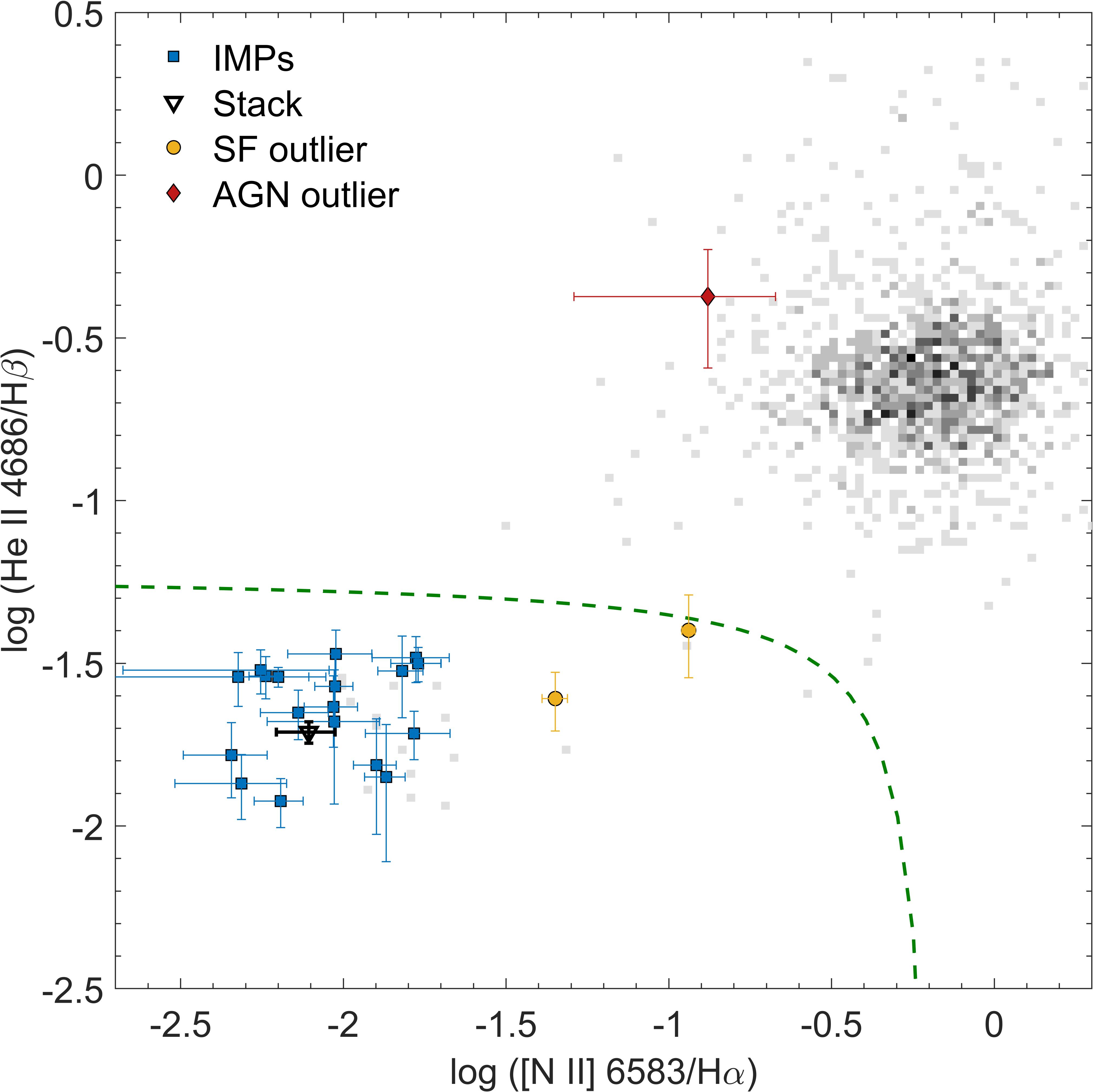}
\caption{\citet{Shirazi12} diagram including their AGN diagnostic (green dashed line). Blue squares represent the IMPs, whereas the position of their stack is shown as a black downward-facing open triangle. Red diamonds and yellow circles mark the AGN contaminated galaxies and star-forming objects classified as outliers (see text). The grey scale density diagram represents all eligible galaxies from the SDSS DR12 with AoN>2 in the utilised emission lines, binned in $0.02\times0.02~\mathrm{dex}$ bins for the line ratios shown. Lower contrast indicates lower bin occupancy.}
\label{fig_SB}
\end{figure}

\begin{figure}
\centering
\includegraphics[width=\hsize]{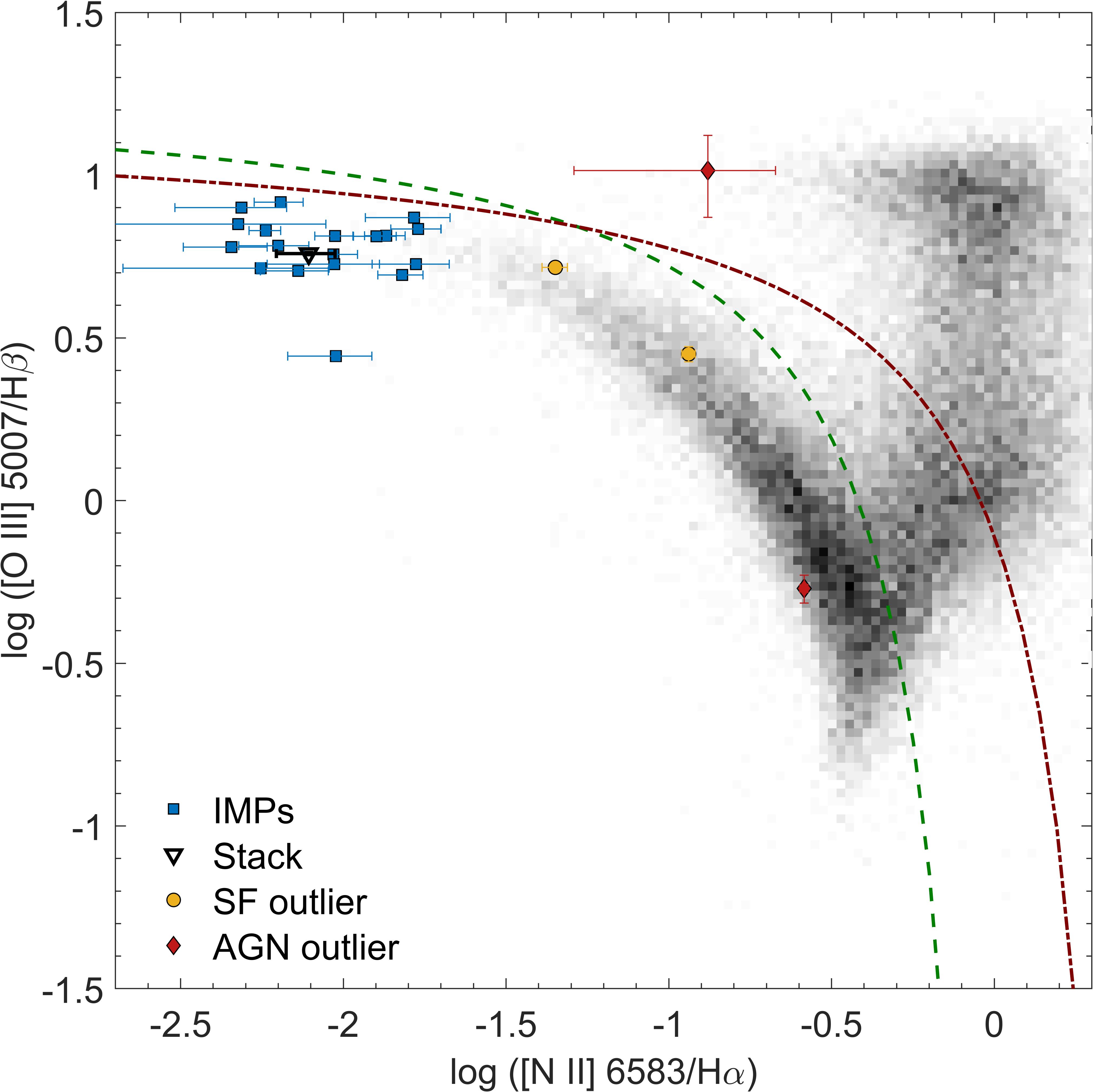}
\caption{BPT diagram including the photoionisation limit from \citet{Kewley01} (red dash-dotted line) and the AGN separation line from \citet{Kauffmann03} (green dashed line). Symbols are the same as in Fig. \ref{fig_SB}.}
\label{fig_BPT}
\end{figure}

\begin{figure}
\centering
\includegraphics[height=\hsize]{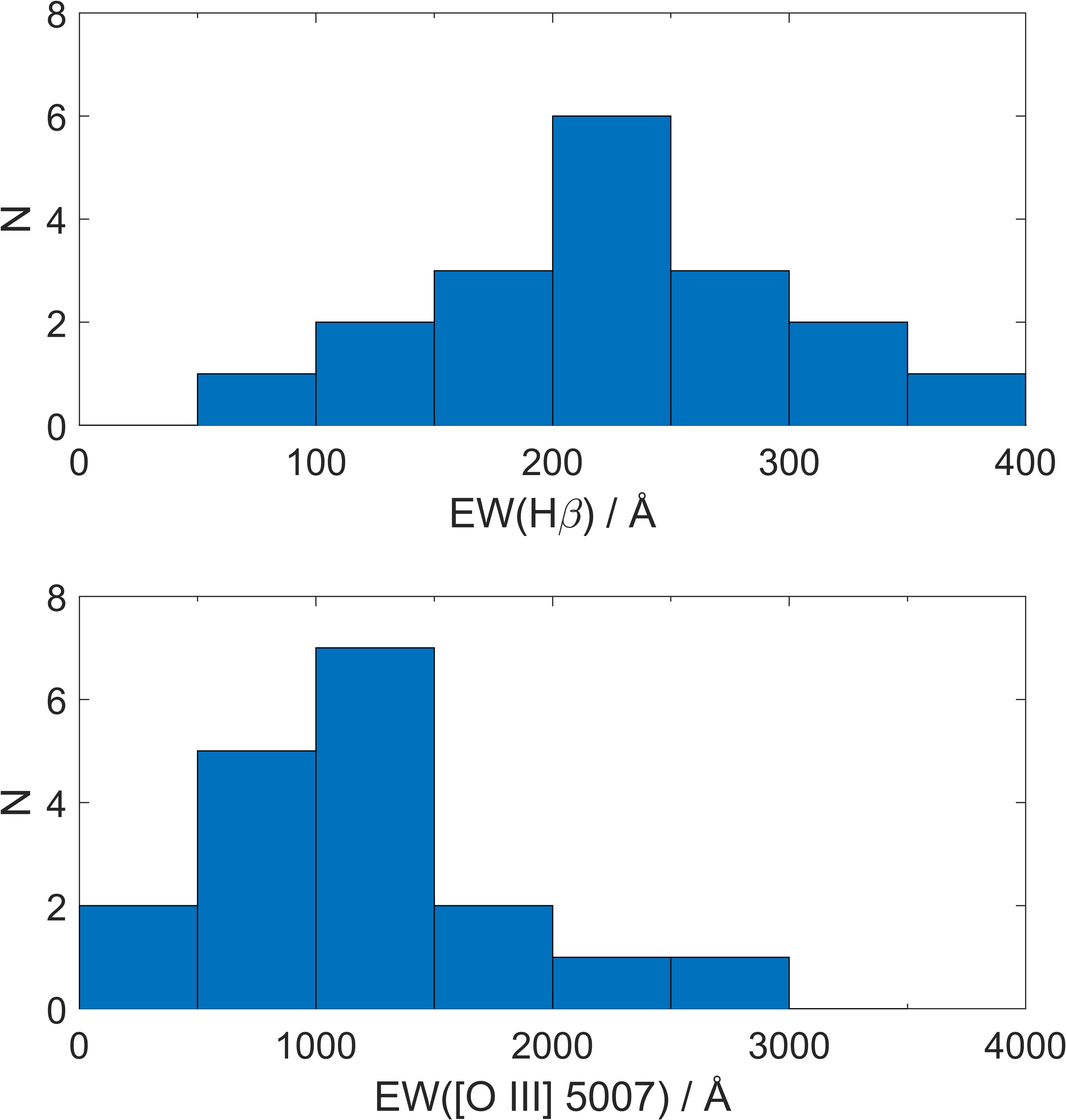}
\caption{Distribution of EW$\left(\mathrm{H}\beta\right)$ (top) and EW$\left(\ion{[O}{III]}~5007\right)$ (bottom) for the IMPs, binned in $50~\AA$ and $500~\AA$ bins, respectively.}
\label{fig_EW}
\end{figure}

Of a total of $1\,381\,398$ objects present in the dataset, $1\,504$ ($\sim0.01\%$) satisfied the AoN limitation imposed upon the \ion{He}{II}~4686, H$\beta$, H$\alpha$, and \ion{[N}{II]}~6583 diagnostic lines. Within this subset, an initial sample of 22 galaxies satisfied our selection criterion given by Eq. \ref{eqn_SB12}. After our renewed emission-line fitting, we found that for two of those objects, the \ion{He}{II}~4686~/~H$\beta$ ratios had been understimated in the catalogue of \citet{Thomas13}, and the diagnostic by \citet{Shirazi12} reveals these to comprise a significant AGN contribution (red diamonds in Fig. \ref{fig_SB}). Furthermore, when considering the diagnostic diagram first introduced by \citet[][hereafter BPT]{Baldwin81} (Fig. \ref{fig_BPT}), commonly used to separate SFGs from AGN, we find the majority of the remaining objects to occupy the same region in the high-ionisation tail of the diagram. Two galaxies are found to be clearly offset from this population (yellow circles in Fig. \ref{fig_BPT}) and their line ratios are more in line with those found in typical SFGs, which justifies their exclusion from our final sample. For completeness, and also as each of these outliers exhibit interesting characteristics in their own right, we discuss them in Appendix \ref{app_outl}. Our final sample now consists of 18 objects, which we dub the ionised, metal poor galaxies (IMPs). Within the diagnostic diagram used for the selection of our sample (Fig. \ref{fig_SB}), these reside in a well-defined locus significantly offset from the bulk of \ion{He}{II}~4686 emitters, suggesting a similar ionisation mechanism other than an AGN is prevalent in these galaxies.

To the best of our knowledge, none of these objects have previously been studied individually in the literature. We list their coordinates and redshifts along with several general properties in Table \ref{tab_galprops1}. For reasons of legibility, we reduce the galaxy name to the first four+four digits of the SDDS identifier.

\begin{table*}
\caption{Global properties of the galaxies studied in this work.}             
\label{tab_galprops1}      
\centering          
\begin{tabular}{c c c c c c c c c c c}
\hline\hline       
ID & RA & DEC & $z\tablefootmark{a}$ & $r_\mathrm{petro}^\mathrm{ang}\tablefootmark{b}$ & $r_\mathrm{petro}^\mathrm{lin}\tablefootmark{c}$ & $M_\star\tablefootmark{d}$ & SFR$\left(\mathrm{H}\alpha\right)\tablefootmark{e}$ & sSFR\tablefootmark{f} & $\mathrm{SFR}_\mathrm{B}\tablefootmark{g}$ & $\mathrm{sSFR}_\mathrm{B}\tablefootmark{g}$ \\ 
\hline
J0006+0255 & 00:06:06.53 & 02:55:04.13 & 0.09382828 & 1.59 & 2.85 & 8.48 & 1.118 & -8.43 & 1.158 & -8.42 \\
J0028+3035 & 00:28:33.55 & 30:35:54.14 & 0.07451165 & 2.65 & 3.88 & 8.59 & 0.707 & -8.74 & 0.660 & -8.77 \\
J0131+0210 & 01:31:38.15 & 02:10:14.58 & 0.05806014 & 2.12 & 2.44 & 6.66 & 0.564 & -6.91 & 0.577 & -6.90 \\
J0138+1114 & 01:38:27.65 & 11:14:34.45 & 0.06219218 & 1.43 & 1.77 & 6.95 & 1.228 & -6.86 & 1.271 & -6.85 \\
J0150+1643 & 01:50:46.25 & 16:43:26.43 & 0.05909079 & 1.65 & 1.94 & 7.82 & 1.316 & -7.70 & 1.281 & -7.71 \\
J0744+1858 & 07:44:52.39 & 18:58:30.01 & 0.05227381 & 3.35 & 3.53 & 8.25 & 1.272 & -8.14 & 1.265 & -8.14 \\
J0753+2820 & 07:53:25.26 & 28:20:12.74 & 0.06752457 & 3.94 & 5.24 & 8.35 & 0.867 & -8.41 & 0.821 & -8.43 \\
J0809+4918 & 08:09:42.74 & 49:18:21.46 & 0.07809167 & 1.72 & 2.63 & 8.40 & 1.470 & -8.23 & 1.447 & -8.24 \\
J1037+2325 & 10:37:28.61 & 23:25:29.81 & 0.01058074 & 5.27 & 1.21 & 7.10 & 0.009 & -9.13 & 0.009 & -9.13 \\
J1109+3429 & 11:09:20.89 & 34:29:02.90 & 0.06697157 & 2.48 & 3.31 & 8.83 & 0.662 & -9.01 & 0.649 & -9.02 \\
J1141+6059 & 11:41:46.74 & 60:59:43.13 & 0.01096911 & 6.93 & 1.46 & 6.88 & 0.027 & -8.45 & 0.026 & -8.47 \\
J1311+3750 & 13:11:35.82 & 37:50:58.73 & 0.05381261 & 3.95 & 4.24 & 7.70 & 1.657 & -7.48 & 1.643 & -7.48 \\
J1313+6044 & 13:13:31.26 & 60:44:54.52 & 0.07113630 & 2.52 & 3.52 & 8.11 & 3.104 & -7.62 & 2.871 & -7.65 \\
J1338+4213 & 13:38:42.16 & 42:13:38.20 & 0.00863322 & 2.83 & 0.52 & 6.20 & 0.011 & -8.16 & 0.011 & -8.16 \\
J1411+0550 & 14:11:13.40 & 05:50:35.07 & 0.04940260 & 1.61 & 1.58 & 7.99 & 1.715 & -7.75 & 1.607 & -7.78 \\
J1528+2318 & 15:28:26.53 & 23:18:43.10 & 0.05948849 & 2.97 & 3.56 & 8.54 & 0.705 & -8.69 & 0.665 & -8.71 \\
J1556+1818 & 15:56:47.51 & 18:18:25.57 & 0.09052273 & 2.28 & 3.94 & 7.82 & 1.964 & -7.53 & 2.041 & -7.51 \\
J1608+0413 & 16:08:01.16 & 04:13:23.85 & 0.06322286 & 1.50 & 1.89 & 7.42 & 0.690 & -7.58 & 0.640 & -7.61 \\
\hline
\end{tabular}
\tablefoot{
\tablefoottext{a}{Redshifts $z$ adopted from the SDSS SkyServer at \href{http://skyserver.sdss.org/}{http://skyserver.sdss.org/}.}
\tablefoottext{b}{$r$-band petrosian radii in arcsec taken from \href{http://skyserver.sdss.org/}{http://skyserver.sdss.org/}.}
\tablefoottext{c}{$r$-band petrosian radii converted to linear scales (kpc).}
\tablefoottext{d}{Logarithm of the stellar mass in units of $\mathrm{M}_\sun$ as derived by \textsc{fado} assuming a \citet{Kroupa01} IMF.}
\tablefoottext{e}{Star formation rate in units of $\mathrm{M}_\sun~\mathrm{yr}^{-1}$.}
\tablefoottext{f}{Logarithm of the sSFR, i.e. the SFR divided by stellar mass.}
\tablefoottext{g}{Based on corrected H$\alpha$ values, see Sect. \ref{sec_bdec}.}
}
\end{table*}

We find the IMPs ubiquitously reside at low redshifts (0.01$\la z\la0.09$), which can be attributed to a selection effect as we require a reliably measured \ion{He}{II}~4686 line. All of them are readily classified as DGs, with stellar masses ranging from $6.2\la \log\left(M_\star/\mathrm{M}_\sun\right)\la8.8$. They are characterised by moderate to large sSFRs of $\sim 0.74-137~\mathrm{Gyr}^{-1}$ with a median of $\sim 7.09~\mathrm{Gyr}^{-1}$, which serves as a first indicator that the IMPs blend well into the population of LCGs and CSFGs \citep{Izotov11,Izotov21a}, where similar values are reported. This likeness is further supported by the large values we find for EW$\left(\mathrm{H}\beta\right)$ and EW$\left(\ion{[O}{III]}~5007\right)$ (Fig. \ref{fig_EW}), the former of which already points towards the IMPs harbouring a young starburst.

\subsection{Morphology}\label{sec_morpho}

\begin{figure*}[hpbt]
\centering
\includegraphics[width=\hsize]{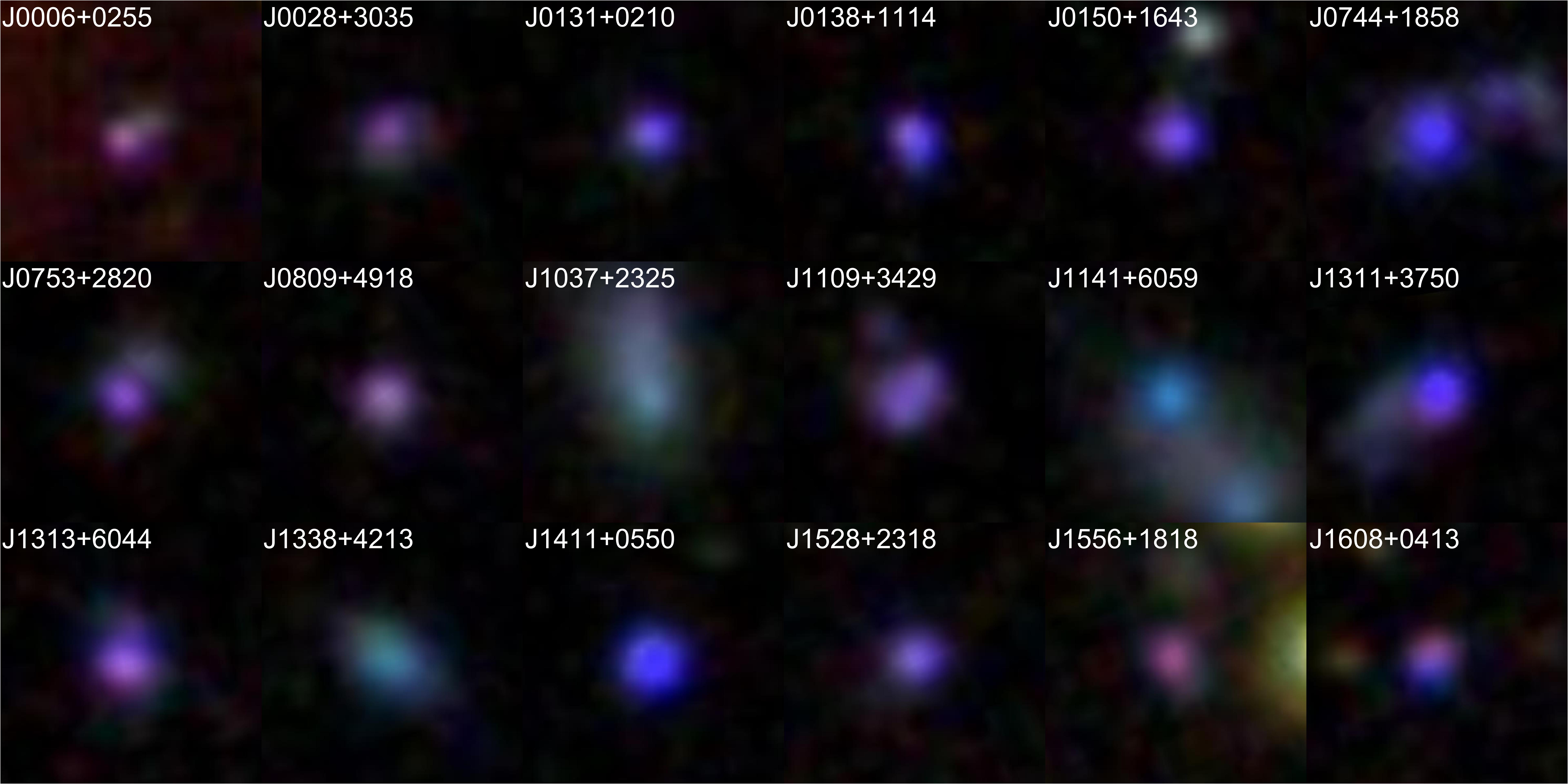}
\caption{Mosaic of the SDSS $gri$ cutouts of our final sample obtained from the SDSS SkyServer at \href{http://skyserver.sdss.org/}{http://skyserver.sdss.org/}. Each cutout spans a region of $12.5\arcsec\times12.5\arcsec$, where north is up and east is to the left.}
\label{fig_SDSSmosaic}
\end{figure*}

The SDSS image cutouts of the final sample are shown in Fig. \ref{fig_SDSSmosaic}. While most of the IMPs are only marginally resolved, they exhibit diverse morphological features. A few of them appear almost circular in projection (arguably most pronounced in J1411+0550), whereas others display signs of anisotropic extended emission. A variety of possible causes may serve to explain this, such as the galaxies' intrinsic irregular morphology, large-scale outflows, or a recent merger, which is difficult to deduce from the SDSS images alone. Occassionally, we see more than one bright, star-forming knot (e.g. in J0006+0255, J1141+6059) which lends further credibility to a merging scenario, albeit it is equally plausible these are isolated regions of increased star-forming activity within the same galactic system.

One feature all IMPs certainly appear to share is their compactness. The SDSS pipeline lists petrosian radii ($r$-band) between $1.43\arcsec$ and $6.93\arcsec$ for the IMPs, which translate to sizes of $r_\mathrm{petro}\sim0.52 - 5.2~\mathrm{kpc}$ at their given redshifts (calculated using Ned Wright's Cosmology Calculator \citep{Wright06} with cosmological parameters from \citealt{Planck20}), with a median value of $r_\mathrm{petro}\sim2.74~\mathrm{kpc}$. Note that most of these objects have been flagged \textit{MANYPETRO} by the SDSS pipeline, meaning several possible values for $r_\mathrm{petro}$ were found and those listed are the maximum values. Consequently, these are best interpreted as upper limits to the true (petrosian) sizes. We further point out that for several (5/18) of the IMPs, the radii (see Table \ref{tab_galprops1}) exceed the $3\arcsec$ fibre diameter of the BOSS spectrograph \citep{Smee13}, hence the properties derived for these objects exclude the contribution from the galactic outskirts (to varying degrees).

Their compactness further reinforces the notion that the IMPs can be considered a subclass of the LCGs, discovered by \citet{Izotov11}. For most of the objects, the same conclusion can immediately be drawn regarding the CSFGs \citep{Izotov21a}, which (amongst other criteria) were selected as galaxies with $r_\mathrm{petro}\leq3\arcsec$. The conversion to linear scales of the IMPs' radii (see Table \ref{tab_galprops1}) reveals these likewise are suitable objects to fulfil the compactness criterion imposed on the CSFGs.

\subsection{Integrated nebular characteristics}

\begin{figure}[hptb]
\centering
\includegraphics[width=\hsize]{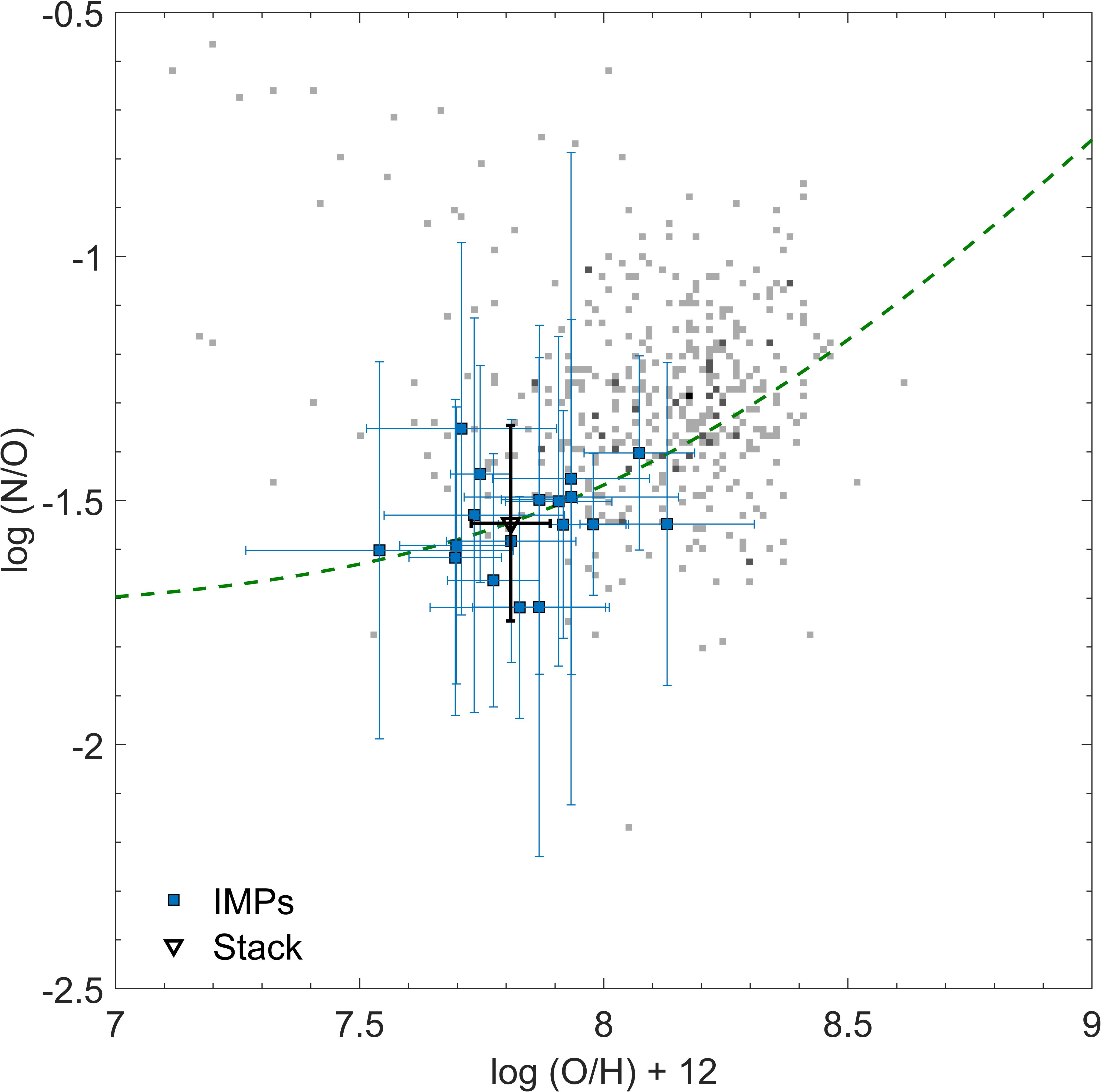}
\caption{Nitrogen vs. oxygen abundance, with symbols representing the same objects as in Fig. \ref{fig_SB}. The SDSS DR12 galaxies (black scatter diagram) are limited to SFGs with appropriate AoN in the line fluxes relevant for the determination of N/O and O/H with the direct method. The green dahed line is the generic scaling relation from \citet{Nicholls17}, i.e. $\log\left(\mathrm{N}/\mathrm{O}\right)=\log\left(10^{-1.732}+10^{\log\mathrm{O}/\mathrm{H}+2.19}\right)$.}
\label{fig_no}
\end{figure}

\begin{figure}[hpbt]
\centering
\includegraphics[width=\hsize]{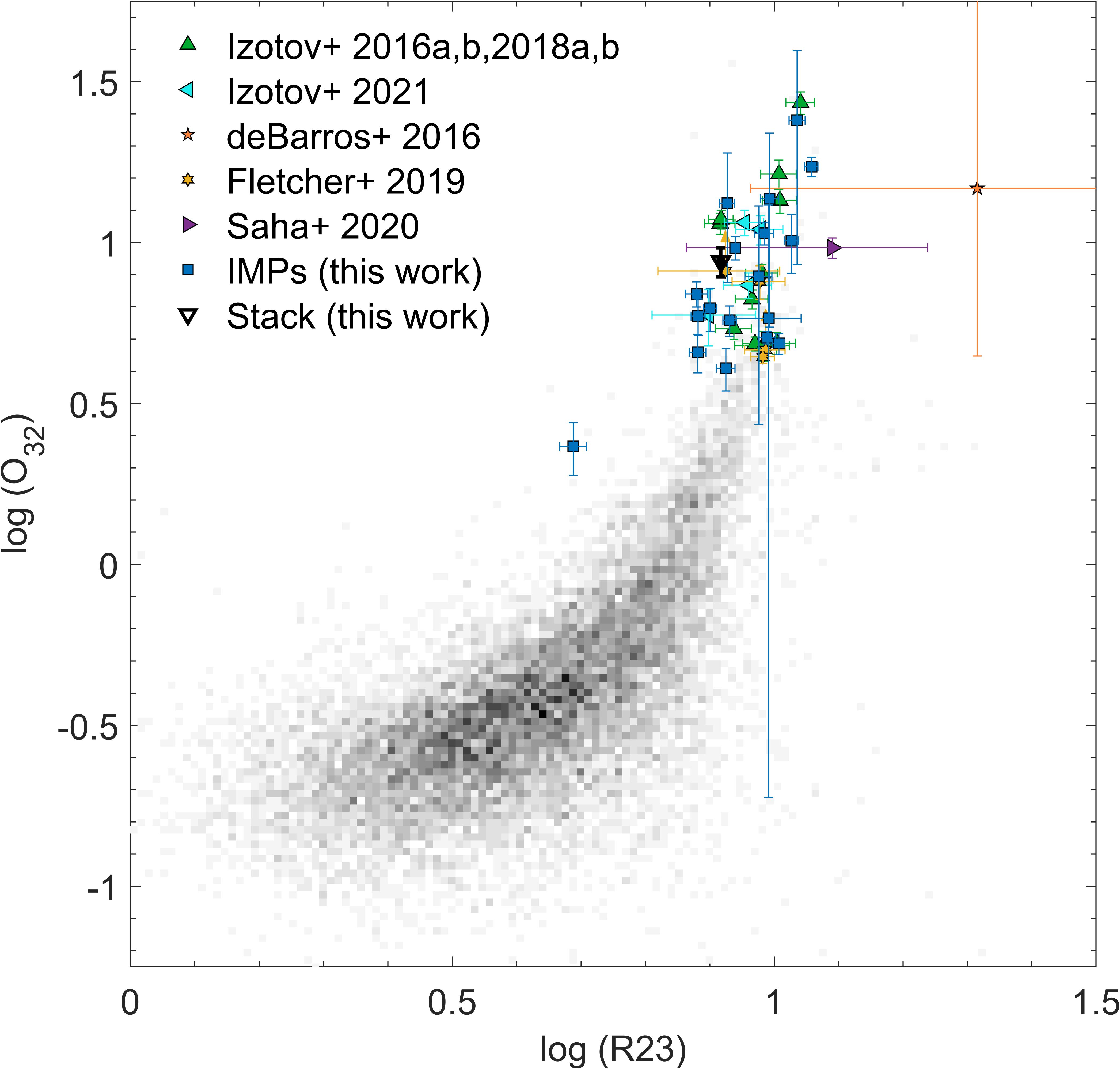}
\caption{Ionisation parameter and metallicity sensitive $\log\left(\mathrm{O}_{32}\right)=\log\left(\ion{[O}{III]}~5007/\ion{[O}{II]}~3727\right)$ vs. $\log\left(\mathrm{R23}\right)=\log\left(\left(\ion{[O}{II]}~3727+\ion{[O}{III]}~4959+\ion{[O}{III]}~5007\right)/\mathrm{H}\beta\right)$ diagram. Symbols are the same as in Fig. \ref{fig_SB}. Additional markers were added for several low-$z$ LyC leakers, i.e. the GPs from \citet{Izotov16a,Izotov16b,Izotov18a,Izotov18b} (green upward triangles) and low-mass galaxies from \citet{Izotov21b} (cyan left-facing triangles). Also shown are the intermediate-$z$ LyC leakers \textit{Ion2} from \citet{deBarros16} (orange pentagram), four galaxies from \citet{Fletcher19} (dark red hexagrams, where we converted their listed \ion{[O}{III]}~4959,5007 values to \ion{[O}{III]}~5007 assuming \ion{[O}{III]}~5007/\ion{[O}{III]}~4959=2.98, \citealt{Storey00}), and AUDFs01 from \citet{Saha20} (purple right-facing triangle). The SDSS DR12 galaxies (black scatter diagram) are limited to those identified as SFGs with the BPT diagnostics from \citet{Kewley01} and \citet{Kauffmann03}.}
\label{fig_R23O32}
\end{figure}

\begin{table*}
\caption{Derived properties for the galaxies in our sample as indicated in the first column.}             
\label{tab_galprops2}      
\centering          
\begin{tabular}{l c c c c c c}
\hline\hline

$ $ & J0006+0255 & J0028+3035 & J0131+0210 & J0138+1114 & J0150+1643 & J0744+1858 \\
\hline

$t_\mathrm{e}\left(\ion{[O}{III]}\right) / \mathrm{K}$ & $16530\pm610$ & $17670\pm550$ & $14550\pm810$ & $16230\pm610$ & $16770\pm650$ & $15320\pm460$ \\
$t_\mathrm{e}\left(\ion{[O}{II]}\right) / \mathrm{K}$ & $15780\pm10000$ & $10960\pm4100$ & $15040\pm14000$ & $11700\pm5000$ & $11570\pm2700$ & $11510\pm2500$ \\
$n_\mathrm{e}\left(\ion{[S}{II]}\right) / \mathrm{cm}^{-3}$ & $50\pm310$ & $380\pm570$ & $30\pm400$ & $260\pm480$ & $290\pm280$ & $250\pm250$ \\
$\left(\mathrm{O}^{+}/\mathrm{H}^{+}\right)\times10^5$ & $0.6\pm1.1$ & $1.6\pm2.4$ & $0.9\pm2.6$ & $1\pm1.7$ & $0.9\pm0.9$ & $0.98\pm0.82$ \\
$\left(\mathrm{O}^{2+}/\mathrm{H}^{+}\right)\times10^5$ & $4.42\pm0.37$ & $5.07\pm0.37$ & $7.7\pm1.1$ & $6.36\pm0.56$ & $5.05\pm0.44$ & $8.59\pm0.63$ \\
$y^+\times10^2$ & $7.56\pm0.16$ & $5.34\pm0.02$ & $7.59\pm0.18$ & $6.217\pm0.009$ & $7.731\pm0.013$ & $6.932\pm0.002$ \\
$y^{2+}\times10^2$ & $0.1\pm1.2$ & $0.1\pm1.3$ & $0.1\pm1.6$ & $0.1\pm1.5$ & $0.1\pm1.3$ & $0.05\pm0.92$ \\
$\mathrm{ICF}\left(\mathrm{O}^{+}+\mathrm{O}^{2+}\right)$ & $1\pm0.14$ & $1.01\pm0.21$ & $1\pm0.17$ & $1\pm0.2$ & $1\pm0.14$ & $0.99\pm0.11$ \\
$\log\left(\mathrm{O}/\mathrm{H}\right)+12$ & $7.7\pm0.12$ & $7.83\pm0.18$ & $7.93\pm0.16$ & $7.87\pm0.14$ & $7.77\pm0.09$ & $7.98\pm0.07$ \\
$\log\left(\mathrm{He}/\mathrm{H}\right)$ & $-1.12\pm0.07$ & $-1.26\pm0.11$ & $-1.12\pm0.09$ & $-1.2\pm0.1$ & $-1.11\pm0.07$ & $-1.16\pm0.06$ \\
$\log\left(\mathrm{N}/\mathrm{O}\right)$ & $-1.59\pm0.28$ & $-1.72\pm0.23$ & $-1.46\pm0.67$ & $-1.72\pm0.51$ & $-1.66\pm0.26$ & $-1.55\pm0.15$ \\
$\mathrm{O}_{32}$ & $5.91\pm0.77$ & $10.69\pm0.85$ & $5.8\pm7.3$ & $13.7\pm8.2$ & $13.2\pm5.7$ & $17.2\pm1.2$ \\
$\Delta\ion{[S}{II]}$ & $-0.34\pm0.04$ & $-0.33\pm0.07$ & $-0.13\pm0.05$ & $-0.56\pm0.06$ & $-0.54\pm0.02$ & $-0.17\pm0.03$ \\
$f_\mathrm{esc}\left(\mathrm{LyC}\right)$\tablefootmark{b} & $(0.064)$ & $(0.199)$ & $(0.062)$ & $(0.323)$ & $(0.303)$ & $(0.509)$ \\
\multicolumn{7}{c}{} \\

$ $ & J0753+2820 & J0809+4918 & J1037+2325 & J1109+3429 & J1141+6059 & J1311+3750 \\
\hline
$t_\mathrm{e}\left(\ion{[O}{III]}\right) / \mathrm{K}$ & $14970\pm590$ & $15760\pm820$ & $16300\pm1100$ & $14100\pm490$ & $12750\pm360$ & $16650\pm360$ \\
$t_\mathrm{e}\left(\ion{[O}{II]}\right) / \mathrm{K}$ & $13910\pm6800$ & $14080\pm7200$ & $16010\pm15000$ & $12940\pm8900$ & $12120\pm7400$ & $11660\pm3200$ \\
$n_\mathrm{e}\left(\ion{[S}{II]}\right) / \mathrm{cm}^{-3}$ & $80\pm300$ & $90\pm310$ & $30\pm430$ & $100\pm500$\tablefootmark{a} & $100\pm500$\tablefootmark{a} & $270\pm320$ \\
$\left(\mathrm{O}^{+}/\mathrm{H}^{+}\right)\times10^5$ & $0\pm2$ & $1\pm1.6$ & $0.7\pm1.9$ & $0\pm4$ & $2.3\pm5.1$ & $0.64\pm0.79$ \\
$\left(\mathrm{O}^{2+}/\mathrm{H}^{+}\right)\times10^5$ & $7.03\pm0.69$ & $5.52\pm0.67$ & $4.4\pm0.7$ & $6.87\pm0.63$ & $11.16\pm0.88$ & $6.77\pm0.36$ \\
$y^+\times10^2$ & $7.737\pm0.005$ & $8.206\pm0.023$ & $6.88\pm0.34$ & $6.950\pm0.009$ & $7.14\pm0.08$ & $6.775\pm0.003$ \\
$y^{2+}\times10^2$ & $0.1\pm1.5$ & $0.1\pm1.5$ & $0.1\pm1.8$ & $0.1\pm1.5$ & $0.1\pm1.3$ & $0.1\pm1.1$ \\
$\mathrm{ICF}\left(\mathrm{O}^{+}+\mathrm{O}^{2+}\right)$ & $1\pm0.16$ & $1\pm0.15$ & $1\pm0.22$ & $1\pm0.18$ & $1\pm0.15$ & $1\pm0.13$ \\
$\log\left(\mathrm{O}/\mathrm{H}\right)+12$ & $7.92\pm0.13$ & $7.81\pm0.13$ & $7.71\pm0.19$ & $7.93\pm0.22$ & $8.13\pm0.18$ & $7.87\pm0.08$ \\
$\log\left(\mathrm{He}/\mathrm{H}\right)$ & $-1.11\pm0.09$ & $-1.08\pm0.08$ & $-1.16\pm0.11$ & $-1.15\pm0.09$ & $-1.14\pm0.08$ & $-1.17\pm0.07$ \\
$\log\left(\mathrm{N}/\mathrm{O}\right)$ & $-1.55\pm0.23$ & $-1.58\pm0.25$ & $-1.35\pm0.38$ & $-1.49\pm0.36$ & $-1.55\pm0.33$ & $-1.5\pm0.36$ \\
$\mathrm{O}_{32}$ & $5.06\pm0.23$ & $5.73\pm0.62$ & $4.56\pm0.62$ & $4.06\pm0.61$ & $4.85\pm0.36$ & $24\pm15$ \\
$\Delta\ion{[S}{II]}$ & $-0.09\pm0.04$ & $-0.23\pm0.03$ & $-0.26\pm0.05$ & $-0.14\pm0.04$ & $0.03\pm0.03$ & $-0.41\pm0.04$ \\
$f_\mathrm{esc}\left(\mathrm{LyC}\right)$\tablefootmark{b} & $(0.049)$ & $(0.061)$ & $(0.040)$ & $(0.033)$ & $(0.045)$ & $(0.983)$ \\
\multicolumn{7}{c}{} \\
\hline\hline

$ $ & J1313+6044 & J1338+4213 & J1411+0550 & J1528+2318 & J1556+1818 & J1608+0413 \\
\hline
$t_\mathrm{e}\left(\ion{[O}{III]}\right) / \mathrm{K}$ & $16610\pm240$ & $15810\pm910$ & $14510\pm310$ & $14980\pm550$ & $16500\pm910$ & $16640\pm550$ \\
$t_\mathrm{e}\left(\ion{[O}{II]}\right) / \mathrm{K}$ & $16650\pm8300$ & $15890\pm14000$ & $9870\pm1900$ & $13490\pm5000$ & $14260\pm12000$ & $15230\pm6300$ \\
$n_\mathrm{e}\left(\ion{[S}{II]}\right) / \mathrm{cm}^{-3}$ & $20\pm230$ & $30\pm410$ & $460\pm420$ & $100\pm250$ & $100\pm500$\tablefootmark{a} & $70\pm220$ \\
$\left(\mathrm{O}^{+}/\mathrm{H}^{+}\right)\times10^5$ & $0.36\pm0.51$ & $0\pm2$ & $2.9\pm2.6$ & $0.9\pm1.3$ & $0.8\pm2.1$ & $0.6\pm0.7$ \\
$\left(\mathrm{O}^{2+}/\mathrm{H}^{+}\right)\times10^5$ & $5.22\pm0.18$ & $2.67\pm0.36$ & $8.89\pm0.51$ & $7.16\pm0.64$ & $4.66\pm0.58$ & $4.4\pm0.36$ \\
$y^+\times10^2$ & $7.51\pm0.04$ & $6.54\pm0.28$ & $6.919\pm0.005$ & $7.559\pm0.003$ & $9.23\pm0.04$ & $7.93\pm0.03$ \\
$y^{2+}\times10^2$ & $0.11\pm0.92$ & $0.1\pm1.6$ & $0.1\pm1.2$ & $0.1\pm1.6$ & $0.1\pm1.9$ & $0.1\pm1.4$ \\
$\mathrm{ICF}\left(\mathrm{O}^{+}+\mathrm{O}^{2+}\right)$ & $1\pm0.1$ & $1.01\pm0.21$ & $1\pm0.14$ & $1\pm0.18$ & $1\pm0.17$ & $1\pm0.15$ \\
$\log\left(\mathrm{O}/\mathrm{H}\right)+12$ & $7.75\pm0.06$ & $7.54\pm0.27$ & $8.07\pm0.11$ & $7.91\pm0.11$ & $7.73\pm0.18$ & $7.70\pm0.10$ \\
$\log\left(\mathrm{He}/\mathrm{H}\right)$ & $-1.12\pm0.05$ & $-1.18\pm0.11$ & $-1.16\pm0.07$ & $-1.12\pm0.09$ & $-1.03\pm0.09$ & $-1.10\pm0.08$ \\
$\log\left(\mathrm{N}/\mathrm{O}\right)$ & $-1.45\pm0.22$ & $-1.6\pm0.39$ & $-1.4\pm0.2$ & $-1.5\pm0.34$ & $-1.5\pm0.4$ & $-1.62\pm0.32$ \\
$\mathrm{O}_{32}$ & $9.62\pm0.81$ & $2.32\pm0.43$ & $10.1\pm2.1$ & $7.9\pm5.1$ & $6.24\pm0.96$ & $6.92\pm0.62$ \\
$\Delta\ion{[S}{II]}$ & $-0.41\pm0.04$ & $-0.82\pm0.05$ & $-0.34\pm0.04$ & $-0.20\pm0.03$ & $-0.45\pm0.03$ & $-0.40\pm0.04$ \\
$f_\mathrm{esc}\left(\mathrm{LyC}\right)$\tablefootmark{b} & $(0.163)$ & $(0.014)$ & $(0.179)$ & $(0.110)$ & $(0.071)$ & $(0.086)$ \\

\hline                  
\end{tabular}
\tablefoot{
\tablefoottext{a}{Assumed value.}
\tablefoottext{b}{Inferred from $\mathrm{O}_{32}$.}
}
\end{table*}

The properties we derived from the observed emission lines (listed in Table \ref{tab_lines}) of our sample are presented in Table \ref{tab_galprops2}. Particularly noteworthy is the fact that the choice of stellar population models allows \textsc{fado} to successfully reproduce the \ion{He}{II}~4686 line for all objects studied here, without any need of invoking `exotic' radiative sources as discussed in the Sect. \ref{sec_intro}.

The IMPs are ubiquitously characterised by large electron temperatures in the high-ionisation regime, with values typically found in the range of $t_\mathrm{e}\left(\ion{[O}{III]}\right)\sim 12\,750 - 17\,670~\mathrm{K}$.
As we selected the IMPs as highly ionised galaxies evident from \ion{He}{II}~4686 emission, this is reflected in the relatively large values of the ionisation parameter-sensitive line ratio $\mathrm{O}_{32}$, spanning a range of $\mathrm{O}_{32}=2.3 - 24$. Considering their respective error ranges, 16 of the IMPs are consistent with a ratio of $\mathrm{O}_{32}>5$, a value large enough to indicate a potential escape of LyC \citep{Izotov16a,Izotov16b,Izotov18a}. A galaxy's ionisation parameter can primarily be elevated by two means, them being a large intrinsic flux of ionising photons or low densities in the ISM. While the former criterion is a given here, the large scatter in $\mathrm{O}_{32}$ is suggestive of an additional mechanism influencing these values. An obvious candidate for this are density-bounded conditions, where \ion{[O}{II]} emission is diminished. We explore this scenario further in Sect. \ref{sec_lycleakage}.

The electron densities derived from \ion{[S}{II]} are found to be consistent with the low values typically found in SFGs and lie between $n_\mathrm{e}=20 - 470~\mathrm{cm}^{-3}$ when accounting for the full error range, subsequently justifying our assumption of $n_\mathrm{e}=100~\mathrm{cm}^{-3}$ for the galaxies where no physically meaningful value for $n_\mathrm{e}$ could be derived.
In terms of oxygen abundance the IMPs are markedly sub-solar, with abundances ranging between 7\% and 28\%~$Z_\sun$ at a median of 14.4\%~$Z_\sun$, assuming $\log\left(\mathrm{O}/\mathrm{H}\right)_\sun+12=8.69$ \citep{Asplund09}. This is comparable to, yet slightly lower than the LCGs and CSFGs from \citet{Izotov11,Izotov21a}, who find median values in $\log\left(\mathrm{O}/\mathrm{H}\right)+12$ of 8.11 ($26\%~Z_\sun$) and 8.0 ($20\%~Z_\sun$), respectively.

The chemical enrichment history as traced by N/O exhibits no apparent peculiarities with respect to values typically found in SFGs. The relative nitrogen abundance is strongly intertwined with a galaxy's SFH, as SNe of young, massive stars expel large amounts of oxygen into the ISM, thereby decreasing N/O. Evolved low- and intermediate-mass stars, on the other hand, can substantially increase N/O via hot bottom burning during their AGB phase \citep{Vincenzo16}. Here, as evident from Fig. \ref{fig_no}, the values we derived are broadly consistent with $\log\left(\mathrm{N}/\mathrm{O}\right)~\sim-1.6$, where $\log\left(\mathrm{N}/\mathrm{O}\right)$ plateaus for low-metallicity SFGs \citep{Vincenzo16}. In particular, the absence of a deviation towards elevated N/O values with respect to other galaxies with similar O/H suggests no dominant older stellar component is present in the IMPs. However, we note that with the large errors associated with N/O alongside the fact that other mechanisms such as gas accretion and galactic winds can alter this ratio, this conclusion remains fairly uncertain at this point.

The fact that the IMPs are otherwise quite extraordinary objects with respect to typical SFGs is reflected in their position in the $\mathrm{O}_{32}$ vs. R23 diagram (Fig. \ref{fig_R23O32}), a diagnostic diagram tracing the ionisation parameter and metallicity. Not only are they clearly offset from the bulk of the DR12 SFGs, but they also share roughly the same parameter space as many known LyC leakers, which we explore further in Sect. \ref{sec_lycleakage}.
Lastly, we remark that the visual inspection of the spectra suggests that at least two of the IMPs (J1311+3750 and J1608+0413) are contenders for studying the rarely detected \ion{[Fe}{V]}~4227 emission line \citep[e.g.][]{Thuan05,Izotov17}, but as the SDSS spectra are increasingly noisy towards the blue end, we cannot claim a reliable detection with the data at hand.

\subsection{Balmer decrements}\label{sec_bdec}

\begin{figure}
\centering
\includegraphics[width=\hsize]{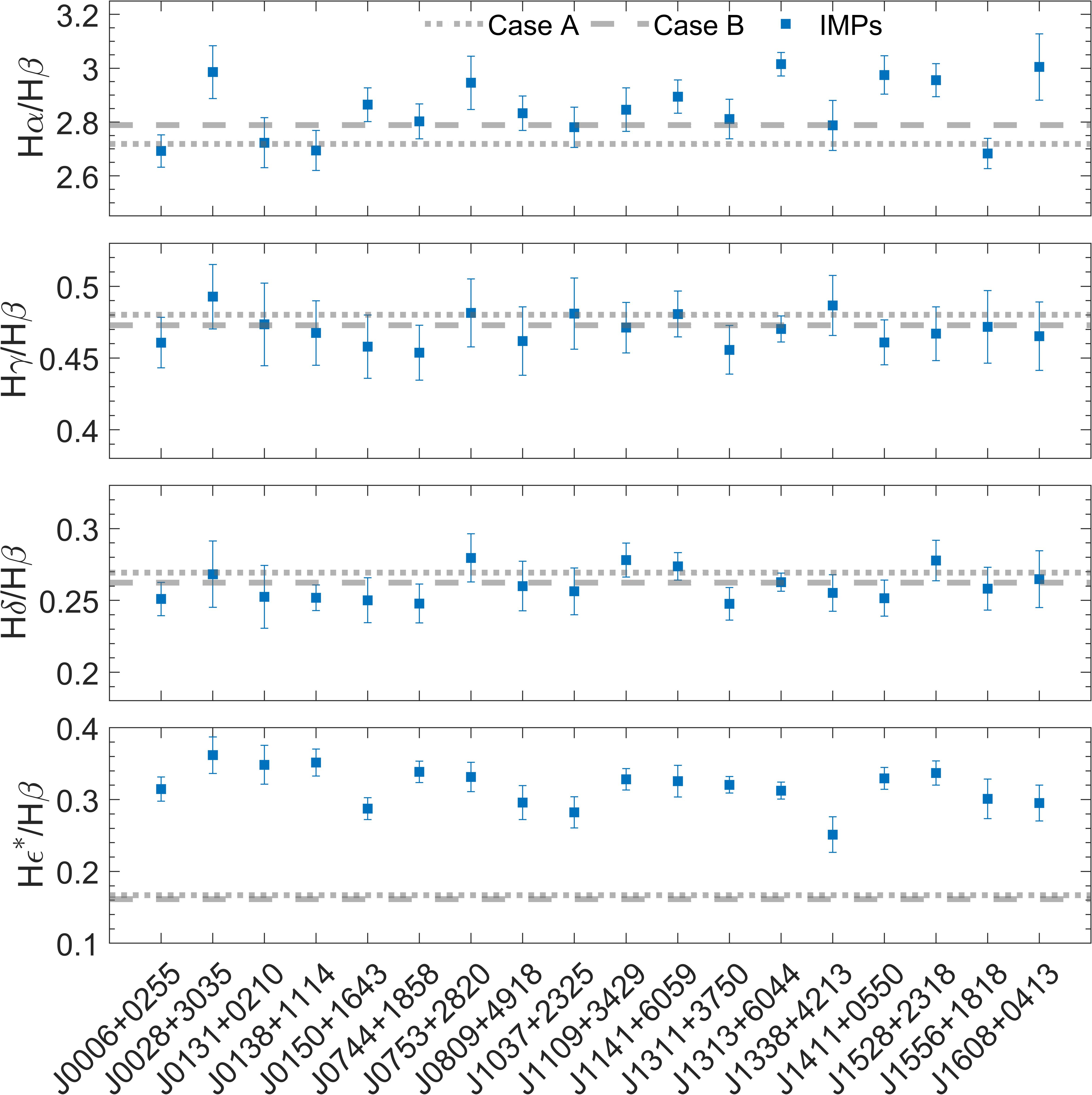}
\caption{Balmer decrements derived for the IMPs after accounting for foreground and internal extinction. The horizontal lines mark the expected theoretical values calculated by \citet{Storey95} for $n_\mathrm{e}=100~\mathrm{cm}^{-3}$ and $t_\mathrm{e}=15\,000~\mathrm{K}$, assuming Case A (dotted line) and Case B (dashed line) conditions. The $^\ast$ symbol signifies a likely blend of multiple lines (see text).}
\label{fig_bdec}
\end{figure}

A seemingly peculiar behaviour of the IMPs is found in their Balmer decrements (the line ratios of hydrogen Balmer lines relative to H$\beta$). Typically, one would expect these to be in good agreement to the theoretical values after accounting for internal reddening, as it is common procedure to scale the extinction law of choice by the H$\beta$ extinction coefficient, which itself is a mean value derived from comparison of theoretically expected to observed Balmer decrements.

As evident from Fig. \ref{fig_bdec}, several of the Balmer decrements in the IMPs disagree with the theoretical predictions. For reference, we show the Case A and B values derived by \citet{Storey95} for $n_\mathrm{e}=100~\mathrm{cm}^{-3}$ and $t_\mathrm{e}=15\,000~\mathrm{K}$, which is broadly consistent with what we found for the IMPs. We note that for the low densities present here, the assumption of tenfold higher values for $n_\mathrm{e}$ ($1\,000~\mathrm{cm}^{-3}$) has virtually no effect on the line ratios. Likewise, varying the temperature by $\pm5\,000~\mathrm{K}$ only moderately affects the line ratios and does not alleviate the tension presented here.

A marked deviation to higher relative fluxes is found in H$\alpha$/H$\beta$, but let us consider the Balmer lines of higher order first. 
A significant deviation is observed in H$\epsilon$/H$\beta$, where the values measured are a factor of $\sim2$ larger than expected. This finding has a rather straightforward explanation, as H$\epsilon$ at $3970~\AA$ is located at roughly the same wavelength as \ion{[Ne}{III]}~3967 and \ion{Ca}{II}~H at $3969~\AA$. The line fitting catalogue contained in \textsc{fado} does not encompass \ion{Ca}{II} emission, and its fitting routine would not fit the \ion{[Ne}{III]}~3967 line in any of the objects studied here, which we ascribe to a resolution effect. The values we report here for H$\epsilon$ are thus likely a blend of H$\epsilon$, \ion{[Ne}{III]}~3967, and interstellar \ion{Ca}{II}~H emission, whereas the latter's contribution is expected to be small. An additional source of uncertainty that becomes evident here is the neglection of interstellar \ion{Ca}{II}~H absorption, whose omission may negatively impact our stellar population fit. While the spectra at hand do not have the resolution sufficient to reliably disentangle these lines, by comparison to the fiducial Case B values we predict relative emission line fluxes of $100\times\left(\ion{[Ne}{III]}~3967+\ion{Ca}{II}~\mathrm{H}_{\mathrm{em+abs}}\right)/\mathrm{H}\beta=9 - 20$ with a median of $16$ in the IMPs.

H$\delta$/H$\beta$ and H$\gamma$/H$\beta$ appear to be skewed to slightly lower line ratios, though the median deviation from the theoretical value is reasonably small at $4.1\,\%$ and $1.8\,\%$, respectively. Considering the non-quantified error sources present here, such as the choice of extinction law and its application to an integrated galactic spectrum, implicitly assuming the same extinction properties independent of the line of sight, some deviations are not unexpected. With the strongest outliers deviating in the line ratios by a mere $6.5\%$ and $4.2\%$, we consider these values to be in decent agreement with the expectations.

Revisiting H$\alpha$/H$\beta$, we find these line ratios to be up to 8.1\% larger than the reported Case B values. While the mean deviation is moderately low at $\sim3.4\%$, about a third of our sample shows line ratios of $\mathrm{H}\alpha/\mathrm{H}\beta\ga2.9$, reminiscent of what is typically found in AGN. We note that as we find indications for an optically thin ISM throughout this paper, Case A conditions might even be the better approximation here, but this only worsens the disagreement to a median deviation of 4.5\% at a maximum of 10.9\%. These values are highly suggestive of an additional low-energy phenomenon that is able to elevate the electron bound by hydrogen to $n=3$, resulting in a net increase in H$\alpha$ flux without altering the higher-order line ratios. We speculate that collisional excitation in a large-scale galactic wind or conversely, during the accretion of large volumes of extragalactic gas, may account for this observation. Ultimately, deeper high-resolution spectra are needed to appropriately characterise the ISM dynamics that would be supportive of this argument. Alternatively, it could turn out to be an effect of technical rather than astrophysical nature, and may be the result of the best fit from population spectral synthesis being a mix of stellar populations that underpredict the total amount of stellar H$\alpha$ absorption.

In Table \ref{tab_galprops1} we list the recalculated (s)SFR(H$\alpha$) values denoted as $\mathrm{(s)SFR}_\mathrm{B}$ based on the assumption that H$\alpha$ shows such an flux excess and the Case B values better represent the star-formation induced H$\alpha$ flux. While this slightly diminishes the values we find (to a median sSFR of $\sim 7.08~\mathrm{Gyr}^{-1}$), it does not qualitatively alter our conclusion. The same is true for the diagnostics incorporating H$\alpha$ (Figs. \ref{fig_SB} and \ref{fig_BPT}).

\subsection{Star-formation histories}\label{sec_sfh}

\begin{figure*}
\centering
\includegraphics[width=0.825\hsize]{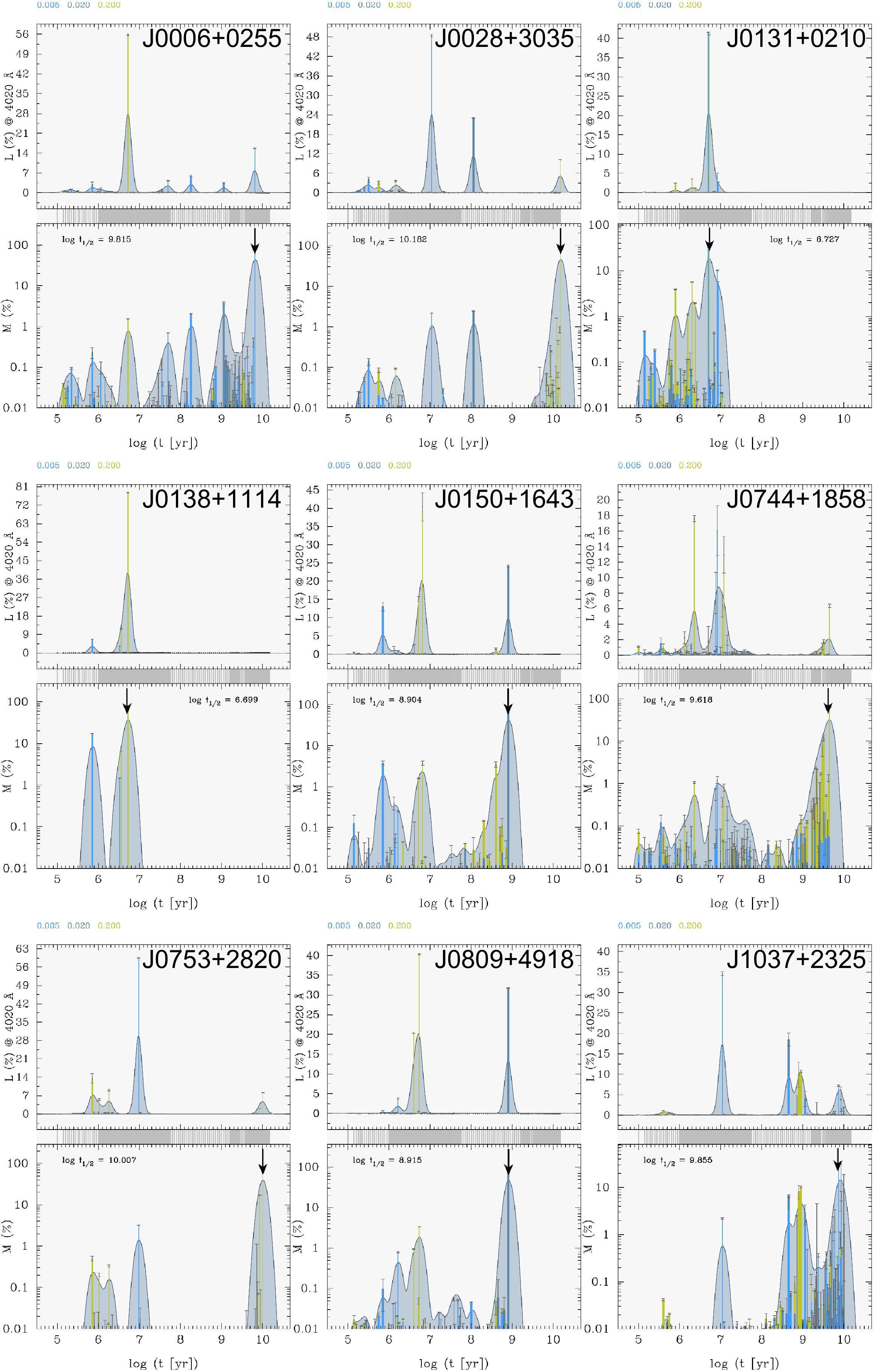}
\caption{SFHs of the first half of the IMPs. The top panel in each plot pair shows the SFH weighted by the stellar population's luminosity at $4020~\AA$, whereas the bottom panel shows the mass-weighted SFH. Here, the black arrow marks the time where half of the galaxy's mass was assembled, as indicated by the $t_{1/2}$ inset. The bar colours correspond to the stellar metallicity as indicated by the numbers above each plot pair, in units of $Z_\sun$.}
\label{fig_sfh1}
\end{figure*}

\begin{figure*}
\centering
\includegraphics[width=0.825\hsize]{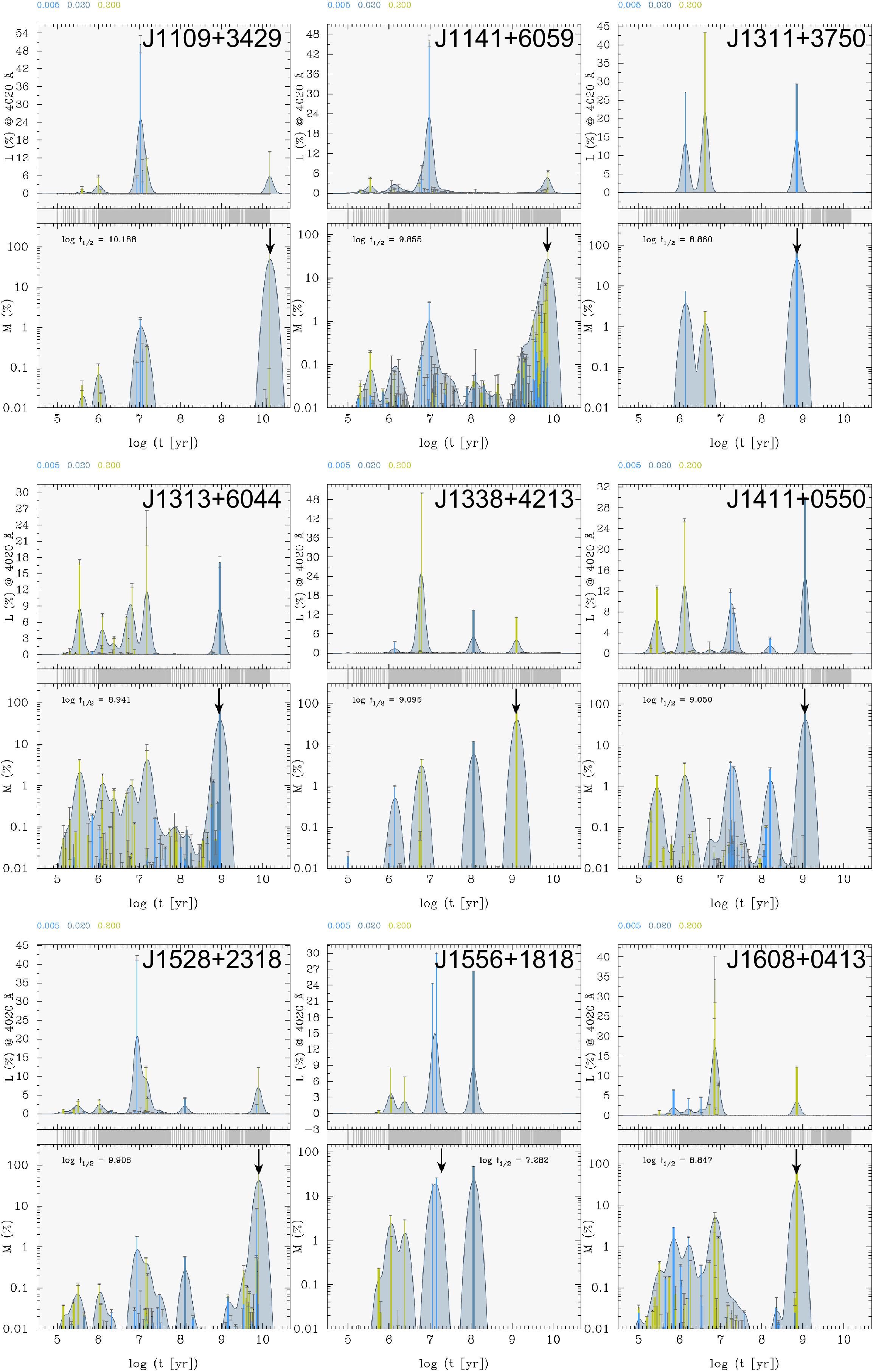}
\caption{Same as Fig. \ref{fig_sfh1} for the second half of the IMPs.}
\label{fig_sfh2}
\end{figure*}

The population synthesis by \textsc{fado} reveals the IMPs to have undergone similar star formation episodes in the past. We show the mass- and luminosity-weighted SFHs of the IMPs in Figs. \ref{fig_sfh1} and \ref{fig_sfh2}.
As evident from their SFHs, the IMPs typically contain a population of old stars at $\sim1-10~\mathrm{Gyr}$ that dominate in mass. This is different for J0131+0210, J0138+1114, and J1556+1818, whose SEDs are consistent with that of a galaxy no older than $\sim5-120~\mathrm{Myr}$.

The SFHs subsequent to the formation of the old population, where present, vary from object to object. In about a third of the sample, the majority of the recent Gyr is a period of relative quiescence, whereas the other galaxies have undergone one or more bursts of star-formation during that time.

The most recent SFH, however, appears to be virtually the same for most of the IMPs: roughly $5-15~\mathrm{Myr}$ ago, star formation in the IMPs had a major resurgence, where a young, typically luminous stellar population was formed. This then is followed by one or several bursts several Myr later, in which the currently youngest stellar generation with ages of $0.1-1.3~\mathrm{Myr}$ was formed. In two of the IMPs (J1037+2325, J1109+3429), the youngest stellar population is less dominant with respect to those of the remaining galaxies, providing less than $\sim0.1\%$ of the galaxies' total mass. The overall pattern, however, is the same as in the other objects.

The situation we find in the majority of the IMPs is remarkably similar to that in other metal-poor DGs, where this two-stage starburst scenario appears to be linked to superbubbles and galactic outflows, for instance in NGC~1569 \citep{Heckman95,Vallenari96,Origlia01} or NGC~5253 \citep{MonrealIbero10,Calzetti15}. Several LCGs, such as the GPs studied by \citet{Amorin12}, have been shown to have a similar SFH. In particular, we highlight the likeness to some well-studied (candidate) LCEs. A prominent example is ESO~338-IG04, whose SFH has been reconstructed from the study of 124 star clusters by \citet{Oestlin03}. While they found a significant number of Gyr old clusters to be present, a population of massive star clusters was formed $~8 - 11~\mathrm{Myr}$ ago, followed by the formation of numerous young stars within the last 2~Myr. A similar situation is found in the GP analogue NGC~2366, where a 3~Myr star cluster has carved a superbubble into the ISM which is now ionised by the young <1~Myr stellar population in the \ion{H}{II} region Mrk~71 \citep{Micheva17}. Tololo~1247-232 likewise was found to contain a young, two-stage starburst at 12~Myr and <4~Myr by \citet{Micheva18}, who suggest this feature might be common in LCEs. \citet{Zastrow13} similarly conclude that a multi-stage, young starburst may be required for the transmission of LyC radiation into the IGM.

\subsection{Ionising sources}

\begin{figure*}
\centering
\includegraphics[width=\hsize]{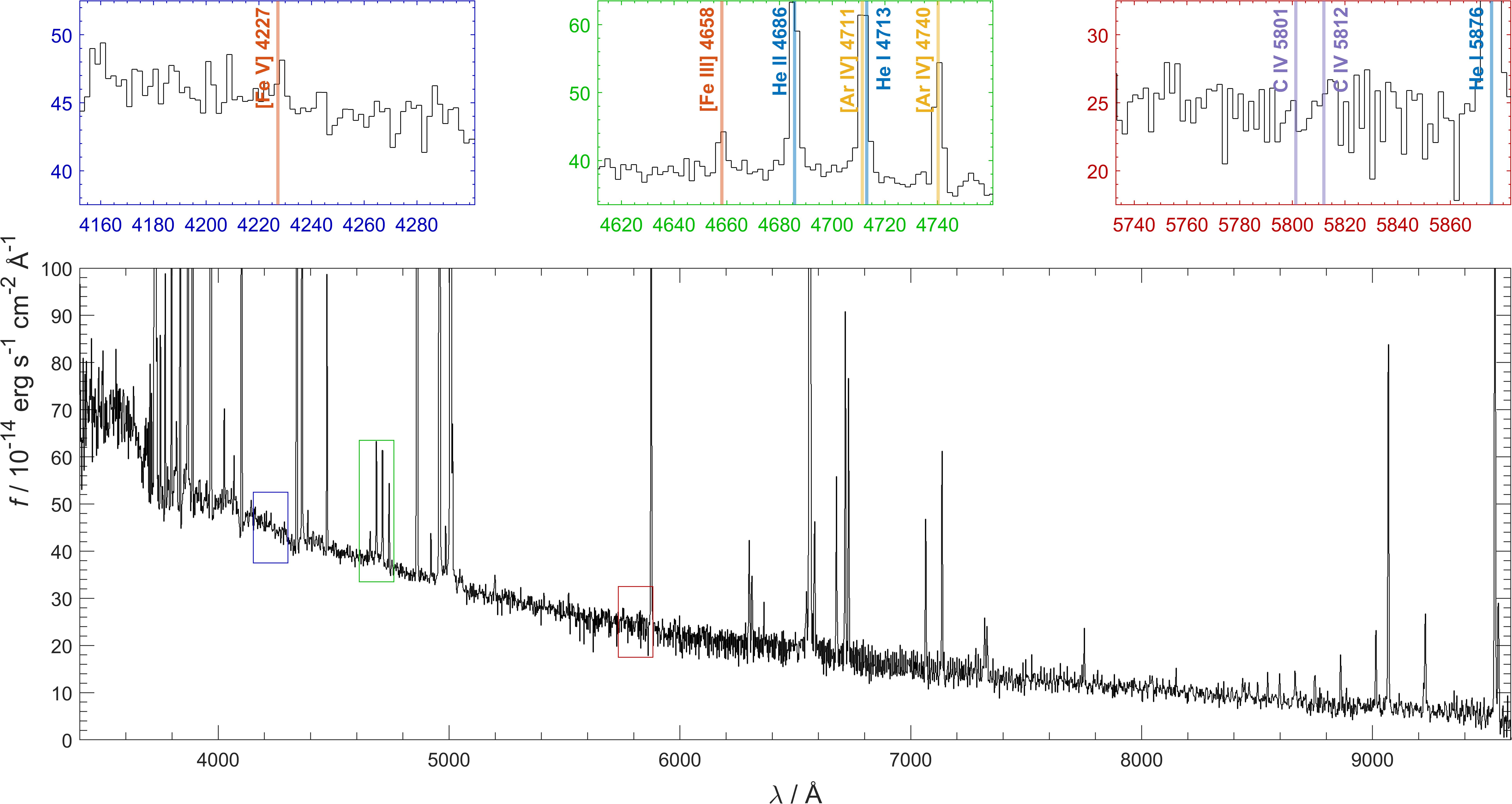}
\caption{$3400~\AA - 9600~\AA$ region of the stack of eighteen redshift-corrected SDSS spectra of the IMP sample, binned in $1~\AA$ bins and truncated in flux density $f$ to enhance the visibility of weak spectral features (bottom row). The coloured boxes indicate the regions shown in the blow-ups in the top row, centred at \ion{[Fe}{V]}~4227, \ion{He}{II}~4686, and \ion{C}{IV}~5808 (blue, green, and red insets, respectively). Noteworthy emission lines (both present and absent) are marked by coloured vertical lines. The SDSS wavelengths were converted to vacuum wavelengths following \citet{Morton91}.}
\label{fig_specstack}
\end{figure*}

\begin{figure*}
\centering
\includegraphics[width=\hsize]{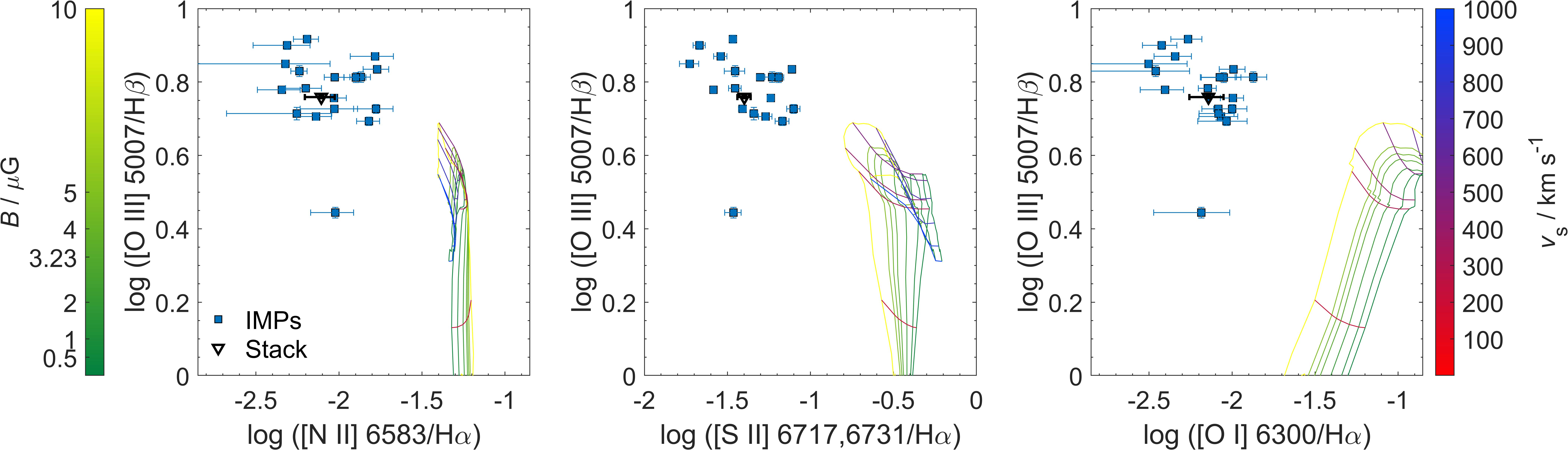}
\caption{Three commonly used BPT diagrams for shock diagnostics; symbols are the same as in Fig. \ref{fig_SB}. Solid lines show the shock+precursor SMC model grids from \citet{Allen08} with their parameters (shock velocity $v_\mathrm{s}$ and transverse magnetic field $B$) coloured as indicated by the colour bars.}
\label{fig_shockdiags}
\end{figure*}

While strong \ion{He}{II}~4686 emission is a feature typically seen in galaxies with an AGN producing a hard radiation field, we are fairly confident in excluding this possibility for the IMPs as per our selection criterion. This conclusion is further supported by the absence of spectral features characteristic of an AGN, such as broadened emission lines. Hence, it is most worthwhile to investigate the origin driving this highly energetic emission, as it may shed light on the mechanisms facilitating LyC escape in low-metallicity SFGs during the EoR.

Judging from their low oxygen abundances and large SFRs, it stands to reason that the stellar contribution to the high-energy part of the spectrum mainly comes from a young starburst. Our reconstructed SFHs from Sect. \ref{sec_sfh} confirm this, thus we are confident that OB stars contribute significantly to the ionisation of the ISM. However, similar to other LCGs \citep[e.g.][]{Amorin12,Clarke21}, the IMPs appear to harbour older stellar populations (Sect. \ref{sec_sfh}). Due to the exposed, hot cores of remnant stars one may reasonably expect to find here, the intrinsic LyC production in the IMPs is likely elevated by these older populations.

Features associated with WR stars, such as the `blue WR bump' that one might expect at $\sim4650~\AA$ \citep[e.g.][]{Schaerer99}, are remarkably absent even in the stacked spectrum of the IMPs (Fig. \ref{fig_specstack}, 2nd panel in the top row). There is a tentative detection of \ion{[Fe}{III]}~4658 in this wavelength regime, but as the WR bump typically is a broad spectral feature and we furthermore detect a rather strong \ion{[Fe}{III]}~4986 signal (see Fig. \ref{fig_specstack}, bottom panel), we deem this feature likely is of nebular origin. In essence, unless a significant population of low-metallicity nitrogen WR stars \citep[in which this feature is pronouncedly fainter than in higher-metallicity WRs,][]{Crowther06} is present in our sample, WR stars as the origin of ionising radiation can be excluded. Further affirmation for this is found in the absence of the `red WR-bump' (Fig. \ref{fig_specstack}, 3rd panel in the top row), a carbon Wolf-Rayet feature of broad \ion{C}{IV}~5808 emission \citep{Schaerer99}. This conclusion is in line with other studies of SFGs, where the relative amount of WR stars reportedly decreases with metallicity \citep[e.g.][]{Guseva00,Shirazi12}. Alternatively, as WR stars constitute a late evolutionary stage of massive stars, this again hints at the fact that the most recent starburst in the IMP galaxies is still young, which in turn further affirms that OB stars producing ionising radiation are present here.

A similar argument can be made for the presence of HMXBs, as these objects require one of the binary partners to have collapsed into a compact object, such as a neutron star or a black hole. Then again, the most massive of stars evolve rapidly and are known to leave behind such compact objects at the end of their lives. Considering the most recent starbursts shown in Figs. \ref{fig_sfh1} and \ref{fig_sfh2}, a large fraction of OB stars created at the onset of the recent star-forming episodes have already evolved into their remnant form. Thus, given a sufficient fraction of massive stars have been born in a binary configuration with an orbit close enough so that mass overflow from their companion can take place, their accretion disks can in principle contribute significant amounts of high-energy photons even in a relatively young starburst. In the case of the IMPs no archival data from Chandra or XMM-Newton is available, so we currently cannot deduce anything about the presence or absence of HMXBs. Future investigations should take this into account, as HMXBs appear to be more abundant in metal-poor dwarf galaxies and thus may contribute significantly to the ionising radiation field \citep{Brorby14}.

Higher resolution and at best, spatially resolved data will be required to conclusively study the contribution of shocks to the ionisation of the ISM. The principle requirements for the occurence of shocks are given in star-forming regions, as kinematic phenomena with supersonic motion can for example be found in stellar winds and the expanding shells of SNe. For a preliminary evaluation, Fig. \ref{fig_shockdiags} shows the IMPs in three commonly used BPT diagrams, where we overplotted the model calculations from \citet{Allen08}. Shown are the tracks from their SMC model (shock front plus photoionised precursor at a pre-shock density of $n=1~\mathrm{cm}^{-3}$), as that model at $\log\left(\mathrm{O}/\mathrm{H}\right)+12=8.03$ most closely matches the metallicity found in the IMPs. As evident from these diagrams, even the most favoured models with shock velocities $v_\mathrm{s}=400~\mathrm{km}~\mathrm{s}^{-1}$ and a magnetic field of $B=10~\mu\mathrm{G}$ predict line ratios strongly differing from the ones found in the IMPs. Given that we are considering integrated spectra, this is not surprising, as the ISM as a whole is obviously affected by a superposition of many ionising mechanisms and shocks can be expected to be more localised phenomena. Still, we can conclude that the ISM's ionisation is unlikely to be dominated by shocks, although their contribution in supposed galactic outflow channels may be more pronounced than suggested in the diagnostics used here.
Future infrared observations may aid in disentangling the shock component from the other ionisation mechanisms and provide another avenue for comparisons to high-$z$ objects \citep[e.g.][and references therein]{Brinchmann22}.

\subsection{LyC leakage}\label{sec_lycleakage}

\begin{figure*}
\centering
\includegraphics[width=\hsize]{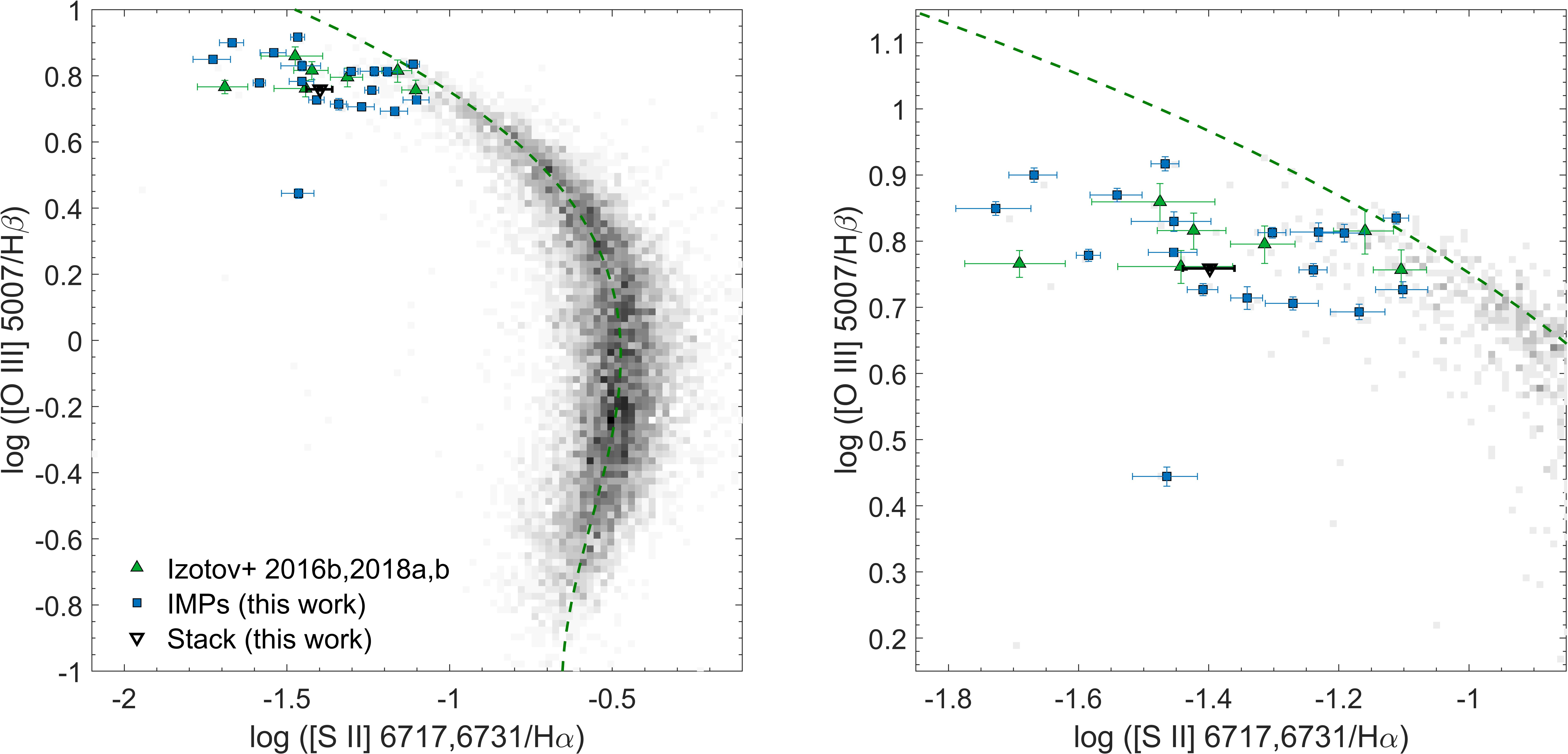}
\caption{$\log\left(\ion{[O}{III]}~5007 / \mathrm{H}\beta\right)$ vs. $\log\left(\ion{[S}{II]}~6717,6731 / \mathrm{H}\alpha\right)$ diagram including the \ion{[S}{II]} ridge line from \citet{Wang21} (green dashed line). Symbols are the same as in Fig. \ref{fig_R23O32}. Left: Panel highlighting the IMPs' positions in the parameter space covered by SDSS DR12 galaxies. Right: Zoom-in on the region occupied by the IMPs.}
\label{fig_SIIdef}
\end{figure*}

\begin{figure}
\centering
\includegraphics[width=\hsize]{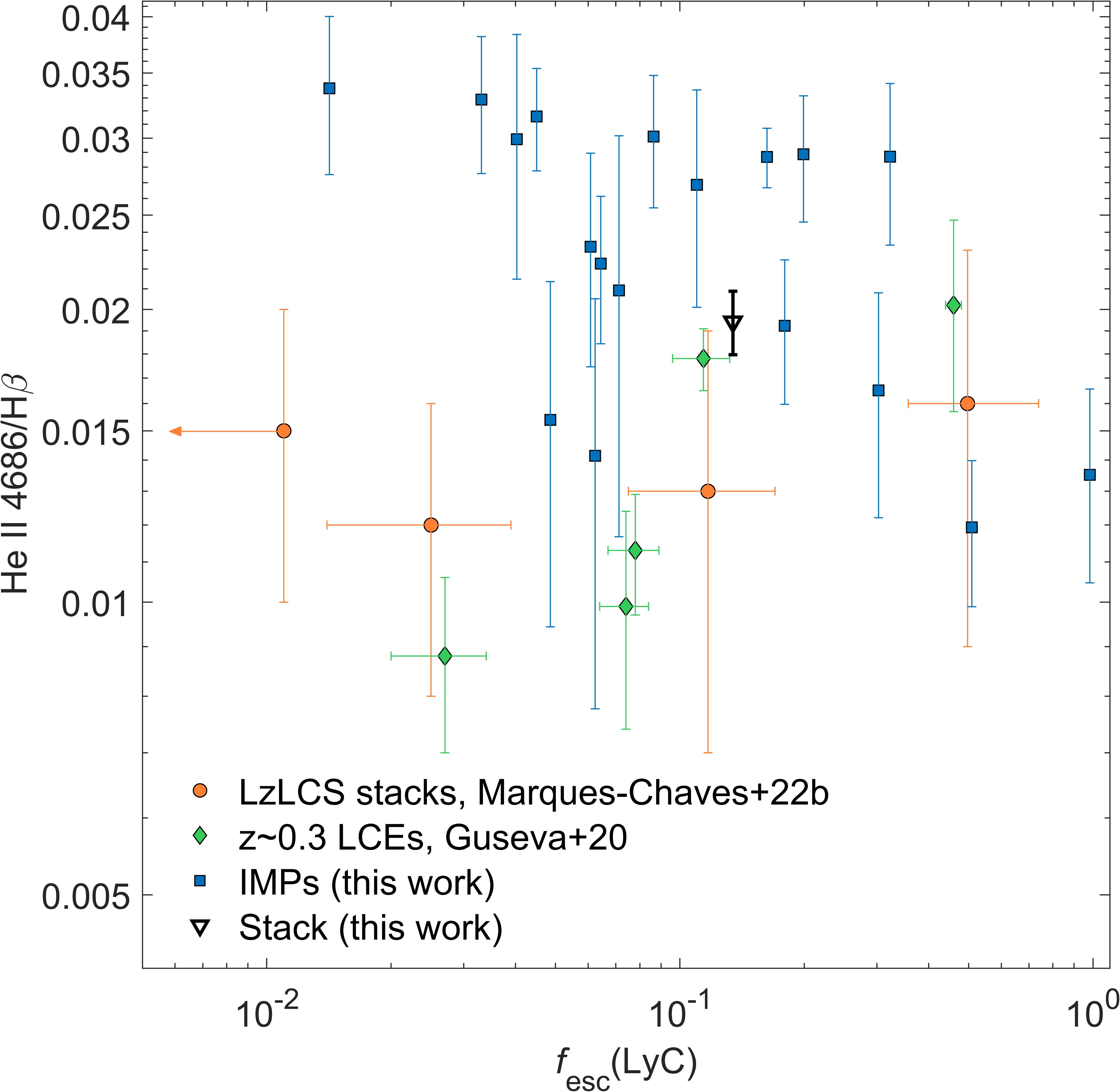}
\caption{Spectral hardness as traced by \ion{He}{II}~4686/H$\beta$ vs. LyC escape fraction $f_\mathrm{esc}\left(\mathrm{LyC}\right)$. Blue squares represent the IMPs, whereas the position of their stack is shown as a black downward-facing open triangle. We show no error bars in $f_\mathrm{esc}\left(\mathrm{LyC}\right)$, as the values we infer are mere estimates from $\mathrm{O}_{32}$. The stacks from the Low-redshift Lyman Continuum Leaker Survey as provided in \cite{MarquesChaves22b} are shown as orange circles, and the LCEs studied in \cite{Guseva20} as green diamonds.}
\label{fig_HeIIHB}
\end{figure}

\begin{figure}
\centering
\includegraphics[width=\hsize]{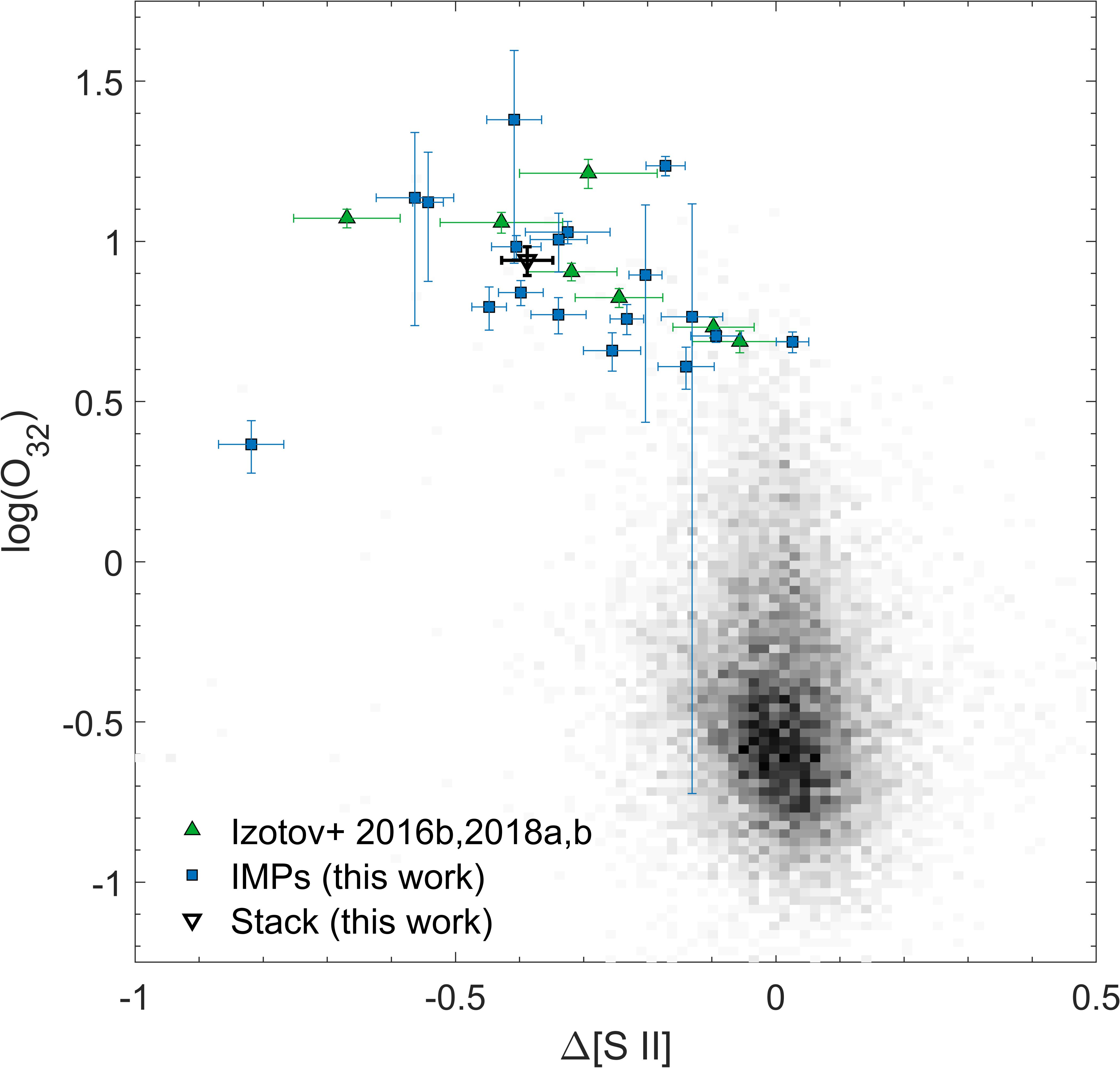}
\caption{$\log\left(\mathrm{O}_{32}\right)=\log\left(\ion{[O}{III]}~5007 / \ion{[O}{II]}~3727\right)$ vs. $\Delta\ion{[S}{II]}$ diagram. Symbols are the same as in Figure \ref{fig_R23O32}.}
\label{fig_deltaSIIO32}
\end{figure}

The lack of UV spectra for the objects studied here makes a direct determination of their LyC escape fraction impossible. That said, the IMPs as \ion{He}{II}~4686 emitters clearly have a significant intrinsic production of photons beyond the Lyman edge $<912~\AA$. This by itself already qualifies these galaxies as remarkable objects for astrophysical research, and in conjunction with the large $\mathrm{O}_{32}$ values as previously discussed makes them suitable candidates for studies concerning LyC leakage, hence we explore the possibility a little further with the optical data at hand.

To get a feel for the relevance in terms of LyC leakage, a first guess for $f_\mathrm{esc}\left(\mathrm{LyC}\right)$ can be obtained from Eq. \ref{eqn_fescO32}. As pointed out in Sect. \ref{sec_lycdiag}, $\mathrm{O}_{32}$ as a LyC leakage diagnostic is problematic, and the calibration from \citet{Chisholm18} is based on low sample statistics, so we caution against an overinterpretation of the results. Performing the calculation yields LyC escape fractions between 1.4\% and 98\% in the IMP sample, the upper bound constituted by J1311+3750. Taken at face value, this would mean that eight of the eighteen IMPs were to be characterised by a cosmologically significant escape fraction >10\%, with eight more candidates residing at 4-10\%. The fraction of strong LCEs increases to 10/18 if one utilises the $\mathrm{O}_{32} - f_\mathrm{esc}\left(\mathrm{LyC}\right)$ relation from \citet{Izotov18a}, and all but two galaxies are found to have $f_\mathrm{esc}\left(\mathrm{LyC}\right)>4~\%$.

We note that the inferred LyC escape fraction of our sample shows no correlation with the relative strength of \ion{He}{II}~4686 (Fig. \ref{fig_HeIIHB}), which is in line with the few other studies considering this potential connection \citep{Guseva20,MarquesChaves22b}. As their authors have stated, this implies that the physical properties traced by \ion{He}{II}~4686 do not govern the LyC escape fraction. However, we remark that the IMPs show considerably larger \ion{He}{II}~4686/H$\beta$ ratios than most established LCEs. Considering these galaxies selected as \ion{He}{II} emitters show strong indications for appreciable escape fractions, we suggest that a mechanism responsible for the \emph{detectability} of \ion{He}{II} may be related to LyC escape. Considering that the bulk of \ion{He}{II} emission will originate close to the sources of hard radiation, an efficient removal of neutral gas is a strong contender here, which we explore further below.

The indications for LyC leakage in the IMPs are further reinforced by the strong deficit in $\ion{[S}{II]}~6717,6731 / \mathrm{H}\alpha$ with respect to typical SFGs, and following the definition of \citet{Wang21}, they span a range between $\Delta\ion{[S}{II]}=0.026$ and $-0.82$ (Table \ref{tab_galprops2}). The markedly negative values indicate that large volumes in the IMPs may be density-bounded, and as such provide conditions strongly favourable for LyC leakage. Figure \ref{fig_SIIdef} shows their position in the plot as worked out in \citet{Wang21} alongside with a few other known LyC leakers, demonstrating their similarity in this parameter space.

In principle, such conditions could likewise arise in an ionisation-bounded nebula, given a hard enough radiation field such that the inner regions where the higher-ionisation emission lines originate span a larger spatial extent and as such, dominate the line flux ratio relative to that of a lower ionisation species. If this were the case, the \ion{[S}{II]} deficiency could be the result of most of the sulphur being doubly ionised. In the spectra of the IMPs, we detect both $\ion{[S}{III]}~9069$ and $\ion{[S}{III]}~9531$ lines (redshift permitting). While some variation is present, the overall ratio of $\mathrm{S}_{32}=\left(\ion{[S}{III]}~9069 + \ion{[S}{III]}~9531\right) / \ion{[S}{II]}~6717,6731$ shows no obvious excess. For our stacked spectrum, we find $\log\mathrm{S}_{32} \sim 0.3$, which at $\log\left(\ion{[O}{III]}~5007 / \mathrm{H}\beta\right) \sim 0.94$ appears to be compatible with the extrapolation of the running median of $z=0$ SFGs as presented in \citet[their Fig. 4]{Sanders20}. Judging from this, we conclude that our previous findings are primarily governed by the ISM structure rather than the hardness of the radiation field.

Equally remarkable is the area the IMPs reside in when comparing the two best optical LyC leakage indications, $\mathrm{O}_{32}$ and $\Delta\ion{[S}{II]}$ (Fig. \ref{fig_deltaSIIO32}), also adapted from \citet{Wang21}. Here, the IMPs are well offset from the bulk of the DR12 galaxies, and roughly occupy the same region as many confirmed local LyC leakers. While \citet{Wang21} report no statistically significant correlation between a galaxy's $f_\mathrm{esc}\left(\mathrm{LyC}\right)$ and $\log\left(\mathrm{O}_{32}\right)$ vs. $\Delta\ion{[S}{II]}$, it is evident that galaxies with large escape fractions generally reside in the upper left of this diagram \citep[][their Fig. 3]{Wang21}, which is equally true for the IMPs.

Our reconstruction of the SFHs (Sect. \ref{sec_sfh}) offers a tempting explanation how these conditions may have come to be. While the IMPs appear to actively form young stars, a major burst in star formation has occured within the last $\sim10~\mathrm{Myr}$ in a majority of these objects. Within the timespan following this burst, all newly formed massive stars have since exploded as SNe, which constitutes a kinematic phenomenon that may have driven a significant fraction of the galaxies' ISM out of their shallow potential wells. The ionising photons provided by the younger stellar generation, whose formation process may well have been triggered by or benefitted from these SN shock waves, will then find ISM conditions strongly favourable for their transmission into the IGM.

While some of the IMPs apparently lack the 10~Myr population, their youngest stellar generation is systematically more massive than that of the two-stage starburst galaxies. Here, the increased ionising flux due to the larger number of OB stars (relative to galaxy mass) may be responsible for a reduction in the neutral gas fraction in these galaxies. Alternatively, the larger number of SNe from massive (O) stars may already have provided ISM conditions similar to what we deduced for the rest of the IMPs.

Our attempts to fit an additional broad Gaussian component to some of the stronger emission lines (e.g. H$\alpha$, \ion{[O}{III]}~5007) did not yield any appreciable results. If such a component, indicative of a large-scale outflow \citep[e.g.][]{Castaneda90}, is present in the spectra of the IMPs, deeper observations are required for its detection. In about half of our sample, we can reproduce the H$\alpha$ and \ion{[O}{III]}~5007 lines with a double-Gaussian fit, where the peak separations indicate a velocity of $\sim 80\,\mathrm{km}\,\mathrm{s}^{-1}$ in the blue-shifted component. For an unambiguous detection of these components, however, higher resolution spectra are needed.

\section{Summary and conclusion}\label{sec_conc}

By applying a single selection criterion based on the \ion{He}{II}~4686 recombination line (Eq. \ref{eqn_SB12}), we have extracted a sample of twenty galaxies from the SDSS DR12. Their position in the BPT diagram (Fig. \ref{fig_BPT}) reveals that eighteen (90\%) of these are clustered together in the respective parameter space, suggesting simliar inherent properties, which our subsequent analysis confirms. These previously undiscussed galaxies, dubbed 'IMPs', are revealed to be metal-poor, star-forming dwarf galaxies. Their highly ionised ISM appears to be predominantly photoionised by a young stellar population, with an additional contribution of an older ($1-10~\mathrm{Gyr}$), evolved population in most of the galaxies. A dominant contribution from AGN, shocks, and WR stars can be ruled out at this stage, whereas the presence or absence of HMXBs remains an open question for now.

We find the IMPs with their low stellar masses, large sSFRs and markedly sub-solar metallicity blend well into the local populations of LCGs and CSFGs \citep{Izotov11,Izotov21a}, which share many properties with high-$z$ SFGs. This likeness is further reinforced by their large values in EW$\left(\mathrm{H}\beta\right)$ and EW$\left(\ion{[O}{III]}~5007\right)$ along with their compact morphology. In all of these properties, however, the IMPs typically occupy the tail of the distribution found in the CSFGs, them being low-mass, low-metallicity objects with emission lines of large EWs, implying these galaxies are fairly extreme objects among the population of local analogues to high-$z$ SFGs. This is notably similar to the CSFG subset constituted by the `blueberry galaxies', which themselves are low-$z$ LCE candidates with $\log\left(M_\star/\mathrm{M}_\sun\right) \sim 7$ and $\log\left(\mathrm{O}/\mathrm{H}\right)+12 \sim 7.7$ \citep{Yang17}.

Judging from the elevated $\mathrm{O}_{32}$ values and pronounced $\ion{[S}{II]}$ deficiency, our results strongly point towards large density-bounded volumes of ISM within the IMPs. This leads us to suspect that a majority of them are very likely LyC leakers, making them excellent candidates for follow-up studies with respect to LyC photon loss. Ultimately, future acquisition of spectroscopic FUV data is inevitable to unequivocally confirm this. Direct detection of LyC radiation will require the next generation of FUV instruments, though, as the sensitivity curve of HST's COS does not permit the detection of faint LyC at these low redshifts.

The IMPs' recent SFHs are found to be similar and offer a tempting scenario that may explain how conditions favourable for LyC leakage have come to be. Most of the galaxies have formed a stellar population which constitutes $\sim 0.1\% - 10\%$ of their total stellar mass some $5 - 15~\mathrm{Myr}$ ago. This timescale is sufficient for the massive stars of this population to have exploded as SNe by now, and the kinetic energy injected into the surrounding matter by these may have served as a mechanism for clearing low-density cavities in the ISM, through which LyC photons then may ultimately escape into the IGM.

If found to be true, these galaxies then would constitute the most close-by LyC leakers, in parts only superseded by Haro~11 \citep{Bergvall06}, Mrk~54 \citep{Leitherer16} and Tol~1247-232 \citep{Leitet13}. Moreover, it would demonstrate the validity of using high-ionisation nebular lines as a selection criterion in search of LCEs. Our $\mathrm{O}_{32}$-based estimates of $f_\mathrm{esc}\left(\mathrm{LyC}\right)$ suggest a detection rate of significant (>10\%) escape fractions for at least eight (40\%) of the studied SFGs, though the \ion{[S}{II]} deficient nebular emission indicates that the remaining galaxies likewise are good LCE candidates.

\begin{acknowledgements}

Many thanks go to our anonymous referee, whose diligent report greatly improved the quality of this paper.

The authors would like to express their gratitude towards P. Papaderos for providing further insights into the inner workings of \textsc{fado}. Additionally, we thank L. Dirks for many helpful discussions.

AUE, DJB and AW acknowledge funding from the German Science Foundation DFG, via the Collaborative Research Center SFB1491 `Cosmic Interacting Matters - From Source to Signal'.

This research has been made possible due to the immense data collected in the SDSS project.

Funding for SDSS-III has been provided by the Alfred P. Sloan Foundation, the Participating Institutions, the National Science Foundation, and the U.S. Department of Energy Office of Science. The SDSS-III web site is \href{http://www.sdss3.org/}{http://www.sdss3.org/}.

SDSS-III is managed by the Astrophysical Research Consortium for the Participating Institutions of the SDSS-III Collaboration including the University of Arizona, the Brazilian Participation Group, Brookhaven National Laboratory, Carnegie Mellon University, University of Florida, the French Participation Group, the German Participation Group, Harvard University, the Instituto de Astrofisica de Canarias, the Michigan State/Notre Dame/JINA Participation Group, Johns Hopkins University, Lawrence Berkeley National Laboratory, Max Planck Institute for Astrophysics, Max Planck Institute for Extraterrestrial Physics, New Mexico State University, New York University, Ohio State University, Pennsylvania State University, University of Portsmouth, Princeton University, the Spanish Participation Group, University of Tokyo, University of Utah, Vanderbilt University, University of Virginia, University of Washington, and Yale University.

This research has made use of the NASA/IPAC Extragalactic Database (NED), which is funded by the National Aeronautics and Space Administration and operated by the California Institute of Technology.

All plots in this paper have been created with MATLAB R2021b (\href{https://mathworks.com/products/matlab.html}{https://mathworks.com/products/matlab.html}), except those in Figs. \ref{fig_sfh1} and \ref{fig_sfh2}, which are part of the default \textsc{fado} output.
\end{acknowledgements}

\bibliographystyle{aa}
\bibliography{bibliogr.bib}

\begin{appendix}\twocolumn
\section{Outlier galaxies}\label{app_outl}

\begin{figure}[hpt]
\centering
\includegraphics[width=\hsize]{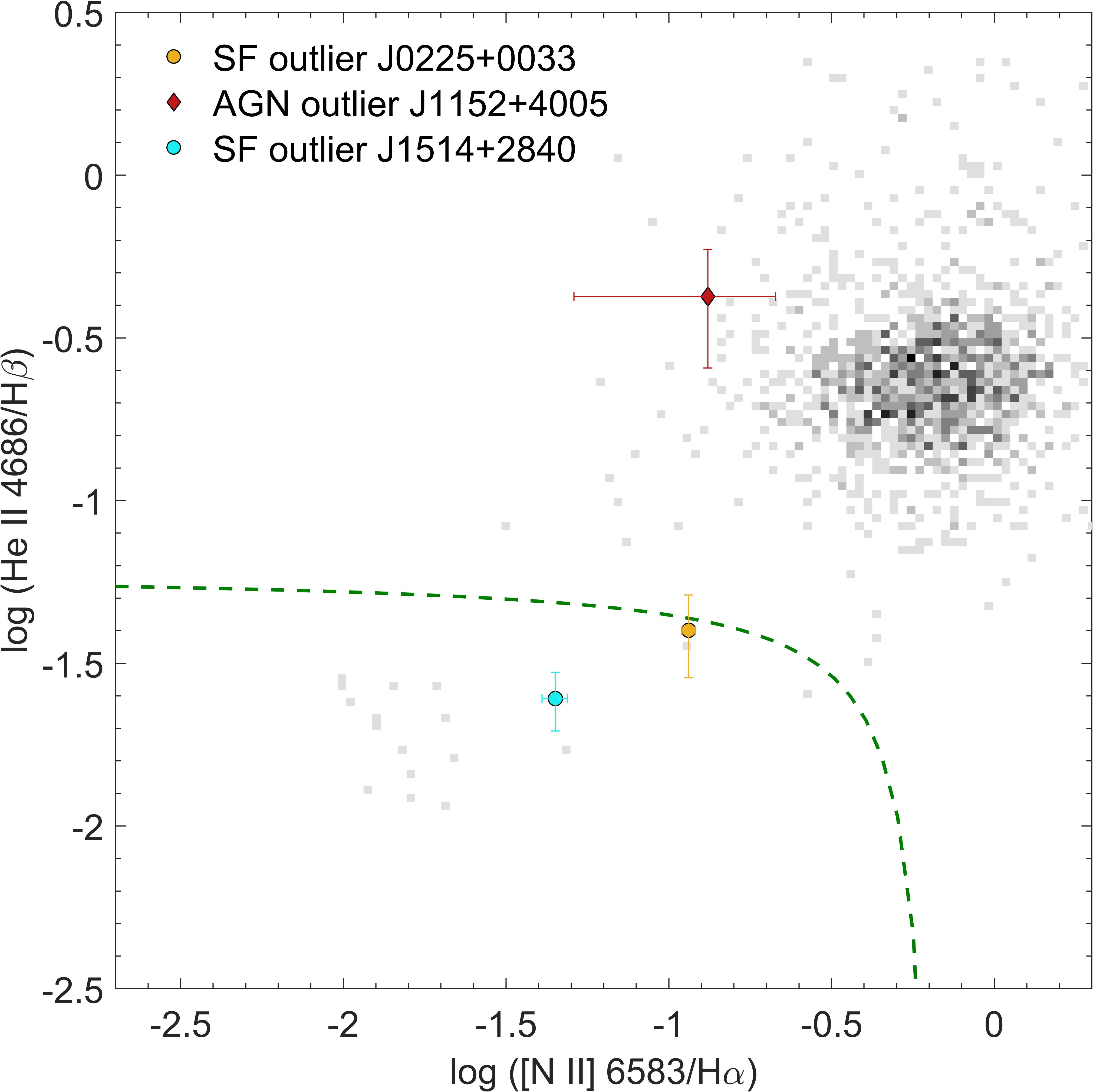}
\caption{Same as Fig. \ref{fig_SB}, highlighting three of the four outlier galaxies removed from our sample as indicated in the plot legend.}
\label{fig_SBoutl}
\end{figure}

\begin{figure}[hpt]
\centering
\includegraphics[width=\hsize]{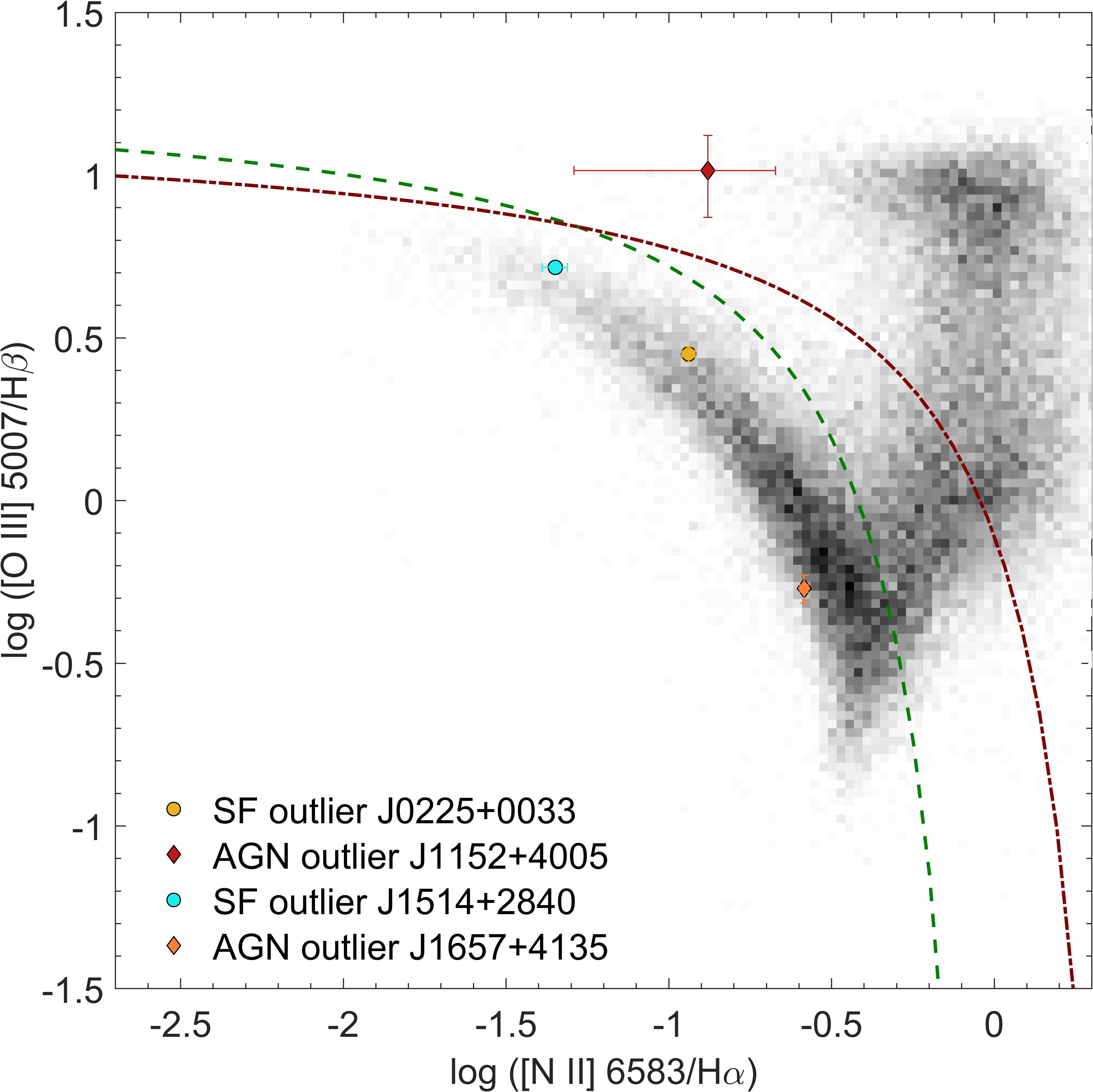}
\caption{Same as Fig. \ref{fig_BPT}, highlighting the four outlier galaxies removed from our sample as indicated in the plot legend.}
\label{fig_BPToutl}
\end{figure}

In this section we briefly discuss the four outlier galaxies we removed from our final sample. In Figs. \ref{fig_SBoutl} and \ref{fig_BPToutl} we reproduce our SFG/AGN diagnostic diagrams (Figs. \ref{fig_SB} and \ref{fig_BPT}) to clarify the positions of the individual objects in the plot. Their general characteristics are listed in Table \ref{tab_galprops1outl}, while Table \ref{tab_galprops2outl} contains the derived nebular characteristics considered in this work.

\begin{figure}[hpt]
\centering
\includegraphics[width=\hsize]{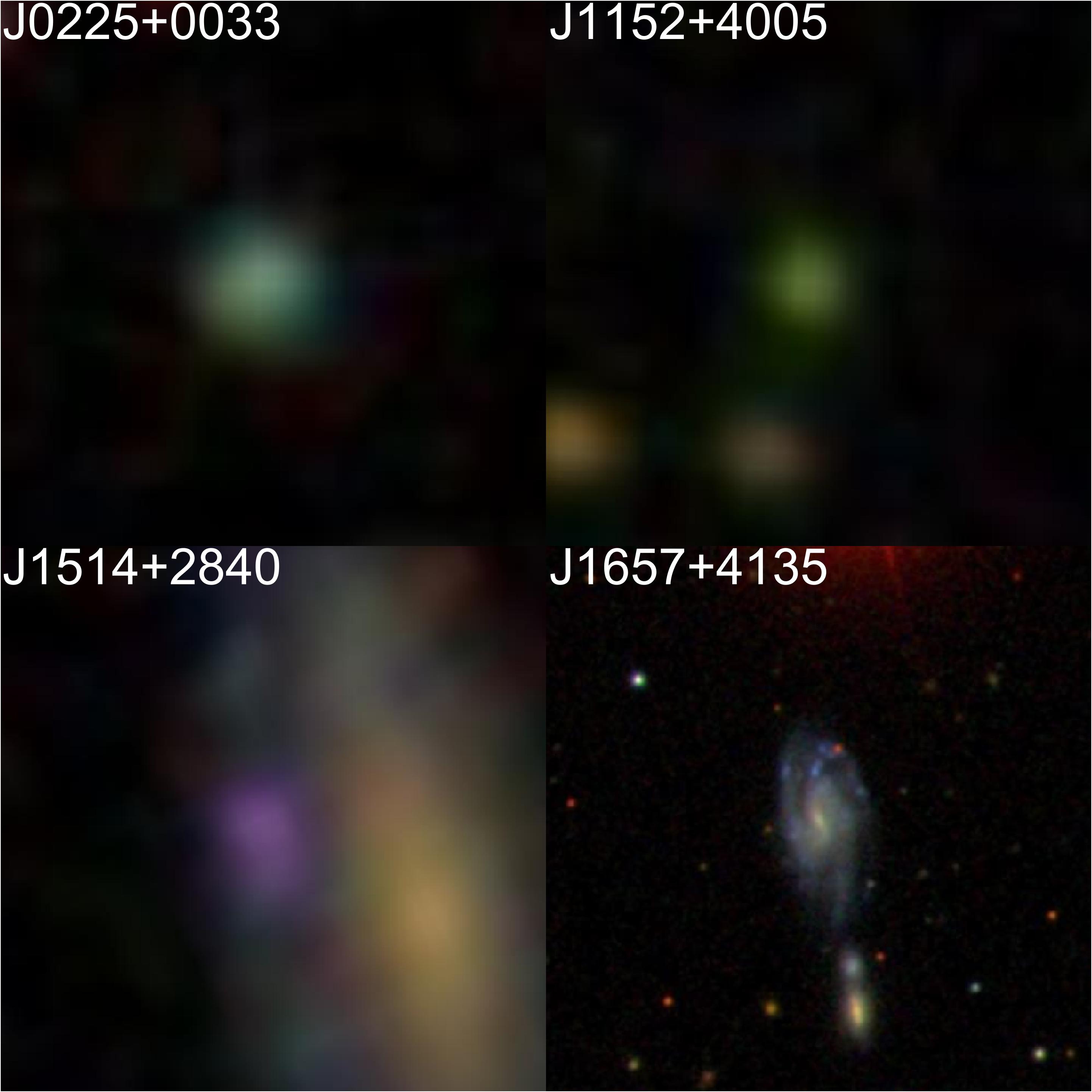}
\caption{Same as Fig. \ref{fig_SDSSmosaic} for the four galaxies identified as outliers. The last panel is zoomed out by a factor of 10, i.e. spanning a region of $125\arcsec\times125\arcsec$.}
\label{fig_SDSSoutl}
\end{figure}

\begin{figure*}
\centering
\includegraphics[width=\hsize]{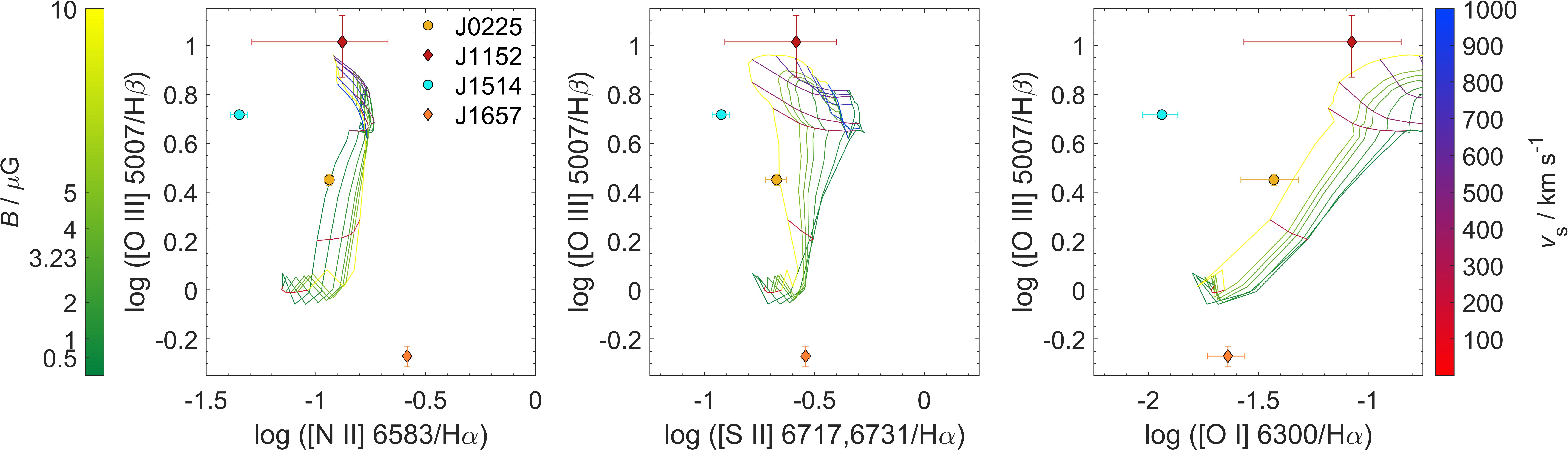}
\caption{Same as Fig. \ref{fig_shockdiags} for the four galaxies identified as outliers. We shortened the object IDs in the legend for reasons of legibility, but the symbols are the same as in Fig. \ref{fig_SBoutl}. The model grids are those for the LMC from \citet{Allen08}, as they are a better match in metallicity for the galaxies considered.}
\label{fig_shockdiagsoutl}
\end{figure*}

Additionally, we reproduce the shock diagnostic plot for the outliers in Fig. \ref{fig_shockdiagsoutl} alongside the LMC tracks from \citet{Allen08}, as they are a better match in metallicity ($\log\left(\mathrm{O}/\mathrm{H}\right)+12=8.35$) for most objects considered here. Due to these higher metallicities, we have additionally incorporated 200 simple stellar populations of ages between $0.1~\mathrm{Myr}$ and $15~\mathrm{Gyr}$ at a metallicity of $Z/Z_\sun=0.4$ during the analysis with \textsc{fado}. The outliers' emission line fluxes can be found in Table \ref{tab_linesoutl}.
Figure \ref{fig_SDSSoutl} shows the SDSS $gri$ composite images of the galaxies in the same order as we discuss them in the subsequent sections.

\begin{table*}
\caption{Same as Table \ref{tab_galprops1}, but for the four galaxies identified as outliers.}             
\label{tab_galprops1outl}      
\centering          
\begin{tabular}{c c c c c c c c c c c}
\hline\hline       
ID & RA & DEC & $z\tablefootmark{a}$ & $r_\mathrm{petro}^\mathrm{ang}\tablefootmark{b}$ & $r_\mathrm{petro}^\mathrm{lin}\tablefootmark{c}$ & $M_\star\tablefootmark{d}$ & SFR$\left(\mathrm{H}\alpha\right)\tablefootmark{e}$ & sSFR\tablefootmark{f} & $\mathrm{SFR}_\mathrm{B}\tablefootmark{g}$ & $\mathrm{sSFR}_\mathrm{B}\tablefootmark{g}$ \\ 
\hline
J0225+0033 & 02:25:52.89 & 00:33:35.33 & 0.30574408 & 1.67 & 7.80 & 9.37 & 3.739 & -8.80 & 3.724 & -8.80 \\
J1152+4005 & 11:52:48.54 & 40:05:27.09 & 0.30773446 & 1.39 & 6.52 & 9.94 & 1.285 & -9.83 & 1.098 & -9.90 \\
J1514+2840 & 15:14:41.66 & 28:40:14.83 & 0.07265168 & 11.31 & 16.23 & 8.47 & 0.501 & -8.77 & 0.483 & -8.79 \\
J1657+4135 & 16:57:33.53 & 41:35:18.73 & 0.02866961 & 13.66 & 8.16 & 9.07 & 0.191 & -9.79 & 0.171 & -9.84 \\

\hline
\end{tabular}
\tablefoot{
\tablefoottext{a}{Redshifts $z$ adopted from the SDSS SkyServer at \href{http://skyserver.sdss.org/}{http://skyserver.sdss.org/}.}
\tablefoottext{b}{$r$-band petrosian radii in arcsec taken from \href{http://skyserver.sdss.org/}{http://skyserver.sdss.org/}.}
\tablefoottext{c}{$r$-band petrosian radii converted to linear scales (kpc).}
\tablefoottext{d}{Logarithm of the stellar mass in units of $\mathrm{M}_\sun$ as derived by \textsc{fado} assuming a \citet{Kroupa01} IMF.}
\tablefoottext{e}{Star formation rate in units of $\mathrm{M}_\sun~\mathrm{yr}^{-1}$.}
\tablefoottext{f}{Logarithm of the sSFR, i.e. the SFR divided by stellar mass.}
\tablefoottext{g}{Based on corrected H$\alpha$ values, see Sect. \ref{sec_bdec}.}
}
\end{table*}

\begin{table*}
\caption{Same as Table \ref{tab_galprops2}, but for the four galaxies identified as outliers.}             
\label{tab_galprops2outl}
\centering          
\begin{tabular}{l c c c c}
\hline\hline

$ $ & J0225+0033 & J1152+4005 & J1514+2840 & J1657+4135 \\
\hline

$t_\mathrm{e}\left(\ion{[O}{III]}\right) / \mathrm{K}$ & $...$ & $16280\pm5700$ & $12510\pm700$ & $...$ \\
$t_\mathrm{e}\left(\ion{[O}{II]}\right) / \mathrm{K}$ & $...$ & $...$ & $10810\pm4500$ & $...$ \\
$n_\mathrm{e}\left(\ion{[S}{II]}\right) / \mathrm{cm}^{-3}$ & $...$ & $...$ & $210\pm530$ & $...$ \\
$\left(\mathrm{O}^{+}/\mathrm{H}^{+}\right)\times10^5$ & $...$ & $...$ & $4.9\pm8.2$ & $...$ \\
$\left(\mathrm{O}^{2+}/\mathrm{H}^{+}\right)\times10^5$ & $...$ & $9.3\pm7.8$ & $9.3\pm1.5$ & $...$ \\
$y^+\times10^2$ & $2.73\pm5.17$ & $...$ & $7.009\pm0.007$ & $...$ \\
$y^{2+}\times10^2$ & $0.15\pm2.17$ & $1.6\pm9.2$ & $0.1\pm1.5$ & $...$ \\
$\mathrm{ICF}\left(\mathrm{O}^{+}+\mathrm{O}^{2+}\right)$ & $...$ & $1.01\pm0.14$ & $1\pm0.18$ & $...$ \\
$\log\left(\mathrm{O}/\mathrm{H}\right)+12$ & $8.25\tablefootmark{1}$ & $8.25?\tablefootmark{2}$ & $8.15\pm0.27$ & $8.16\tablefootmark{1}$ \\
$\log\left(\mathrm{He}/\mathrm{H}\right)$ & $-1.54\pm0.84$ & $...$ & $-1.15\pm0.09$ & $...$ \\
$\log\left(\mathrm{N}/\mathrm{O}\right)$ & $-1\pm0.8$ & $-0.9\pm1.2$ & $-1.33\pm0.25$ & $-0.21\pm0.38$ \\
$\mathrm{O}_{32}$ & $0.9\pm0.6$ & $4.7\pm0.6$ & $2.7\pm0.2$ & $0.19\pm0.03$ \\
$\Delta\ion{[S}{II]}$ & $-0.02\pm0.05$ & $0.9\pm0.4$ & $0.02\pm0.04$ & $-0.044\pm0.019$ \\

\hline                  
\end{tabular}
\tablefoot{
\tablefoottext{1}{Derived with the R23 method \citep{Pagel79}.}
\tablefoottext{2}{Median of three highly uncertain values (see text).}
}
\end{table*}

\subsection{SF outlier: J0225+0033}

\begin{figure}
\centering
\includegraphics[width=\hsize]{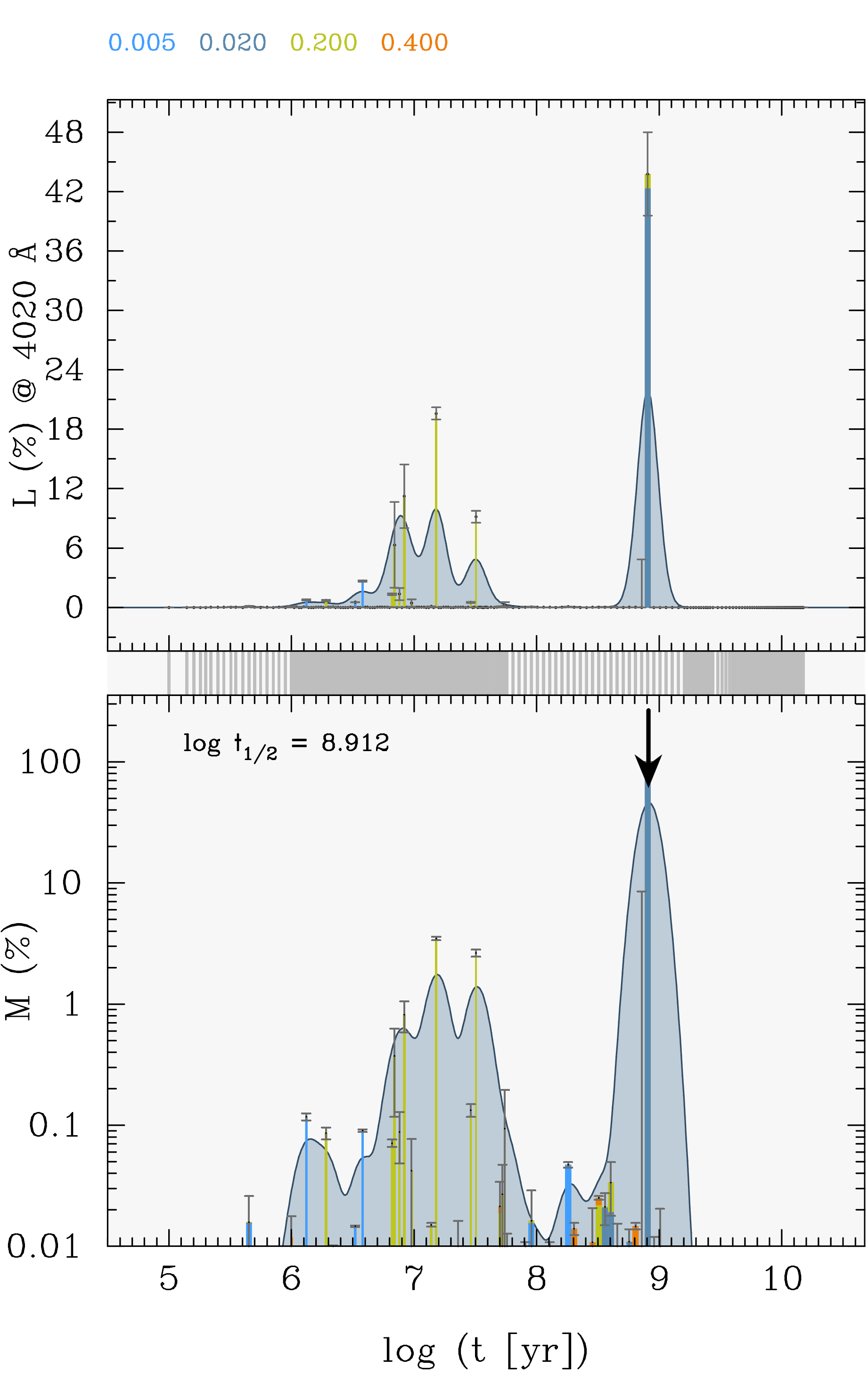}
\caption{Same as Fig. \ref{fig_sfh1} for J0225+0033.}
\label{fig_sfhJ0225}
\end{figure}

J0225+0033 is one of the two GPs found during our sample creation, as readily seen from its SDSS image (Fig. \ref{fig_SDSSoutl}, top left panel). It is physically small ($r_\mathrm{petro}=7.8~\mathrm{kpc}$) but with respect to the IMPs comparatively massive ($\log M_\star/\mathrm{M}_\sun=9.37$) and metal-rich ($\log\left(\mathrm{O}/\mathrm{H}\right)+12=8.25$). Most of the derived properties that would be presented in Table \ref{tab_galprops2outl} cannot be constrained to physically meaningful values due to the large errors associated with the measurements of the faint emission lines present here.

While readily classified as a SFG in the BPT diagnostic diagram (Fig. \ref{fig_BPToutl}), roughly 10\% of its \ion{He}{II}~4686 emission may be attributed to an AGN, as indicated by its position close to the demarcation line in Fig. \ref{fig_SBoutl}.
Its SFH (Fig. \ref{fig_sfhJ0225}) reveals the majority of stars in J0225+0033 has been formed $\sim$ 1~Gyr ago. While some star-formation has taken place during the last 10~Myr, the stellar populations' contribution in terms of luminosity is less pronounced than in the IMPs.
Fig. \ref{fig_shockdiagsoutl} reveals the line ratios of J0225+0033 to align well with the LMC shock models from \citet{Allen08} at $v_\mathrm{s}=200-300~\mathrm{km}~\mathrm{s}^{-1}$, and two out of three diagnostics suggest a strong magnetic field of $B=10~\mu\mathrm{G}$.

\subsection{AGN outlier: J1152+4005}

J1152+4005, the second GP in our initial sample, is the most massive galaxy ($\log M_\star/\mathrm{M}_\sun=9.94$) studied in this paper, while being a compact object at $r_\mathrm{petro}\sim6.52~\mathrm{kpc}$. Both diagnostic diagrams (Figs. \ref{fig_SBoutl} and \ref{fig_BPToutl}) strongly suggest that an AGN is the dominant ionisation mechanism in this object. With no notable line broadening evident in this galaxy's spectrum, it is most reminiscent of a Seyfert 2 type nucleus.

Its faint emission lines allow no conclusive determination of the oxygen abundance, though within their large error ranges of $1.3-1.7$~dex, the direct, emprical $t_\mathrm{e}$ and N2 methods (see Sect. \ref{sec_abund}) agree with a value between $\log\left(\mathrm{O}/\mathrm{H}\right)=7.8-8.4$, whereas the R23 method would predict an unlikely value of $\log\left(\mathrm{O}/\mathrm{H}\right)+12=10.3$.
The comparison of the line ratios found in J1152+4005 to the LMC shock models from \citet{Allen08} suggests that shocks may play a significant role in the ionisation of the galaxy's ISM, but the large errors in the line ratios leave this finding somewhat unconclusive. Fig. \ref{fig_shockdiagsoutl} suggests that models of strong magnetic fields $B \sim 10~\mu\mathrm{G}$ at velocities $v_\mathrm{s}>400~\mathrm{km}~\mathrm{s}^{-1}$ are favoured here.

\subsection{SF outlier: J1514+2840}

\begin{figure}
\centering
\includegraphics[width=\hsize]{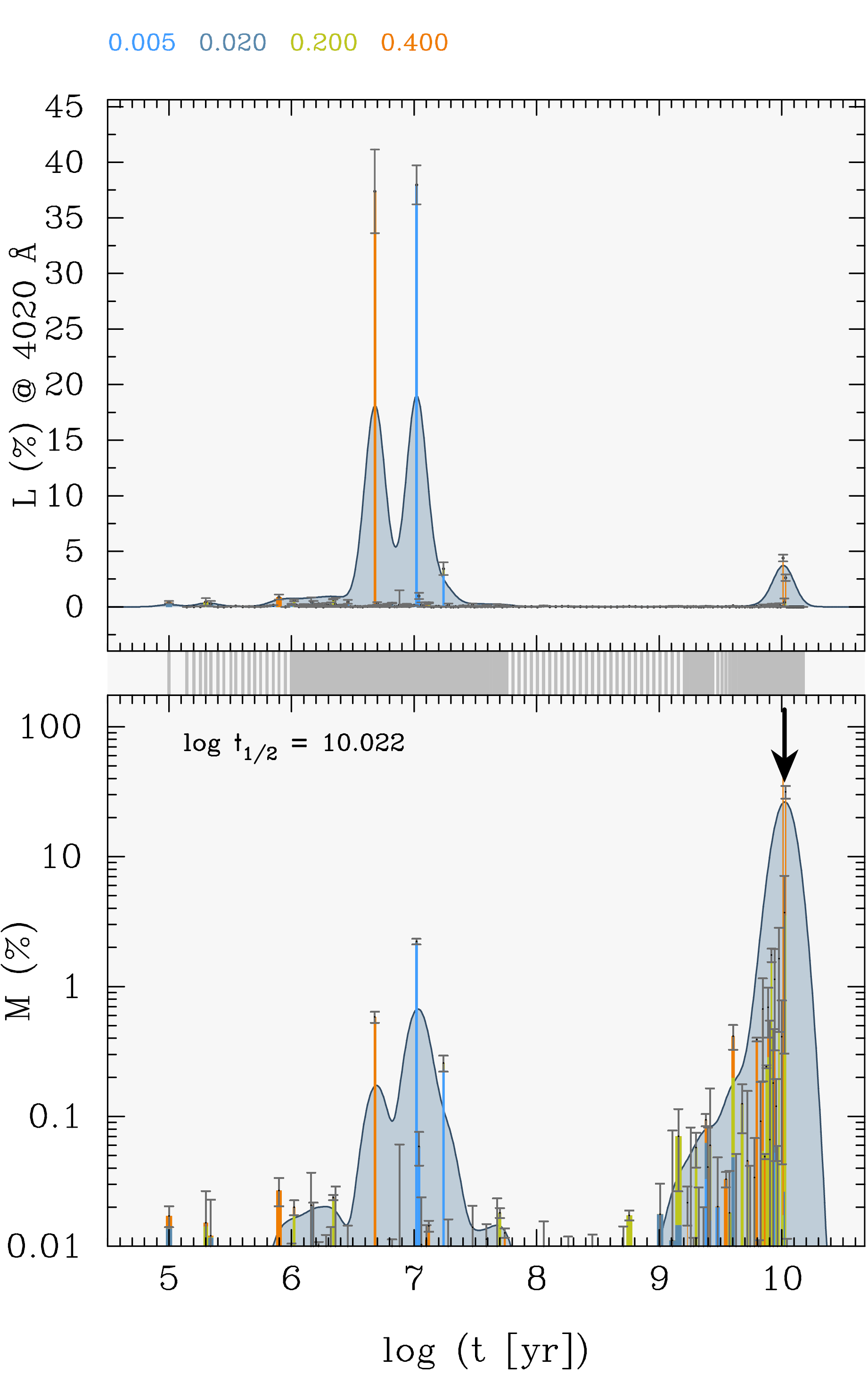}
\caption{Same as Fig. \ref{fig_sfh1} for J1514+2840.}
\label{fig_sfhJ1514}
\end{figure}

The SDSS image of J1514+2840 (Fig. \ref{fig_SDSSoutl}, bottom left panel) reveals this object is of an ambiguous nature. Its (projected) proximity to a larger galaxy suggests it may be an \ion{H}{II} region within the galaxy in question, which by itself has no spectroscopic data published in the SDSS. As it appears to be similar to the IMPs in terms of colour and morphology, another possible classification is a `true' DG which by chance is aligned with a larger background galaxy. In either case, the spectroscopic properties derived for this object are highly uncertain, as the SED is bound to include parts of both the SF object and the apparently quiescent, larger galaxy.

Even in the case of this object being a \ion{H}{II} region, it may still be a worthwhile contender regarding the study of LyC photon loss, as the $\mathrm{O}_{32}$ value of $\sim2.7$ is rather large, considering the `contamination' of the spectrum. Similar situations of localised ionised emission from star-forming regions or stellar clusters have previously been observed in other galaxies - to name one example, in case of Haro 11, its LyC leakage appears to be confined to one of its star-forming knots \citep{Keenan17}.

The notion that J1514+2840 may be relevant in terms of LyC leakage is further supported by the SFH similar to that found in the IMPs, where a young and luminous metal-poor starburst occured at 10~Myr. The younger starburst episode, however, is somewhat older than the typical young population of the IMPs at $5~\mathrm{Myr}$, and its elevated metallicity of $0.4~Z_\sun$ may imply this stellar population is associated with the background source (Fig. \ref{fig_sfhJ1514}).

\subsection{AGN outlier: J1657+4135}

\begin{figure*}[htpb]
\centering
\includegraphics[width=\hsize]{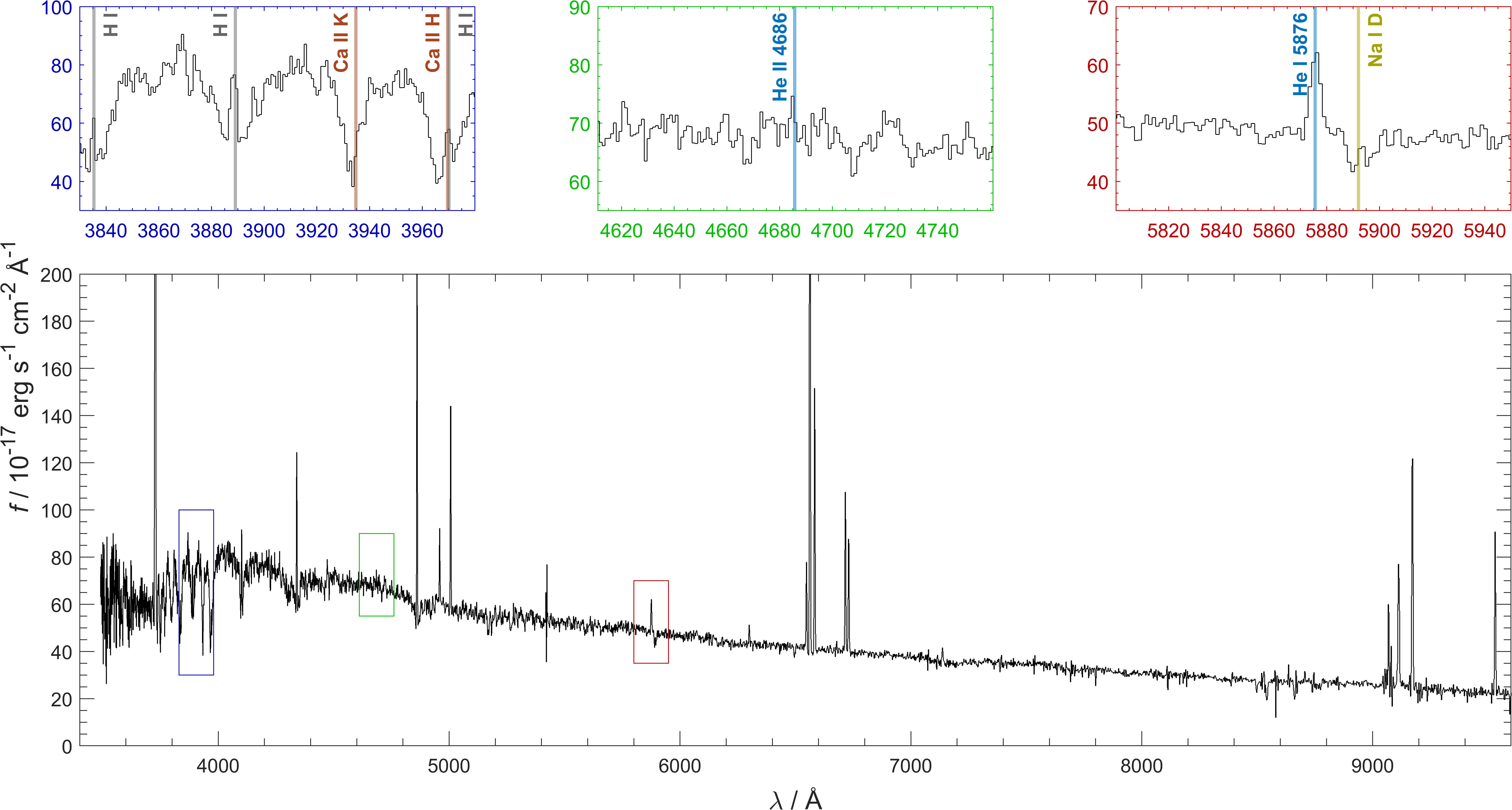}
\caption{$3400~\AA - 9600~\AA$ region of the redshift-corrected spectrum of J1657+4135, truncated in flux density $f$ to enhance the visibility of weak spectral features (bottom row). The coloured boxes indicate the regions shown in the blow-ups in the top row and were selected to depict several features discussed in the text. The wavelengths were converted to vacuum wavelengths following \citet{Morton91}.}
\label{fig_dSpspec}
\end{figure*}

J1657+4135 is a true outlier in the sense that it is the only object considered in this paper where parts of the internal structure are resolved. Its SDSS image (Fig. \ref{fig_SDSSoutl}, bottom right panel) reveals it to be a spiral galaxy with at least three spiral arms, which contain blue knots of enhanced star-forming activity. Its companion in the south (SDSS J165732.76+413434.9, visible as an elongated yellow shape at the bottom of the SDSS image in Fig. \ref{fig_SDSSoutl}) is located at a similar redshift ($z=0.02869666$) and their centres are separated by $\sim27~\mathrm{kpc}$ in projection, so a likely scenario for triggering the current star formation episode is their gravitational interaction.

While classified as a SFG in the BPT diagram (Fig. \ref{fig_BPToutl}), we cannot draw further conclusions concerning the presence of an AGN from \ion{He}{II}~4686. As seen in its spectrum (Fig. \ref{fig_dSpspec}), the \ion{He}{II}~4686 line is not properly detected. Noteworthy are the visibly strong absorption features, such as \ion{Ca}{II}~H\&K at $3969~\AA$ and $3934~\AA$, \ion{Mg}{I} at $5177~\AA$, \ion{Na}{I}~D at $5892~\AA$, and likewise, several \ion{H}{I} lines, particularly pronounced in the features around $\sim3900~\AA$. However, at their given rest-frame wavelengths, the hydrogen lines show a strong, narrow emission feature superimposed on the absorption trough. This indicates that along a few lines of sight, the density of neutral gas is low, either from kinematic effects or due to a high degree of ionisation. As discussed before, star-forming regions have the means to be a driver of both these effects. The SDSS fibre, however, only covers $3\arcsec$ around the galaxy's centre, and the low value of SFR$\left(\mathrm{H}\alpha\right)\sim0.191~\mathrm{M}_\sun~\mathrm{yr}^{-1}$ suggests no region of significantly enhanced star-formation is contained in the SDSS spectrum. We hence deem it likely that these features are related to the galaxy's nucleus, whose activity has a strong impact on the surrounding ISM.

\onecolumn

\section{Emission lines}
\centering
\begin{longtable}[c]{p{0.16\textwidth-2\tabcolsep}*{6}{p{0.14\textwidth-2\tabcolsep}}}
\caption{\label{tab_lines} Emission line fluxes relative to H$\beta$, their equivalent widths and the derived H$\beta$ extinction coefficients of the IMPs.} \\

\hline\hline
$ $ & \multicolumn{2}{c}{J0006+0255} & \multicolumn{2}{c}{J0028+3035} & \multicolumn{2}{c}{J0131+0210} \\
line & $I\left(\lambda\right)$ & EW$\left(\lambda\right)$ & $I\left(\lambda\right)$ & EW$\left(\lambda\right)$ & $I\left(\lambda\right)$ & EW$\left(\lambda\right)$ \\
\hline

\endfirsthead
\caption{continued.}\\

line & $I\left(\lambda\right)$ & EW$\left(\lambda\right)$ & $I\left(\lambda\right)$ & EW$\left(\lambda\right)$ & $I\left(\lambda\right)$ & EW$\left(\lambda\right)$ \\
\hline

\endhead

\hline
\endfoot
\multicolumn{7}{c}{\tablefoot{
Line fluxes are scaled to H$\beta$ as $I\left(\lambda\right)=100\times F\left(\lambda\right)/F\left(\mathrm{H}\beta\right)$. Equivalent widths EW$\left(\lambda\right)$ are given in $\AA$.
\tablefoottext{$\ast$}{Likely blend with \ion{[Ne}{III]}~3967, which constitutes $\sim50\%$ of the reported flux (Sect. \ref{sec_bdec}).}
\tablefoottext{$\dag$}{H$\beta$ line flux in $10^{-17}~\mathrm{erg}~\mathrm{s}^{-1}~\mathrm{cm}^{-2}$ for reference.}
}}
\endlastfoot

\ion{[O}{II]}~3726 & $3.4\pm9.1$ & $3.3\pm8.8$ & $31.2\pm2.4$ & $28.1\pm2.1$ & $7\pm28$ & $7\pm28$ \\
\ion{[O}{II]}~3729 & $82.6\pm2.6$ & $78\pm2.5$ & $32.1\pm3.1$ & $28.9\pm2.6$ & $110\pm110$ & $110\pm110$ \\
H$\epsilon$~3970\tablefootmark{$\ast$} & $31.5\pm1.7$ & $42.9\pm2.5$ & $36.2\pm2.5$ & $43.8\pm3.1$ & $34.8\pm2.7$ & $55\pm4.7$ \\
\ion{He}{I}~4026 & $2.8\pm1.2$ & $2.9\pm1.2$ & $3.5\pm1.3$ & $3.4\pm1.3$ & $0.33\pm0.51$ & $0.38\pm0.59$ \\
\ion{[S}{II]}~4069 & $0.97\pm0.52$ & $0.99\pm0.53$ & $5.4\pm6.3$ & $5.4\pm6.2$ & $0\pm1$ & $1.9\pm1.2$ \\
\ion{[S}{II]}~4076 & $1.15\pm0.72$ & $1.19\pm0.75$ & $0.73\pm0.59$ & $0.73\pm0.59$ & $0\pm1$ & $0.8\pm1.2$ \\
H$\delta$~4102 & $25.1\pm1.2$ & $35\pm1.8$ & $26.8\pm2.3$ & $34.1\pm3.1$ & $25.2\pm2.2$ & $37.5\pm3.8$ \\
H$\gamma$~4341 & $46.1\pm1.8$ & $71.2\pm3.2$ & $49.3\pm2.3$ & $70\pm3$ & $47.3\pm2.9$ & $70\pm5$ \\
\ion{[O}{III]}~4363 & $12.22\pm0.87$ & $10\pm1$ & $18.4\pm1.2$ & $21.1\pm1.2$ & $12.2\pm1.5$ & $16.6\pm2.1$ \\
\ion{He}{I}~4472 & $4.36\pm0.93$ & $5.6\pm1.2$ & $3.7\pm1.3$ & $4.7\pm1.7$ & $4.1\pm1.2$ & $6.1\pm1.8$ \\
\ion{[Fe}{III]}~4659 & $0.86\pm0.31$ & $1.18\pm0.43$ & $0.26\pm0.56$ & $0.34\pm0.74$ & $0.7\pm0.6$ & $1.19\pm0.98$ \\
\ion{He}{II}~4686 & $2.23\pm0.39$ & $3.2\pm0.55$ & $2.89\pm0.43$ & $4.05\pm0.59$ & $1.41\pm0.64$ & $2.4\pm1.1$ \\
\ion{[Ar}{IV]}~4711 & $1.5\pm1.6$ & $2.2\pm2.3$ & $2.14\pm0.43$ & $3\pm0.6$ & $0.59\pm0.86$ & $1\pm1.5$ \\
\ion{He}{I}~4713 & $1.57\pm0.32$ & $2.29\pm0.46$ & $2.43\pm0.77$ & $3.4\pm1.1$ & $1.83\pm0.59$ & $0\pm1$ \\
\ion{[Ar}{IV]}~4740 & $1.33\pm0.38$ & $1.95\pm0.57$ & $2\pm0.5$ & $2.81\pm0.72$ & $0.93\pm0.78$ & $1.6\pm1.4$ \\
H$\beta$~4861 & $100\pm2.9$ & $213.5\pm8.8$ & $100\pm4.6$ & $196\pm13$ & $100\pm4.4$ & $241\pm16$ \\
\ion{He}{I}~4922 & $1.62\pm0.91$ & $2.6\pm1.5$ & $2.2\pm1.5$ & $3.5\pm2.3$ & $1.41\pm0.55$ & $2.8\pm1.1$ \\
\ion{[O}{III]}~4959 & $167.7\pm4.4$ & $270.3\pm6.9$ & $226.6\pm8.2$ & $363\pm14$ & $217.6\pm7.5$ & $422\pm16$ \\
\ion{[Fe}{VII]}~4989 & ... & ... & ... & ... & ... & ... \\
\ion{[O}{III]}~5007 & $508\pm12$ & $851\pm19$ & $676\pm23$ & $1115\pm42$ & $651\pm21$ & $1294\pm46$ \\
\ion{[N}{II]}~5755 & $1.2\pm1.3$ & $2.6\pm2.9$ & $0.3\pm0.73$ & $0.6\pm1.5$ & $0.44\pm0.44$ & $1.1\pm1.1$ \\
\ion{He}{I}~5876 & $10.4\pm0.9$ & $24.1\pm2.2$ & $8.7\pm1.3$ & $19.3\pm2.9$ & $10.79\pm0.81$ & $29.7\pm2.3$ \\
\ion{[Fe}{VII]}~6086 & ... & ... & ... & ... & ... & ... \\
\ion{[O}{I]}~6300 & $2.31\pm0.61$ & $6.1\pm1.6$ & $1.04\pm0.62$ & $2.6\pm1.6$ & $3.65\pm0.77$ & $11.7\pm2.5$ \\
\ion{[S}{III]}~6312 & $0.9\pm0.51$ & $2.4\pm1.4$ & $1.13\pm0.69$ & $2.9\pm1.8$ & $1.75\pm0.74$ & $5.6\pm2.4$ \\
\ion{[O}{I]}~6364 & $0.63\pm0.46$ & $1.7\pm1.2$ & $0.3\pm0.8$ & $0\pm2$ & $0.75\pm0.51$ & $2.4\pm1.7$ \\
\ion{[N}{II]}~6548 & $0.8\pm0.4$ & $2.5\pm1.2$ & $0.7\pm0.15$ & $1.87\pm0.39$ & $1.51\pm0.51$ & $5.3\pm1.8$ \\
H$\alpha$~6563 & $270\pm6$ & $970\pm33$ & $298.5\pm9.8$ & $943\pm13$ & $272.3\pm9.3$ & $1079\pm49$ \\
\ion{[N}{II]}~6583 & $1.96\pm0.46$ & $5.7\pm1.3$ & $1.7\pm0.2$ & $4.62\pm0.52$ & $3.69\pm0.54$ & $12.9\pm1.9$ \\
\ion{He}{I}~6678 & $3.01\pm0.44$ & $9.1\pm1.3$ & $3.12\pm0.22$ & $8.74\pm0.54$ & $3.31\pm0.52$ & $10\pm2$ \\
\ion{[S}{II]}~6716 & $8.35\pm0.56$ & $24.9\pm1.7$ & $5.48\pm0.83$ & $15.4\pm2.4$ & $9.33\pm0.68$ & $34\pm2.5$ \\
\ion{[S}{II]}~6731 & $6.1\pm1.2$ & $18.2\pm3.7$ & $5\pm1.2$ & $14.1\pm3.5$ & $6.7\pm1.4$ & $24.4\pm5.1$ \\
\ion{He}{I}~7065 & $2.89\pm0.69$ & $9.6\pm2.3$ & $3.9\pm1.4$ & $11.7\pm4.3$ & $1.99\pm0.52$ & $8.1\pm2.2$ \\
\ion{[Ar}{III]}~7136 & $4.4\pm1.1$ & $14.5\pm3.8$ & $3.9\pm1.2$ & $12\pm3.9$ & $5.9\pm1.1$ & $23.9\pm4.6$ \\
\ion{[O}{II]}~7320 & $3.8\pm3.1$ & $13\pm11$ & $1.4\pm1.4$ & $4.6\pm4.3$ & $1.08\pm0.81$ & $4.7\pm3.5$ \\
\ion{[O}{II]}~7330 & $0.55\pm0.45$ & $1.9\pm1.6$ & $2.9\pm3.2$ & $9\pm10$ & $0.97\pm0.64$ & $4.3\pm2.8$ \\
\ion{[Ar}{III]}~7751 & $1.8\pm1.4$ & $7.2\pm5.5$ & $1.6\pm1.4$ & $5.6\pm4.9$ & $1.09\pm0.68$ & $5.3\pm3.4$ \\
\ion{[Fe}{II]}~8617 & $0.62\pm0.96$ & $0\pm5$ & $0.1\pm0.2$ & $0.3\pm0.9$ & $0.04\pm0.23$ & $0.3\pm1.6$ \\
$F\left(\mathrm{H}\beta\right)\tablefootmark{\dag}$ & \multicolumn{2}{c}{$408\pm8$} & \multicolumn{2}{c}{$362\pm12$} & \multicolumn{2}{c}{$512\pm16$} \\
$c\left(\mathrm{H}\beta\right)$ & \multicolumn{2}{c}{$0.045$} & \multicolumn{2}{c}{0.052} & \multicolumn{2}{c}{0.098} \\
\multicolumn{7}{c}{}\\

\hline\hline
$ $ & \multicolumn{2}{c}{J0138+1114} & \multicolumn{2}{c}{J0150+1643} & \multicolumn{2}{c}{J0744+1858} \\
line & $I\left(\lambda\right)$ & EW$\left(\lambda\right)$ & $I\left(\lambda\right)$ & EW$\left(\lambda\right)$ & $I\left(\lambda\right)$ & EW$\left(\lambda\right)$ \\
\hline

\ion{[O}{II]}~3726 & $36\pm15$ & $34\pm14$ & $29.7\pm9.9$ & $38\pm13$ & $34\pm1.6$ & $49.4\pm2.3$ \\
\ion{[O}{II]}~3729 & $16\pm16$ & $15\pm15$ & $15.7\pm9.8$ & $20\pm13$ & $14\pm1.9$ & $20.3\pm2.7$ \\
H$\epsilon$~3970\tablefootmark{$\ast$} & $35.2\pm1.9$ & $45\pm2.4$ & $28.7\pm1.5$ & $51.3\pm2.9$ & $33.9\pm1.5$ & $62.8\pm2.8$ \\
\ion{He}{I}~4026 & $2.22\pm0.66$ & $2.42\pm0.72$ & $1.16\pm0.57$ & $1.51\pm0.74$ & $2.08\pm0.51$ & $3.4\pm0.83$ \\
\ion{[S}{II]}~4069 & $0.46\pm0.28$ & $0.51\pm0.31$ & $0.32\pm0.53$ & $0.42\pm0.71$ & $0.7\pm0.5$ & $1.14\pm0.83$ \\
\ion{[S}{II]}~4076 & $0.39\pm0.23$ & $0.44\pm0.25$ & $0.07\pm0.37$ & $0.1\pm0.5$ & $1.1\pm0.92$ & $1.8\pm1.5$ \\
H$\delta$~4102 & $25.19\pm0.89$ & $36.2\pm1.2$ & $25\pm1.6$ & $45.7\pm3.4$ & $24.8\pm1.4$ & $51.9\pm3.3$ \\
H$\gamma$~4341 & $46.7\pm2.3$ & $78.5\pm5.1$ & $45.8\pm2.2$ & $94.2\pm6.3$ & $45.4\pm1.9$ & $102.6\pm5.7$ \\
\ion{[O}{III]}~4363 & $16.3\pm1.2$ & $21.4\pm1.7$ & $14.9\pm1.1$ & $22.7\pm1.8$ & $17.1\pm1.1$ & $32.9\pm2.3$ \\
\ion{He}{I}~4472 & $0\pm1$ & $6.1\pm1.5$ & $3.95\pm0.74$ & $6.5\pm1.2$ & $3.51\pm0.59$ & $7.2\pm1.2$ \\
\ion{[Fe}{III]}~4659 & $0.4\pm0.51$ & $0.65\pm0.82$ & $1.4\pm1.4$ & $2.4\pm2.6$ & $0.022\pm0.094$ & $0.05\pm0.21$ \\
\ion{He}{II}~4686 & $2.87\pm0.54$ & $4.85\pm0.92$ & $1.65\pm0.43$ & $3.01\pm0.78$ & $1.2\pm0.2$ & $2.8\pm0.48$ \\
\ion{[Ar}{IV]}~4711 & $2.38\pm0.48$ & $4.08\pm0.82$ & $0.48\pm0.21$ & $0.89\pm0.39$ & $1.33\pm0.93$ & $3.2\pm2.2$ \\
\ion{He}{I}~4713 & $0.9\pm1.4$ & $1.6\pm2.4$ & $1.6\pm0.35$ & $2.96\pm0.65$ & $3.66\pm0.23$ & $8.68\pm0.52$ \\
\ion{[Ar}{IV]}~4740 & $1.74\pm0.47$ & $3.02\pm0.81$ & $1.7\pm0.4$ & $3.16\pm0.76$ & $3.09\pm0.39$ & $7.42\pm0.93$ \\
H$\beta$~4861 & $100\pm3.4$ & $254\pm13$ & $100\pm2.9$ & $248\pm11$ & $100\pm3.2$ & $317\pm17$ \\
\ion{He}{I}~4922 & $0.79\pm0.18$ & $1.6\pm0.37$ & $0.71\pm0.31$ & $1.5\pm0.66$ & $0.97\pm0.25$ & $2.62\pm0.66$ \\
\ion{[O}{III]}~4959 & $224.7\pm5.5$ & $445.1\pm6.5$ & $198.9\pm4.4$ & $417\pm10$ & $268.6\pm7.1$ & $706\pm18$ \\
\ion{[Fe}{VII]}~4989 & ... & ... & ... & ... & ... & ... \\
\ion{[O}{III]}~5007 & $707\pm17$ & $1441\pm21$ & $601\pm13$ & $1288\pm30$ & $826\pm20$ & $2264\pm55$ \\
\ion{[N}{II]}~5755 & $0.25\pm0.43$ & $0.7\pm1.2$ & $0.31\pm0.35$ & $0.88\pm0.99$ & $0.37\pm0.68$ & $1.3\pm2.4$ \\
\ion{He}{I}~5876 & $9.13\pm0.45$ & $29.1\pm1.4$ & ... & ... & $11.19\pm0.36$ & $40.9\pm1.1$ \\
\ion{[Fe}{VII]}~6086 & ... & ... & ... & ... & ... & ... \\
\ion{[O}{I]}~6300 & $0.8\pm0.6$ & $3.2\pm2.2$ & $1.13\pm0.33$ & $4\pm1.2$ & $1.52\pm0.32$ & $6.4\pm1.4$ \\
\ion{[S}{III]}~6312 & $0.79\pm0.55$ & $3\pm2.1$ & $0.98\pm0.36$ & $3.4\pm1.3$ & $1.41\pm0.34$ & $5.9\pm1.4$ \\
\ion{[O}{I]}~6364 & $1.3\pm1.8$ & $4.9\pm6.7$ & $0.72\pm0.26$ & $2.5\pm0.9$ & $0.18\pm0.19$ & $0.75\pm0.79$ \\
\ion{[N}{II]}~6548 & $0.7\pm0.51$ & $2.9\pm2.2$ & $0.4\pm0.3$ & $1.5\pm1.2$ & $0.59\pm0.19$ & $2.64\pm0.83$ \\
H$\alpha$~6563 & $269.4\pm7.4$ & $1322\pm76$ & $286.5\pm6.3$ & $1311\pm44$ & $280.3\pm6.5$ & $1452\pm35$ \\
\ion{[N}{II]}~6583 & $1.3\pm1.1$ & $5.4\pm4.6$ & $1.3\pm0.38$ & $5\pm1.5$ & $1.8\pm0.31$ & $8\pm1.4$ \\
\ion{He}{I}~6678 & $2.51\pm0.53$ & $11.1\pm2.4$ & $2.52\pm0.31$ & $10\pm1.2$ & $3.09\pm0.22$ & $10\pm1$ \\
\ion{[S}{II]}~6716 & $2.73\pm0.47$ & $11.8\pm2.1$ & $3.99\pm0.23$ & $15.85\pm0.89$ & $5.18\pm0.28$ & $23.8\pm1.3$ \\
\ion{[S}{II]}~6731 & $2.32\pm0.48$ & $10.1\pm2.1$ & $3.46\pm0.24$ & $13.77\pm0.94$ & $4.4\pm0.4$ & $20.1\pm1.8$ \\
\ion{He}{I}~7065 & $2.75\pm0.75$ & $13.8\pm3.8$ & $2.83\pm0.54$ & $12.5\pm2.5$ & $3.42\pm0.43$ & $17.2\pm2.2$ \\
\ion{[Ar}{III]}~7136 & $3.22\pm0.87$ & $16.2\pm4.6$ & $3.24\pm0.54$ & $14.5\pm2.5$ & $5.02\pm0.47$ & $25.5\pm2.5$ \\
\ion{[O}{II]}~7320 & $0.99\pm0.84$ & $5.3\pm4.5$ & $1.43\pm0.65$ & $6.8\pm3.1$ & $1.08\pm0.35$ & $5.7\pm1.8$ \\
\ion{[O}{II]}~7330 & $0.5\pm1.3$ & $2.9\pm7.2$ & $0.81\pm0.56$ & $3.8\pm2.6$ & $0.91\pm0.33$ & $4.8\pm1.8$ \\
\ion{[Ar}{III]}~7751 & $0.6\pm0.5$ & $4\pm3.2$ & $0.91\pm0.46$ & $4.9\pm2.5$ & $1.4\pm0.31$ & $8.2\pm1.8$ \\
\ion{[Fe}{II]}~8617 & $0.11\pm0.17$ & $1\pm1.5$ & $0.1\pm0.9$ & $1.1\pm6.7$ & ... & ... \\
$F\left(\mathrm{H}\beta\right)\tablefootmark{\dag}$ & \multicolumn{2}{c}{$987\pm24$} & \multicolumn{2}{c}{$1099\pm24$} & \multicolumn{2}{c}{$1377\pm31$} \\
$c\left(\mathrm{H}\beta\right)$ & \multicolumn{2}{c}{$0.220$} & \multicolumn{2}{c}{0.177} & \multicolumn{2}{c}{0.169} \\
\multicolumn{7}{c}{}\\

\hline\hline
$ $ & \multicolumn{2}{c}{J0753+2820} & \multicolumn{2}{c}{J0809+4918} & \multicolumn{2}{c}{J1037+2325} \\
line & $I\left(\lambda\right)$ & EW$\left(\lambda\right)$ & $I\left(\lambda\right)$ & EW$\left(\lambda\right)$ & $I\left(\lambda\right)$ & EW$\left(\lambda\right)$ \\
\hline

\ion{[O}{II]}~3726 & $30.6\pm3.7$ & $37\pm4.4$ & $94.3\pm3.8$ & $69.7\pm2.9$ & $30.1\pm6.9$ & $11.6\pm2.7$ \\
\ion{[O}{II]}~3729 & $97.6\pm3.6$ & $117.8\pm3.6$ & $5.3\pm7.5$ & $3.8\pm5.4$ & $78.2\pm8.2$ & $31.5\pm3.2$ \\
H$\epsilon$~3970\tablefootmark{$\ast$} & $30\pm2$ & $54.3\pm3.4$ & $29.6\pm2.4$ & $28.7\pm2.4$ & $28.2\pm2.2$ & $15.2\pm1.1$ \\
\ion{He}{I}~4026 & $1.9\pm0.46$ & $2.69\pm0.65$ & $1.4\pm0.92$ & $0.98\pm0.64$ & $1.63\pm0.96$ & $0.53\pm0.31$ \\
\ion{[S}{II]}~4069 & $1.42\pm0.66$ & $2.04\pm0.95$ & $1.2\pm2.2$ & $0.8\pm1.5$ & $2.5\pm1.6$ & $0.83\pm0.54$ \\
\ion{[S}{II]}~4076 & $0.6\pm0.5$ & $0.84\pm0.72$ & $1.3\pm4.4$ & $0.9\pm3.1$ & $0.33\pm0.58$ & $0.1\pm0.2$ \\
H$\delta$~4102 & $28\pm1.7$ & $49.8\pm3.5$ & $26\pm1.7$ & $26\pm1.8$ & $25.6\pm1.6$ & $12.97\pm0.79$ \\
H$\gamma$~4341 & $48.1\pm2.4$ & $98\pm5.8$ & $46.2\pm2.4$ & $52.7\pm3.2$ & $48.1\pm2.5$ & $24.9\pm1.3$ \\
\ion{[O}{III]}~4363 & $12.6\pm1.1$ & $21.4\pm1.9$ & $12.4\pm1.3$ & $10.2\pm1.1$ & $11.5\pm1.5$ & $4.36\pm0.58$ \\
\ion{He}{I}~4472 & $4.02\pm0.95$ & $7.5\pm1.8$ & $3.77\pm0.93$ & $3.36\pm0.83$ & $3.7\pm1.2$ & $1.45\pm0.48$ \\
\ion{[Fe}{III]}~4659 & $0.47\pm0.33$ & $0.95\pm0.67$ & $0.83\pm0.86$ & $0.79\pm0.82$ & $0\pm1.3$ & $0\pm0.55$ \\
\ion{He}{II}~4686 & $1.5\pm0.6$ & $3.3\pm1.3$ & $2.32\pm0.57$ & $2.27\pm0.56$ & $2.99\pm0.84$ & $1.3\pm0.37$ \\
\ion{[Ar}{IV]}~4711 & $1.21\pm0.55$ & $2.6\pm1.2$ & $0.81\pm0.41$ & $0.81\pm0.41$ & $10\pm4$ & $3.1\pm1.8$ \\
\ion{He}{I}~4713 & $1.62\pm0.48$ & $0\pm1$ & $0.5\pm0.35$ & $0.5\pm0.35$ & $0.52\pm0.65$ & $0.23\pm0.29$ \\
\ion{[Ar}{IV]}~4740 & $1.91\pm0.62$ & $4.1\pm1.3$ & $0.79\pm0.52$ & $0.8\pm0.53$ & $5.6\pm3.2$ & $2.5\pm1.4$ \\
H$\beta$~4861 & $100\pm4.3$ & $281\pm17$ & $100\pm3.1$ & $160\pm6$ & $100\pm3.6$ & $63.7\pm2.1$ \\
\ion{He}{I}~4922 & $0.81\pm0.25$ & $1.99\pm0.61$ & $0.8\pm0.3$ & $0.9\pm0.34$ & $4\pm5.1$ & $2\pm2.5$ \\
\ion{[O}{III]}~4959 & $197.3\pm6.2$ & $482\pm11$ & $182.6\pm4.2$ & $207.1\pm3.2$ & $159\pm4.7$ & $76.9\pm1.7$ \\
\ion{[Fe}{VII]}~4989 & ... & ... & ... & ... & ... & ... \\
\ion{[O}{III]}~5007 & $649\pm20$ & $1628\pm35$ & $571\pm13$ & $665.2\pm9.3$ & $493\pm13$ & $246.6\pm4.5$ \\
\ion{[N}{II]}~5755 & $0.22\pm0.34$ & $0.7\pm1.1$ & $6\pm3.9$ & $9.5\pm6.2$ & $0.89\pm0.82$ & $0.5\pm0.5$ \\
\ion{He}{I}~5876 & $10.9\pm1.3$ & $37.6\pm4.7$ & ... & ... & $9.57\pm0.91$ & $6.08\pm0.57$ \\
\ion{[Fe}{VII]}~6086 & ... & ... & ... & ... & ... & ... \\
\ion{[O}{I]}~6300 & $2.63\pm0.71$ & $10.4\pm2.8$ & $2.86\pm0.46$ & $5.6\pm0.9$ & $2.58\pm0.85$ & $1.87\pm0.62$ \\
\ion{[S}{III]}~6312 & $1.36\pm0.78$ & $5.3\pm3.1$ & $1.34\pm0.64$ & $2.6\pm1.2$ & $1.24\pm0.74$ & $0.9\pm0.54$ \\
\ion{[O}{I]}~6364 & $0.7\pm0.4$ & $2.7\pm1.6$ & $0.67\pm0.34$ & $1.32\pm0.67$ & $7\pm3.6$ & $5.1\pm2.6$ \\
\ion{[N}{II]}~6548 & $1.6\pm0.49$ & $6.7\pm2.1$ & $0.99\pm0.32$ & $2.17\pm0.71$ & $1.4\pm0.6$ & $1.07\pm0.48$ \\
H$\alpha$~6563 & $294.6\pm9.9$ & $1413\pm77$ & $283.3\pm6.4$ & $716\pm14$ & $278\pm7.5$ & $270\pm5$ \\
\ion{[N}{II]}~6583 & $3.73\pm0.57$ & $15.6\pm2.4$ & $2.6\pm0.5$ & $5.8\pm1.1$ & $4.22\pm0.68$ & $3.31\pm0.53$ \\
\ion{He}{I}~6678 & $3.47\pm0.59$ & $15.1\pm2.6$ & $2.57\pm0.29$ & $5.68\pm0.64$ & $3.27\pm0.74$ & $2.6\pm0.59$ \\
\ion{[S}{II]}~6716 & $10.84\pm0.79$ & $47.3\pm3.7$ & $9.3\pm0.64$ & $20.6\pm1.4$ & $10.94\pm0.77$ & $8.74\pm0.58$ \\
\ion{[S}{II]}~6731 & $8.1\pm1.2$ & $35.3\pm5.3$ & $7.04\pm0.55$ & $15.6\pm1.2$ & $7.9\pm1.7$ & $6.3\pm1.3$ \\
\ion{He}{I}~7065 & $3.9\pm1.2$ & $20\pm6$ & $2.82\pm0.72$ & $7\pm1.8$ & $2.86\pm0.59$ & $2.48\pm0.51$ \\
\ion{[Ar}{III]}~7136 & $7.3\pm1.6$ & $35.3\pm8.3$ & $4.57\pm0.56$ & $11.5\pm1.4$ & $4.71\pm0.96$ & $4.14\pm0.84$ \\
\ion{[O}{II]}~7320 & $1.6\pm0.9$ & $7.9\pm4.5$ & $1.51\pm0.64$ & $4.1\pm1.7$ & $1.45\pm0.43$ & $1.3\pm0.4$ \\
\ion{[O}{II]}~7330 & $2.3\pm1.5$ & $11.5\pm7.3$ & $1.51\pm0.54$ & $4.1\pm1.4$ & $1.99\pm0.62$ & $1.82\pm0.57$ \\
\ion{[Ar}{III]}~7751 & $1.7\pm1.1$ & $9.7\pm6.6$ & $0.95\pm0.39$ & $2.9\pm1.2$ & $4.1\pm3.7$ & $4.2\pm3.8$ \\
\ion{[Fe}{II]}~8617 & $0.06\pm0.25$ & $0.5\pm1.9$ & ... & ... & $0.2\pm0.4$ & $0.2\pm0.5$ \\
$F\left(\mathrm{H}\beta\right)\tablefootmark{\dag}$ & \multicolumn{2}{c}{$544\pm17$} & \multicolumn{2}{c}{$725\pm16$} & \multicolumn{2}{c}{$238\pm6$} \\
$c\left(\mathrm{H}\beta\right)$ & \multicolumn{2}{c}{$0.042$} & \multicolumn{2}{c}{0.242} & \multicolumn{2}{c}{0.041} \\
\multicolumn{7}{c}{}\\

\hline\hline
$ $ & \multicolumn{2}{c}{J1109+3429} & \multicolumn{2}{c}{J1141+6059} & \multicolumn{2}{c}{J1311+3750} \\
line & $I\left(\lambda\right)$ & EW$\left(\lambda\right)$ & $I\left(\lambda\right)$ & EW$\left(\lambda\right)$ & $I\left(\lambda\right)$ & EW$\left(\lambda\right)$ \\
\hline

\ion{[O}{II]}~3726 & $123\pm5.2$ & $81.3\pm3.5$ & $130\pm6$ & $130\pm8$ & $12\pm10$ & $23\pm19$ \\
\ion{[O}{II]}~3729 & $8\pm16$ & $5\pm10$ & $12.4\pm5.1$ & $13\pm5.3$ & $21\pm11$ & $38\pm21$ \\
H$\epsilon$~3970\tablefootmark{$\ast$} & $32.8\pm1.5$ & $32.6\pm1.3$ & $32.6\pm2.2$ & $44.1\pm3.2$ & $32.1\pm1.1$ & $83.5\pm2.7$ \\
\ion{He}{I}~4026 & $1.36\pm0.69$ & $1.01\pm0.52$ & $2.19\pm0.72$ & $2.58\pm0.85$ & $2.1\pm0.33$ & $4\pm0.63$ \\
\ion{[S}{II]}~4069 & $1.18\pm0.58$ & $0.9\pm0.44$ & $2.55\pm0.56$ & $3.06\pm0.67$ & ... & ... \\
\ion{[S}{II]}~4076 & $0.21\pm0.29$ & $0.16\pm0.22$ & $0.52\pm0.52$ & $0.63\pm0.63$ & ... & ... \\
H$\delta$~4102 & $27.8\pm1.2$ & $26.02\pm0.98$ & $27.36\pm0.96$ & $37.3\pm1.3$ & $24.8\pm1.1$ & $60.9\pm3.1$ \\
H$\gamma$~4341 & $47.1\pm1.8$ & $51.5\pm1.8$ & $48.1\pm1.6$ & $74.1\pm2.7$ & $45.6\pm1.7$ & $132.4\pm5.8$ \\
\ion{[O}{III]}~4363 & $9.33\pm0.75$ & $8.19\pm0.64$ & $9.11\pm0.66$ & $12.75\pm0.93$ & $19.3\pm0.91$ & $43.2\pm2.1$ \\
\ion{He}{I}~4472 & $3.09\pm0.93$ & $2.95\pm0.89$ & $4.13\pm0.71$ & $6.2\pm1.1$ & $3.47\pm0.43$ & $8.5\pm1.1$ \\
\ion{[Fe}{III]}~4659 & $1.26\pm0.63$ & $1.3\pm0.65$ & $0.37\pm0.21$ & $0.61\pm0.34$ & $0.43\pm0.37$ & $1.11\pm0.95$ \\
\ion{He}{II}~4686 & $3.29\pm0.53$ & $3.57\pm0.57$ & $3.16\pm0.38$ & $5.49\pm0.66$ & $1.35\pm0.31$ & $3.61\pm0.82$ \\
\ion{[Ar}{IV]}~4711 & $0\pm2$ & $4.4\pm2.2$ & $0.97\pm0.22$ & $1.7\pm0.39$ & $3.16\pm0.38$ & $10\pm1$ \\
\ion{He}{I}~4713 & $0.46\pm0.34$ & $0.5\pm0.37$ & $0.45\pm0.88$ & $0.8\pm1.5$ & $0.9\pm0.39$ & $2.4\pm1.1$ \\
\ion{[Ar}{IV]}~4740 & $1.78\pm0.92$ & $0\pm1$ & $0.93\pm0.43$ & $1.65\pm0.77$ & $2.66\pm0.36$ & $10\pm1$ \\
H$\beta$~4861 & $100\pm3.7$ & $142.6\pm6.1$ & $100\pm2.8$ & $240\pm10$ & $100\pm3.5$ & $384\pm18$ \\
\ion{He}{I}~4922 & $1.04\pm0.74$ & $1.3\pm0.92$ & $0.78\pm0.28$ & $1.55\pm0.55$ & $1.19\pm0.18$ & $3.73\pm0.57$ \\
\ion{[O}{III]}~4959 & $176.8\pm5.4$ & $218.9\pm7.2$ & $192.2\pm4.1$ & $378.5\pm7.5$ & $258.2\pm6.6$ & $797\pm15$ \\
\ion{[Fe}{VII]}~4989 & ... & ... & ... & ... & ... & ... \\
\ion{[O}{III]}~5007 & $533\pm15$ & $683\pm20$ & $684\pm14$ & $1393\pm26$ & $794\pm20$ & $2511\pm44$ \\
\ion{[N}{II]}~5755 & $0.8\pm1.3$ & $0\pm2$ & $0.15\pm0.26$ & $0.38\pm0.67$ & ... & ... \\
\ion{He}{I}~5876 & $10.83\pm0.99$ & $18.3\pm1.7$ & $9.65\pm0.64$ & $26.5\pm1.9$ & $10.91\pm0.42$ & $48.5\pm1.7$ \\
\ion{[Fe}{VII]}~6086 & ... & ... & ... & ... & ... & ... \\
\ion{[O}{I]}~6300 & $2.85\pm0.65$ & $5.5\pm1.3$ & $2.95\pm0.52$ & $9.3\pm1.7$ & $1.06\pm0.25$ & $5.5\pm1.3$ \\
\ion{[S}{III]}~6312 & $1.93\pm0.82$ & $3.7\pm1.6$ & $2.77\pm0.64$ & $8.8\pm2.1$ & $0.81\pm0.22$ & $4.2\pm1.1$ \\
\ion{[O}{I]}~6364 & $0.83\pm0.51$ & $1.61\pm0.98$ & $1.39\pm0.52$ & $4.4\pm1.6$ & $0.18\pm0.18$ & $0.92\pm0.97$ \\
\ion{[N}{II]}~6548 & $2.16\pm0.66$ & $4.4\pm1.3$ & $1.43\pm0.26$ & $4.8\pm0.87$ & $1.2\pm0.3$ & $7\pm1.7$ \\
H$\alpha$~6563 & $284.6\pm8.1$ & $649\pm20$ & $289.4\pm6.2$ & $1142\pm31$ & $281.1\pm7.3$ & $1830\pm72$ \\
\ion{[N}{II]}~6583 & $4.8\pm1.3$ & $9.7\pm2.6$ & $4.91\pm0.87$ & $16.5\pm2.9$ & $1.37\pm0.51$ & $7.8\pm2.9$ \\
\ion{He}{I}~6678 & $2.72\pm0.55$ & $5.8\pm1.2$ & $3.1\pm0.3$ & $10.7\pm1.1$ & $3.14\pm0.32$ & $18.4\pm1.9$ \\
\ion{[S}{II]}~6716 & $13.54\pm0.76$ & $28.6\pm1.6$ & $13.26\pm0.55$ & $50\pm2$ & $3.24\pm0.21$ & $19\pm1.2$ \\
\ion{[S}{II]}~6731 & $10\pm2$ & $19\pm4.1$ & $9.11\pm0.89$ & $31.8\pm3.1$ & $2.79\pm0.48$ & $16.4\pm2.8$ \\
\ion{He}{I}~7065 & $2.46\pm0.61$ & $5.6\pm1.4$ & $2.22\pm0.35$ & $8.5\pm1.4$ & $4.25\pm0.36$ & $27.7\pm2.5$ \\
\ion{[Ar}{III]}~7136 & $10\pm1$ & $15\pm2.5$ & $8.21\pm0.83$ & $31.7\pm3.5$ & $3.4\pm0.37$ & $22.4\pm2.5$ \\
\ion{[O}{II]}~7320 & $1.77\pm0.74$ & $4.2\pm1.8$ & $1.7\pm0.3$ & $6.6\pm1.2$ & $0.63\pm0.21$ & $4.4\pm1.5$ \\
\ion{[O}{II]}~7330 & $1.08\pm0.59$ & $2.6\pm1.4$ & $1.27\pm0.37$ & $5.1\pm1.5$ & $0.59\pm0.19$ & $4.1\pm1.4$ \\
\ion{[Ar}{III]}~7751 & $1.4\pm0.9$ & $3.7\pm2.4$ & $1.84\pm0.82$ & $8.2\pm3.7$ & $0.87\pm0.28$ & $6.8\pm2.2$ \\
\ion{[Fe}{II]}~8617 & $0.45\pm0.46$ & $1.5\pm1.5$ & $0.9\pm1.5$ & $5.2\pm9.2$ & ... & ... \\
$F\left(\mathrm{H}\beta\right)\tablefootmark{\dag}$ & \multicolumn{2}{c}{$437\pm11$} & \multicolumn{2}{c}{$612\pm12$} & \multicolumn{2}{c}{$1691\pm42$} \\
$c\left(\mathrm{H}\beta\right)$ & \multicolumn{2}{c}{$0.015$} & \multicolumn{2}{c}{0.027} & \multicolumn{2}{c}{0.141} \\
\multicolumn{7}{c}{}\\

\hline\hline
$ $ & \multicolumn{2}{c}{J1313+6044} & \multicolumn{2}{c}{J1338+4213} & \multicolumn{2}{c}{J1411+0550} \\
line & $I\left(\lambda\right)$ & EW$\left(\lambda\right)$ & $I\left(\lambda\right)$ & EW$\left(\lambda\right)$ & $I\left(\lambda\right)$ & EW$\left(\lambda\right)$ \\
\hline

\ion{[O}{II]}~3726 & $60.5\pm2.1$ & $72.7\pm2.9$ & $74\pm6.4$ & $39.8\pm3.7$ & $22\pm10$ & $34\pm16$ \\
\ion{[O}{II]}~3729 & $2.5\pm3.4$ & $3\pm4.1$ & $46\pm16$ & $23.9\pm8.5$ & $51.4\pm5.2$ & $79.1\pm9.8$ \\
H$\epsilon$~3970\tablefootmark{$\ast$} & $31.3\pm1.2$ & $47.8\pm1.9$ & $25.1\pm2.5$ & $17.9\pm1.8$ & $33\pm1.5$ & $69\pm3.5$ \\
\ion{He}{I}~4026 & $1.54\pm0.31$ & $1.94\pm0.39$ & $1.9\pm1.1$ & $0.97\pm0.57$ & $0.69\pm0.33$ & $1.1\pm0.53$ \\
\ion{[S}{II]}~4069 & $0.5\pm1.4$ & $0.7\pm1.8$ & $0.5\pm0.6$ & $0.27\pm0.32$ & $0.88\pm0.54$ & $1.44\pm0.89$ \\
\ion{[S}{II]}~4076 & $0.5\pm0.3$ & $0.6\pm0.4$ & $0.95\pm0.52$ & $0.51\pm0.28$ & $0.7\pm0.7$ & $1.1\pm1.2$ \\
H$\delta$~4102 & $26.27\pm0.63$ & $43.3\pm1.1$ & $25.5\pm1.3$ & $18.56\pm0.79$ & $25.2\pm1.3$ & $54\pm3.1$ \\
H$\gamma$~4341 & $47\pm0.9$ & $90\pm2$ & $48.7\pm2.1$ & $40.5\pm1.4$ & $46.1\pm1.6$ & $101.6\pm3.8$ \\
\ion{[O}{III]}~4363 & $14.73\pm0.43$ & $22.52\pm0.65$ & $6.12\pm0.71$ & $3.66\pm0.41$ & $13.8\pm0.7$ & $26.2\pm1.3$ \\
\ion{He}{I}~4472 & $3.61\pm0.38$ & $5.96\pm0.63$ & $3.4\pm1.5$ & $0\pm1$ & $3.59\pm0.74$ & $7.2\pm1.5$ \\
\ion{[Fe}{III]}~4659 & $0.28\pm0.12$ & $0.5\pm0.22$ & $0.15\pm0.31$ & $0.1\pm0.22$ & $0.43\pm0.23$ & $0.9\pm0.5$ \\
\ion{He}{II}~4686 & $2.9\pm0.2$ & $5.34\pm0.37$ & $3.38\pm0.62$ & $2.43\pm0.44$ & $1.92\pm0.33$ & $4.37\pm0.74$ \\
\ion{[Ar}{IV]}~4711 & $1.99\pm0.18$ & $3.75\pm0.33$ & $0.96\pm0.41$ & $0.7\pm0.3$ & $1.76\pm0.27$ & $4.05\pm0.63$ \\
\ion{He}{I}~4713 & $0.05\pm0.43$ & $0.1\pm0.81$ & $1.4\pm2.8$ & $1\pm2.1$ & $0.52\pm0.35$ & $1.2\pm0.8$ \\
\ion{[Ar}{IV]}~4740 & $0.94\pm0.16$ & $1.79\pm0.32$ & $2\pm1.7$ & $1.5\pm1.3$ & $1.52\pm0.34$ & $3.55\pm0.79$ \\
H$\beta$~4861 & $100\pm1.8$ & $271\pm7.9$ & $100\pm4.5$ & $112.3\pm5.8$ & $100\pm3.3$ & $321\pm18$ \\
\ion{He}{I}~4922 & $0.63\pm0.17$ & $1.35\pm0.36$ & $0\pm1$ & $2.06\pm0.86$ & $0.9\pm0.4$ & $0\pm1$ \\
\ion{[O}{III]}~4959 & $200.2\pm2.8$ & $426.7\pm6.4$ & $89.9\pm3.2$ & $73.7\pm1.7$ & $249.6\pm6.2$ & $642\pm19$ \\
\ion{[Fe}{VII]}~4989 & ... & ... & ... & ... & ... & ... \\
\ion{[O}{III]}~5007 & $606.6\pm8.1$ & $1333\pm20$ & $278.1\pm9.2$ & $237.5\pm4.6$ & $741\pm18$ & $1954\pm56$ \\
\ion{[N}{II]}~5755 & $0.18\pm0.28$ & $0.52\pm0.82$ & ... & ... & $0.13\pm0.26$ & $0.46\pm0.89$ \\
\ion{He}{I}~5876 & $10.35\pm0.44$ & $32.4\pm1.5$ & $10\pm1$ & $10.9\pm1.2$ & $12.08\pm0.54$ & $43.9\pm1.9$ \\
\ion{[Fe}{VII]}~6086 & ... & ... & ... & ... & ... & ... \\
\ion{[O}{I]}~6300 & $2.15\pm0.27$ & $10\pm1$ & $1.82\pm0.88$ & $2.5\pm1.2$ & $1.35\pm0.34$ & $5.6\pm1.4$ \\
\ion{[S}{III]}~6312 & $1.19\pm0.43$ & $4.3\pm1.6$ & $1.01\pm0.77$ & $1.4\pm1.1$ & $1.62\pm0.43$ & $6.7\pm1.8$ \\
\ion{[O}{I]}~6364 & $0.87\pm0.29$ & $3.2\pm1.1$ & $0.21\pm0.32$ & $0.29\pm0.45$ & $0.47\pm0.31$ & $2\pm1.3$ \\
\ion{[N}{II]}~6548 & $0.79\pm0.26$ & $3.2\pm1.1$ & $0.36\pm0.53$ & $0.56\pm0.82$ & $1.58\pm0.31$ & $7.2\pm1.4$ \\
H$\alpha$~6563 & $301.5\pm4.4$ & $1432\pm37$ & $278.7\pm9.3$ & $521\pm14$ & $297.5\pm7.1$ & $1606\pm44$ \\
\ion{[N}{II]}~6583 & $1.9\pm0.46$ & $7.6\pm1.9$ & $2.65\pm0.77$ & $4\pm1.2$ & $4.9\pm1.4$ & $22\pm6.4$ \\
\ion{He}{I}~6678 & $2.98\pm0.25$ & $10\pm1$ & $2.65\pm0.55$ & $4.13\pm0.86$ & $3.09\pm0.26$ & $14.2\pm1.2$ \\
\ion{[S}{II]}~6716 & $6.18\pm0.35$ & $25.4\pm1.5$ & $5.57\pm0.74$ & $8.7\pm1.1$ & $4.38\pm0.35$ & $20.2\pm1.6$ \\
\ion{[S}{II]}~6731 & $4.42\pm0.85$ & $18.2\pm3.5$ & $4\pm0.83$ & $6.2\pm1.3$ & $4.19\pm0.72$ & $19.3\pm3.3$ \\
\ion{He}{I}~7065 & $4\pm0.4$ & $18.2\pm1.9$ & $2.65\pm0.53$ & $4.58\pm0.91$ & $4.85\pm0.61$ & $24.7\pm3.3$ \\
\ion{[Ar}{III]}~7136 & $3.8\pm0.4$ & $17.5\pm1.9$ & $2.56\pm0.67$ & $4.5\pm1.2$ & $5.04\pm0.54$ & $25.9\pm2.9$ \\
\ion{[O}{II]}~7320 & $1.18\pm0.39$ & $10\pm2$ & $1.05\pm0.53$ & $1.94\pm0.98$ & $1.42\pm0.42$ & $7.7\pm2.3$ \\
\ion{[O}{II]}~7330 & $1.06\pm0.33$ & $5.2\pm1.6$ & $1.26\pm0.71$ & $2.3\pm1.3$ & $0.95\pm0.37$ & $10\pm2$ \\
\ion{[Ar}{III]}~7751 & $1.24\pm0.37$ & $6.9\pm2.1$ & $1.49\pm0.95$ & $0\pm2$ & $1.24\pm0.41$ & $7.5\pm2.5$ \\
\ion{[Fe}{II]}~8617 & $0.046\pm0.078$ & $0.36\pm0.61$ & $1.1\pm2.3$ & $2.8\pm5.9$ & $0.38\pm0.61$ & $3.4\pm5.4$ \\
$F\left(\mathrm{H}\beta\right)\tablefootmark{\dag}$ & \multicolumn{2}{c}{$1720\pm22$} & \multicolumn{2}{c}{$424\pm13$} & \multicolumn{2}{c}{$1953\pm45$} \\
$c\left(\mathrm{H}\beta\right)$ & \multicolumn{2}{c}{$0.170$} & \multicolumn{2}{c}{0.080} & \multicolumn{2}{c}{0.190} \\
\multicolumn{7}{c}{}\\

\hline\hline
$ $ & \multicolumn{2}{c}{J1528+2318} & \multicolumn{2}{c}{J1556+1818} & \multicolumn{2}{c}{J1608+0413} \\
line & $I\left(\lambda\right)$ & EW$\left(\lambda\right)$ & $I\left(\lambda\right)$ & EW$\left(\lambda\right)$ & $I\left(\lambda\right)$ & EW$\left(\lambda\right)$ \\
\hline

\ion{[O}{II]}~3726 & $6\pm14$ & $5\pm12$ & $2.7\pm9.1$ & $2.6\pm8.7$ & $31.6\pm3.1$ & $33.5\pm3.2$ \\
\ion{[O}{II]}~3729 & $76\pm40$ & $62\pm32$ & $82.7\pm4.3$ & $79.4\pm4.7$ & $43.3\pm4.2$ & $45.1\pm4.2$ \\
H$\epsilon$~3970\tablefootmark{$\ast$} & $33.7\pm1.7$ & $42.2\pm2.2$ & $30.1\pm2.8$ & $40\pm4$ & $29.5\pm2.5$ & $46.8\pm4.1$ \\
\ion{He}{I}~4026 & $2.06\pm0.77$ & $1.83\pm0.69$ & $1.73\pm0.82$ & $1.81\pm0.87$ & $1.09\pm0.82$ & $1.17\pm0.88$ \\
\ion{[S}{II]}~4069 & $1.09\pm0.62$ & $0.99\pm0.57$ & $0.69\pm0.55$ & $0.73\pm0.58$ & ... & ... \\
\ion{[S}{II]}~4076 & $1.1\pm0.99$ & $1.01\pm0.91$ & $0.7\pm3.5$ & $0.7\pm3.8$ & ... & ... \\
H$\delta$~4102 & $27.8\pm1.4$ & $33\pm1.8$ & $25.8\pm1.5$ & $38\pm2.5$ & $30\pm2$ & $39.7\pm3.1$ \\
H$\gamma$~4341 & $46.7\pm1.9$ & $63.6\pm3.1$ & $47.2\pm2.5$ & $76.8\pm5.5$ & $46.5\pm2.4$ & $80.3\pm3.9$ \\
\ion{[O}{III]}~4363 & $12.89\pm0.99$ & $13.5\pm1.1$ & $12.8\pm1.3$ & $16.6\pm1.8$ & $12.52\pm0.92$ & $15.7\pm1.1$ \\
\ion{He}{I}~4472 & $4.5\pm0.85$ & $5.15\pm0.98$ & $0\pm1$ & $6.2\pm1.5$ & $3.5\pm1.3$ & $4.8\pm1.8$ \\
\ion{[Fe}{III]}~4659 & $0.2\pm0.38$ & $0.24\pm0.46$ & $0.55\pm0.99$ & $0.9\pm1.6$ & $0.54\pm0.34$ & $0.8\pm0.5$ \\
\ion{He}{II}~4686 & $2.69\pm0.68$ & $3.44\pm0.87$ & $2.09\pm0.92$ & $3.5\pm1.5$ & $3.01\pm0.47$ & $4.53\pm0.69$ \\
\ion{[Ar}{IV]}~4711 & $1.89\pm0.95$ & $2.4\pm1.2$ & $1.66\pm0.67$ & $2.8\pm1.1$ & $0\pm2$ & $0\pm3$ \\
\ion{He}{I}~4713 & $0.6\pm0.46$ & $0.77\pm0.59$ & $0.13\pm0.49$ & $0.23\pm0.83$ & $1.1\pm0.41$ & $1.67\pm0.63$ \\
\ion{[Ar}{IV]}~4740 & $0.7\pm0.4$ & $0.87\pm0.52$ & $1.03\pm0.51$ & $1.79\pm0.88$ & $5.5\pm4.1$ & $8.5\pm6.3$ \\
H$\beta$~4861 & $100\pm2.6$ & $180\pm6$ & $100\pm2.8$ & $239\pm10$ & $100\pm5.5$ & $230\pm19$ \\
\ion{He}{I}~4922 & $1.14\pm0.43$ & $1.67\pm0.63$ & $0.8\pm0.4$ & $1.61\pm0.78$ & $0.94\pm0.29$ & $1.65\pm0.51$ \\
\ion{[O}{III]}~4959 & $213.8\pm4.2$ & $304.3\pm5.1$ & $180\pm4$ & $350.2\pm9.3$ & $165.5\pm6.6$ & $285.8\pm5.4$ \\
\ion{[Fe}{VII]}~4989 & ... & ... & ... & ... & ... & ... \\
\ion{[O}{III]}~5007 & $650\pm12$ & $962\pm15$ & $533\pm11$ & $1086\pm28$ & $518\pm20$ & $920\pm17$ \\
\ion{[N}{II]}~5755 & $1.3\pm2.7$ & $2.3\pm5.1$ & $0.17\pm0.48$ & $0.5\pm1.4$ & ... & ... \\
\ion{He}{I}~5876 & $10.48\pm0.96$ & $20\pm2$ & ... & ... & ... & ... \\
\ion{[Fe}{VII]}~6086 & ... & ... & ... & ... & ... & ... \\
\ion{[O}{I]}~6300 & $2.49\pm0.59$ & $5.6\pm1.3$ & $2.22\pm0.45$ & $8.4\pm1.7$ & $2.5\pm0.6$ & $7.2\pm1.7$ \\
\ion{[S}{III]}~6312 & $2.04\pm0.75$ & $4.5\pm1.7$ & $1.3\pm0.51$ & $4.9\pm1.9$ & $0.89\pm0.43$ & $2.6\pm1.2$ \\
\ion{[O}{I]}~6364 & $1.8\pm1.4$ & $0\pm3$ & $0.84\pm0.58$ & $3.2\pm2.2$ & $0.52\pm0.46$ & $1.5\pm1.3$ \\
\ion{[N}{II]}~6548 & $1.11\pm0.42$ & $0\pm1$ & $1.4\pm0.3$ & $5.8\pm1.3$ & $1.16\pm0.52$ & $3.8\pm1.7$ \\
H$\alpha$~6563 & $295.6\pm6.1$ & $766\pm19$ & $268.3\pm5.6$ & $1282\pm30$ & $300\pm12$ & $1195\pm57$ \\
\ion{[N}{II]}~6583 & $2.79\pm0.38$ & $6.52\pm0.89$ & $2.52\pm0.95$ & $10\pm4$ & $0\pm1$ & $5.3\pm3.3$ \\
\ion{He}{I}~6678 & $3.49\pm0.42$ & $10\pm1$ & $2.1\pm0.2$ & $9.37\pm0.88$ & $3.51\pm0.56$ & $11.4\pm1.8$ \\
\ion{[S}{II]}~6716 & $8.37\pm0.56$ & $20.4\pm1.4$ & $6.23\pm0.43$ & $27.1\pm1.9$ & $7.87\pm0.65$ & $30\pm2$ \\
\ion{[S}{II]}~6731 & $6.37\pm0.49$ & $15.5\pm1.2$ & $4.2\pm0.4$ & $18.5\pm1.8$ & $5.84\pm0.54$ & $19\pm1.7$ \\
\ion{He}{I}~7065 & $2.72\pm0.59$ & $7.2\pm1.6$ & $2.94\pm0.64$ & $14.6\pm3.3$ & $2.65\pm0.79$ & $9.6\pm2.9$ \\
\ion{[Ar}{III]}~7136 & $5.5\pm0.96$ & $14.7\pm2.6$ & $3.14\pm0.88$ & $15.8\pm4.7$ & $4.2\pm1.3$ & $15.4\pm4.9$ \\
\ion{[O}{II]}~7320 & $1.7\pm0.73$ & $0\pm2$ & $1.31\pm0.59$ & $7\pm3.2$ & $1.38\pm0.74$ & $5.3\pm2.9$ \\
\ion{[O}{II]}~7330 & $1.5\pm0.65$ & $4.1\pm1.8$ & $0.6\pm0.6$ & $3.3\pm3.2$ & $1.56\pm0.65$ & $6\pm2.6$ \\
\ion{[Ar}{III]}~7751 & $1.5\pm0.6$ & $4.7\pm1.8$ & $0.4\pm0.6$ & $2.6\pm3.7$ & $0.71\pm0.67$ & $3.1\pm2.9$ \\
\ion{[Fe}{II]}~8617 & $0.17\pm0.27$ & $0.7\pm1.1$ & $0.01\pm0.25$ & $0.1\pm2.1$ & $0\pm0.15$ & $0.01\pm0.84$ \\
$F\left(\mathrm{H}\beta\right)\tablefootmark{\dag}$ & \multicolumn{2}{c}{$563\pm11$} & \multicolumn{2}{c}{$771\pm15$} & \multicolumn{2}{c}{$482\pm19$} \\
$c\left(\mathrm{H}\beta\right)$ & \multicolumn{2}{c}{$0.065$} & \multicolumn{2}{c}{0.015} & \multicolumn{2}{c}{0.082} \\
\multicolumn{7}{c}{}\\

\hline
\end{longtable}

\begin{longtable}[c]{p{0.16\textwidth-2\tabcolsep}*{6}{p{0.14\textwidth-2\tabcolsep}}}
\caption{\label{tab_linesoutl}Same as Table \ref{tab_lines}, but for the four galaxies identified as outliers.} \\

\hline\hline
$ $ & \multicolumn{2}{c}{J0225+0033} & \multicolumn{2}{c}{J1152+4005} & \multicolumn{2}{c}{J1514+2840} \\
line & $I\left(\lambda\right)$ & EW$\left(\lambda\right)$ & $I\left(\lambda\right)$ & EW$\left(\lambda\right)$ & $I\left(\lambda\right)$ & EW$\left(\lambda\right)$ \\
\hline

\endfirsthead
\caption{continued.}\\

line & $I\left(\lambda\right)$ & EW$\left(\lambda\right)$ &  &  &  &  \\
\hline

\endhead

\hline
\endfoot
\multicolumn{7}{c}{\tablefoot{
Line fluxes are scaled to H$\beta$ as $I\left(\lambda\right)=100\times F\left(\lambda\right)/F\left(\mathrm{H}\beta\right)$. Equivalent widths EW$\left(\lambda\right)$ are given in $\AA$.
\tablefoottext{$\ast$}{Likely blend with \ion{[Ne}{III]}~3967, which constitutes $\sim50\%$ of the reported flux (Sect. \ref{sec_bdec}).}
\tablefoottext{$\dag$}{H$\beta$ line flux in $10^{-17}~\mathrm{erg}~\mathrm{s}^{-1}~\mathrm{cm}^{-2}$ for reference.}
}}
\endlastfoot

\ion{[O}{II]}~3726 & $24\pm51$ & $6\pm13$ & $199\pm57$ & $104.3\pm8.4$ & $175.2\pm8.6$ & $113.5\pm6.4$ \\
\ion{[O}{II]}~3729 & $290\pm140$ & $73\pm36$ & $21\pm18$ & $10.7\pm8.4$ & $14.4\pm6.3$ & $9.2\pm4.1$ \\
H$\epsilon$~3970\tablefootmark{$\ast$} & $20\pm3$ & $7.07\pm0.97$ & $59\pm27$ & $39\pm16$ & $30.2\pm1.6$ & $29.2\pm1.4$ \\
\ion{He}{I}~4026 & $7\pm12$ & $1.5\pm2.5$ & $29\pm23$ & $12.8\pm9.6$ & $2\pm1.2$ & $1.46\pm0.89$ \\
\ion{[S}{II]}~4069 & $2.6\pm1.5$ & $0.56\pm0.32$ & $8.5\pm4.3$ & $3.8\pm1.6$ & $3.5\pm1.4$ & $0\pm1$ \\
\ion{[S}{II]}~4076 & $2.4\pm2.7$ & $0.5\pm0.6$ & $5.9\pm4.2$ & $2.8\pm1.9$ & $4.2\pm3.3$ & $3.2\pm2.5$ \\
H$\delta$~4102 & $25.8\pm3.1$ & $7.98\pm0.88$ & $32\pm10$ & $15.6\pm2.5$ & $26.7\pm1.2$ & $25.1\pm1.1$ \\
H$\gamma$~4341 & $46\pm14$ & $20\pm5$ & $45\pm20$ & $23.9\pm9.1$ & $46.9\pm2.1$ & $48.6\pm2.2$ \\
\ion{[O}{III]}~4363 & $7\pm33$ & $1.7\pm8.1$ & $24\pm17$ & $10.1\pm6.8$ & $6.9\pm0.99$ & $6.06\pm0.87$ \\
\ion{He}{I}~4472 & $21\pm19$ & $5.4\pm4.9$ & $36\pm44$ & $15\pm17$ & $0\pm1$ & $0\pm1$ \\
\ion{[Fe}{III]}~4659 & $1.4\pm3.7$ & $0\pm1$ & $16\pm12$ & $6.5\pm4.4$ & $0.4\pm0.33$ & $0.42\pm0.34$ \\
\ion{He}{II}~4686 & $4\pm1.1$ & $1.12\pm0.32$ & $42\pm17$ & $18.1\pm5.3$ & $2.5\pm0.5$ & $2.73\pm0.56$ \\
\ion{[Ar}{IV]}~4711 & ... & ... & $0\pm30$ & $0\pm13$ & $3\pm1.4$ & $3.3\pm1.5$ \\
\ion{He}{I}~4713 & ... & ... & $25\pm19$ & $10.3\pm7.3$ & $0.93\pm0.42$ & $1.04\pm0.47$ \\
\ion{[Ar}{IV]}~4740 & ... & ... & $29\pm22$ & $12.3\pm8.6$ & $4.2\pm1.6$ & $4.7\pm1.9$ \\
H$\beta$~4861 & $100\pm7.2$ & $44.4\pm2.8$ & $100\pm40$ & $55\pm18$ & $100\pm3.8$ & $149.3\pm7.1$ \\
\ion{He}{I}~4922 & $0\pm1$ & $0.18\pm0.33$ & $10.4\pm9.3$ & $4.5\pm3.9$ & $1.4\pm1.6$ & $1.8\pm2.1$ \\
\ion{[O}{III]}~4959 & $91.4\pm5.6$ & $28.9\pm1.1$ & $344\pm99$ & $145\pm13$ & $174.7\pm5.6$ & $223.7\pm8.4$ \\
\ion{[Fe}{VII]}~4989 & ... & ... & ... & ... & ... & ... \\
\ion{[O}{III]}~5007 & $282\pm15$ & $92.2\pm2.3$ & $1030\pm290$ & $466\pm35$ & $521\pm15$ & $691\pm24$ \\
\ion{[N}{II]}~5755 & $2.8\pm3.4$ & $1.2\pm1.4$ & ... & ... & ... & ... \\
\ion{He}{I}~5876 & $8.1\pm2.2$ & $3.53\pm0.92$ & ... & ... & $10.2\pm1.1$ & $20\pm2$ \\
\ion{[Fe}{VII]}~6086 & ... & ... & ... & ... & ... & ... \\
\ion{[O}{I]}~6300 & $10.4\pm3.1$ & $5.2\pm1.5$ & $27\pm20$ & $13.2\pm8.8$ & $3.32\pm0.62$ & $7.1\pm1.3$ \\
\ion{[S}{III]}~6312 & $0.7\pm1.9$ & $0.37\pm0.97$ & $4.6\pm9.4$ & $2.2\pm4.6$ & $1.65\pm0.73$ & $3.5\pm1.6$ \\
\ion{[O}{I]}~6364 & $1.9\pm2.6$ & $1\pm1.3$ & $27\pm29$ & $13\pm13$ & $1.43\pm0.58$ & $3.1\pm1.3$ \\
\ion{[N}{II]}~6548 & $9.5\pm1.2$ & $5.31\pm0.64$ & $29\pm27$ & $14\pm12$ & $3.79\pm0.73$ & $8.6\pm1.7$ \\
H$\alpha$~6563 & $280\pm15$ & $195.6\pm4.4$ & $330\pm110$ & $198\pm48$ & $289.5\pm8.9$ & $758\pm37$ \\
\ion{[N}{II]}~6583 & $32.2\pm2.2$ & $17.68\pm0.84$ & $43\pm28$ & $20\pm12$ & $13\pm1.2$ & $29.5\pm2.8$ \\
\ion{He}{I}~6678 & $2.8\pm1.1$ & $1.55\pm0.61$ & $21\pm52$ & $10\pm25$ & $3.86\pm0.74$ & $9.2\pm1.8$ \\
\ion{[S}{II]}~6716 & $30\pm4$ & $19.5\pm2.1$ & $45\pm33$ & $23\pm16$ & $18.9\pm1.3$ & $44.7\pm3.4$ \\
\ion{[S}{II]}~6731 & $24.5\pm5.5$ & $10\pm3$ & $40\pm31$ & $20\pm15$ & $15.6\pm2.9$ & $40\pm7$ \\
\ion{He}{I}~7065 & $0\pm1.1$ & $0.01\pm0.66$ & $37\pm43$ & $19\pm22$ & $2.5\pm1.2$ & $6.6\pm3.3$ \\
\ion{[Ar}{III]}~7136 & $6.1\pm9.5$ & $3.8\pm5.9$ & $23\pm26$ & $12\pm15$ & $6.3\pm1.5$ & $20\pm4$ \\
\ion{[O}{II]}~7320 & $2.5\pm4.6$ & $0\pm3$ & $38\pm19$ & $19\pm9.3$ & $2.36\pm0.97$ & $6.5\pm2.7$ \\
\ion{[O}{II]}~7330 & $9.6\pm6.2$ & $6.3\pm4.1$ & $10\pm30$ & $5\pm15$ & $1.4\pm1.2$ & $3.9\pm3.2$ \\
\ion{[Ar}{III]}~7751 & $2.2\pm7.1$ & $1.6\pm5.3$ & $70\pm190$ & $40\pm110$ & $1.6\pm1.2$ & $5\pm3.6$ \\
\ion{[Fe}{II]}~8617 & ... & ... & ... & ... & $0.4\pm2.7$ & $2\pm11$ \\
$F\left(\mathrm{H}\beta\right)\tablefootmark{\dag}$ & \multicolumn{2}{c}{$155\pm8$} & \multicolumn{2}{c}{$45\pm13$} & \multicolumn{2}{c}{$277\pm7$} \\
$c\left(\mathrm{H}\beta\right)$ & \multicolumn{2}{c}{$0.173$} & \multicolumn{2}{c}{0.185} & \multicolumn{2}{c}{0.028} \\
\multicolumn{7}{c}{}\\

\hline\hline
$ $ & \multicolumn{2}{c}{J1657+4135} & \multicolumn{2}{c}{} & \multicolumn{2}{c}{} \\
line & $I\left(\lambda\right)$ & EW$\left(\lambda\right)$ &  &  &  &  \\
\hline

\ion{[O}{II]}~3726 & $167\pm20$ & $16.5\pm1.7$ & $ $ & $ $ \\
\ion{[O}{II]}~3729 & $114\pm22$ & $11.1\pm2.1$ & $ $ & $ $ \\
H$\epsilon$~3970\tablefootmark{$\ast$} & $11.6\pm6.4$ & $1.77\pm0.97$ & $ $ & $ $ \\
\ion{He}{I}~4026 & $5\pm10$ & $0.36\pm0.77$ & $ $ & $ $ \\
\ion{[S}{II]}~4069 & $10\pm4$ & $0.4\pm0.31$ & $ $ & $ $ \\
\ion{[S}{II]}~4076 & $15\pm17$ & $1.2\pm1.4$ & $ $ & $ $ \\
H$\delta$~4102 & $26\pm3.5$ & $3.07\pm0.39$ & $ $ & $ $ \\
H$\gamma$~4341 & $53.5\pm6.1$ & $7.12\pm0.73$ & $ $ & $ $ \\
\ion{[O}{III]}~4363 & $24\pm20$ & $2\pm1.7$ & $ $ & $ $ \\
\ion{He}{I}~4472 & $7.8\pm4.9$ & $0.67\pm0.42$ & $ $ & $ $ \\
\ion{[Fe}{III]}~4659 & ... & ... & $ $ & $ $ \\
\ion{He}{II}~4686 & ... & ... & $ $ & $ $ \\
\ion{[Ar}{IV]}~4711 & ... & ... & $ $ & $ $ \\
\ion{He}{I}~4713 & ... & ... & $ $ & $ $ \\
\ion{[Ar}{IV]}~4740 & ... & ... & $ $ & $ $ \\
H$\beta$~4861 & $100\pm7.9$ & $14.51\pm0.84$ & $ $ & $ $ \\
\ion{He}{I}~4922 & $1.1\pm6.3$ & $0.11\pm0.65$ & $ $ & $ $ \\
\ion{[O}{III]}~4959 & $18\pm4.3$ & $1.82\pm0.42$ & $ $ & $ $ \\
\ion{[Fe}{VII]}~4989 & ... & ... & $ $ & $ $ \\
\ion{[O}{III]}~5007 & $53.7\pm5.2$ & $5.69\pm0.46$ & $ $ & $ $ \\
\ion{[N}{II]}~5755 & $4\pm10$ & $0.5\pm1.2$ & $ $ & $ $ \\
\ion{He}{I}~5876 & $10\pm1$ & $1.3\pm0.1$ & $ $ & $ $ \\
\ion{[Fe}{VII]}~6086 & ... & ... & $ $ & $ $ \\
\ion{[O}{I]}~6300 & $7.1\pm1.4$ & $1\pm0.2$ & $ $ & $ $ \\
\ion{[S}{III]}~6312 & $1\pm1.5$ & $0.14\pm0.22$ & $ $ & $ $ \\
\ion{[O}{I]}~6364 & $3.4\pm2.1$ & $0.5\pm0.3$ & $ $ & $ $ \\
\ion{[N}{II]}~6548 & $28.5\pm2.4$ & $4.33\pm0.27$ & $ $ & $ $ \\
H$\alpha$~6563 & $311\pm18$ & $58.88\pm0.89$ & $ $ & $ $ \\
\ion{[N}{II]}~6583 & $80\pm5$ & $12.11\pm0.32$ & $ $ & $ $ \\
\ion{He}{I}~6678 & $3.2\pm1.4$ & $0.49\pm0.22$ & $ $ & $ $ \\
\ion{[S}{II]}~6716 & $52\pm3.1$ & $8.03\pm0.18$ & $ $ & $ $ \\
\ion{[S}{II]}~6731 & $37.6\pm3.8$ & $5.72\pm0.48$ & $ $ & $ $ \\
\ion{He}{I}~7065 & $0.8\pm2.1$ & $0.14\pm0.34$ & $ $ & $ $ \\
\ion{[Ar}{III]}~7136 & $6.8\pm2.5$ & $1.13\pm0.41$ & $ $ & $ $ \\
\ion{[O}{II]}~7320 & $0.8\pm1.3$ & $0.14\pm0.22$ & $ $ & $ $ \\
\ion{[O}{II]}~7330 & $2\pm6.5$ & $0.3\pm1.1$ & $ $ & $ $ \\
\ion{[Ar}{III]}~7751 & $2\pm3.1$ & $0.37\pm0.59$ & $ $ & $ $ \\
\ion{[Fe}{II]}~8617 & ... & ... & $ $ & $ $ \\
$F\left(\mathrm{H}\beta\right)\tablefootmark{\dag}$ & \multicolumn{2}{c}{$605\pm34$} & \multicolumn{2}{c}{} & \multicolumn{2}{c}{} \\
$c\left(\mathrm{H}\beta\right)$ & \multicolumn{2}{c}{$0.294$} & \multicolumn{2}{c}{} & \multicolumn{2}{c}{} \\
\multicolumn{7}{c}{}\\

\hline                  
\end{longtable}

\end{appendix}

\end{document}